\documentclass[modern]{aastex61}

\usepackage{microtype}
\usepackage{url}
\usepackage{amsmath}
\usepackage{mathtools}
\usepackage{esint}
\usepackage{amssymb}
\usepackage{natbib}
\usepackage{multirow}
\usepackage{graphicx}
\usepackage{scalerel}
\usepackage{calc}
\usepackage{etoolbox}
\usepackage{marginnote}
\usepackage{nicefrac}
\usepackage{tabstackengine}
\usepackage{diagbox}
\usepackage[makeroom]{cancel}
\usepackage{mathdots}
\usepackage{bbm}
\usepackage{booktabs}
\usepackage{xspace}
\usepackage{upgreek}
\usepackage[T1]{fontenc} 
\usepackage{textcomp}
\usepackage{ifxetex}
\ifxetex
\usepackage{fontspec}
\defaultfontfeatures{Extension = .otf}
\fi
\usepackage{fontawesome}
\usepackage{listings}
\usepackage{mathtools}
\stackMath

\definecolor{editcolor}{rgb}{0,0,0}
\newcommand{\edited}[1]{#1}

\newcommand{\starry}{\textsf{starry}\xspace}
\newcommand{\exocartographer}{\textsf{exocartographer}\xspace}

\newcommand{\pybind}{\textsf{pybind11}\xspace}
\newcommand{\batman}{\textsf{batman}\xspace}
\newcommand{\spiderman}{\textsf{spiderman}\xspace}

\newcommand{\Python}{\textsf{Python}\xspace}
\newcommand{\cpp}{\textsf{C}++\xspace}
\newcommand{\Jupyter}{\textsf{Jupyter}\xspace}

\newcommand{\Map}{\href{https://rodluger.github.io/starry/api.html\#starry.Map}{\color{black}\textsf{Map}}\xspace}
\newcommand{\Secondary}{\href{https://rodluger.github.io/starry/api.html\#starry.kepler.Secondary}{\color{black}\textsf{kepler.Secondary}}\xspace}
\newcommand{\Primary}{\href{https://rodluger.github.io/starry/api.html\#starry.kepler.Primary}{\color{black}\textsf{kepler.Primary}}\xspace}
\newcommand{\System}{\href{https://rodluger.github.io/starry/api.html\#starry.kepler.System}{\color{black}\textsf{kepler.System}}\xspace}
\newcommand{\Mapy}{\href{https://rodluger.github.io/starry/api.html\#starry.Map.y}{\color{black}\textsf{Map.y}}\xspace}
\newcommand{\Mapp}{\href{https://rodluger.github.io/starry/api.html\#starry.Map.p}{\color{black}\textsf{Map.p}}\xspace}
\newcommand{\Mapg}{\href{https://rodluger.github.io/starry/api.html\#starry.Map.g}{\color{black}\textsf{Map.g}}\xspace}
\newcommand{\Mapu}{\href{https://rodluger.github.io/starry/api.html\#starry.Map.u}{\color{black}\textsf{Map.u}}\xspace}
\newcommand{\Mapaxis}{\href{https://rodluger.github.io/starry/api.html\#starry.Map.axis}{\color{black}\textsf{Map.axis}}\xspace}
\newcommand{\Mapflux}{\href{https://rodluger.github.io/starry/api.html\#starry.Map.flux}{\color{black}\textsf{Map.flux()}}\xspace}
\newcommand{\SecondaryL}{\href{https://rodluger.github.io/starry/api.html\#starry.kepler.Secondary.L}{\color{black}\textsf{kepler.Secondary.L}}\xspace}
\newcommand{\Systemlightcurve}{\href{https://rodluger.github.io/starry/api.html\#starry.kepler.System.lightcurve}{\color{black}\textsf{kepler.System.lightcurve}}\xspace}
\newcommand{\Systemcompute}{\href{https://rodluger.github.io/starry/api.html\#starry.kepler.System.compute}{\color{black}\textsf{kepler.System.compute()}}\xspace}

\renewcommand{\eqref}[1]{\ref{eq:#1}}
\newcommand{\Eq}[1]{Equation~(\eqref{#1})}
\newcommand{\eq}[1]{\Eq{#1}}

\definecolor{linkcolor}{rgb}{0.1216,0.4667,0.7059}
\newcommand{\codeicon}{{\color{linkcolor}\faFileCodeO}}
\newcommand{\prooficon}{{\color{linkcolor}\faPencilSquareO}}
\newcommand{\animicon}{{\color{linkcolor}\faPlayCircle}}
\newcommand{\codelink}[1]{\href{https://github.com/rodluger/starry/blob/v0.2.2/tex/figures/#1.py}{\codeicon}\,\,}
\newcommand{\animlink}[1]{\href{https://github.com/rodluger/starry/blob/v0.2.2/tex/figures/#1.gif}{\animicon}\,\,}
\newcommand{\prooflink}[1]{\href{https://github.com/rodluger/starry/blob/v0.2.2/docs/proofs/#1.ipynb}{\raisebox{-0.1em}{\prooficon}}}

\newtagform{eqtag}[]{(}{)}
\newcommand{\currentlabel}{None}

\newenvironment{proof}[1]{%
\ifstrempty{#1}{%
\renewtagform{eqtag}[]{\raisebox{-0.1em}{{\color{red}\faPencilSquareO}}\,(}{)}%
}{%
\renewtagform{eqtag}[]{\prooflink{#1}\,(}{)}%
}%
\usetagform{eqtag}%
\renewcommand{\currentlabel}{#1}
\align%
}{%
\endalign%
\renewtagform{eqtag}[]{(}{)}%
\usetagform{eqtag}%
\message{<<<\currentlabel: \theequation>>>}%
}

\newenvironment{proof*}[1]{%
\ifstrempty{#1}{%
\renewtagform{eqtag}[]{\raisebox{-0.1em}{{\color{red}\faPencilSquareO}}\,(}{)}%
}{%
\renewtagform{eqtag}[]{\prooflink{#1}\,(}{)}%
}%
\usetagform{eqtag}%
\renewcommand{\currentlabel}{#1}
\equation%
}{%
\endequation%
\renewtagform{eqtag}[]{(}{)}%
\usetagform{eqtag}%
\message{<<<\currentlabel: \theequation>>>}
}

\newcommand{\ii}{\ensuremath{\mathbf{i}}}
\newcommand{\dd}{\ensuremath{ \mathrm{d}}}

\newcommand{\bvec}[1]{{\ensuremath{\mathbf{#1}}}}

\newcommand{\x}{\ensuremath{\mbox{$x$}}}
\newcommand{\y}{\ensuremath{\mbox{$y$}}}
\newcommand{\z}{\ensuremath{\mbox{$z$}}}
\newcommand{\xhat}{\ensuremath{\mathbf{\hat{x}}}\xspace}
\newcommand{\yhat}{\ensuremath{\mathbf{\hat{y}}}\xspace}
\newcommand{\zhat}{\ensuremath{\mathbf{\hat{z}}}\xspace}
\DeclareMathAlphabet\mathbfcal{OMS}{cmsy}{b}{n}

\DeclarePairedDelimiter\floor{\lfloor}{\rfloor}
\definecolor{dim}{rgb}{0.8,0.8,0.8}
\newcolumntype{L}[1]{>{\raggedright\let\newline\\\arraybackslash\hspace{0pt}}m{#1}}
\newcommand{\bigdot}{\scaleto{\cdot}{6pt}}

\newcommand{\pbasis}{\ensuremath{\tilde{\bvec{p}}}}
\newcommand{\gbasis}{\ensuremath{\tilde{\bvec{g}}}}
\newcommand{\ybasis}{\ensuremath{\tilde{\bvec{y}}}}
\newcommand{\pbasisn}{\ensuremath{\tilde{p}_n}}
\newcommand{\gbasisn}{\ensuremath{\tilde{g}_n}}

\newcommand{\AOne}{\ensuremath{\bvec{A_1}}}
\newcommand{\ATwo}{\ensuremath{\bvec{A_2}}}

\definecolor{codegreen}{rgb}{0,0.6,0}
\definecolor{codegray}{rgb}{0.5,0.5,0.5}
\definecolor{codepurple}{rgb}{0.58,0,0.82}
\definecolor{backcolour}{rgb}{0.95,0.95,0.95}
\lstdefinestyle{mystyle}{
    backgroundcolor=\color{backcolour},
    commentstyle=\color{codegreen},
    keywordstyle=\color{magenta},
    numberstyle=\tiny\color{codegray},
    stringstyle=\color{codepurple},
    basicstyle=\small\ttfamily,
    breakatwhitespace=false,
    breaklines=true,
    captionpos=b,
    keepspaces=true,
    numbers=left,
    numbersep=5pt,
    showspaces=false,
    showstringspaces=false,
    showtabs=false,
    tabsize=2,
    aboveskip=1em,
    belowskip=1em,
    keywords=[2]{map},
    keywordstyle=[2]{\color{black!80!black}},
    upquote=true
}
\makeatletter
\def\Ddots{\mathinner{\mkern1mu\raise\p@
\vbox{\kern7\p@\hbox{.}}\mkern2mu
\raise4\p@\hbox{.}\mkern2mu\raise7\p@\hbox{.}\mkern1mu}}
\makeatother

\renewcommand\quad{\hskip\fontdimen3\font}

\begin{document}

\setlength{\abovedisplayskip}{1.5em}
\setlength{\belowdisplayskip}{1.5em}

\title{%
    \textbf{STARRY}: Analytic Occultation Light Curves
}

\author[0000-0002-0296-3826]{Rodrigo Luger}\altaffiliation{Flatiron Fellow}
\email{rluger@flatironinstitute.org}
\affil{Center~for~Computational~Astrophysics, Flatiron~Institute, New~York, NY}
\affil{Virtual~Planetary~Laboratory, University~of~Washington, Seattle, WA}
\author[0000-0002-0802-9145]{Eric Agol}\altaffiliation{Guggenheim Fellow}
\affil{Department~of~Astronomy, University~of~Washington, Seattle, WA}
\affil{Virtual~Planetary~Laboratory, University~of~Washington, Seattle, WA}
\author[0000-0002-9328-5652]{Daniel Foreman-Mackey}
\affil{Center~for~Computational~Astrophysics, Flatiron~Institute, New~York, NY}
\author{David P. Fleming}
\affil{Department~of~Astronomy, University~of~Washington, Seattle, WA}
\affil{Virtual~Planetary~Laboratory, University~of~Washington, Seattle, WA}
\author{Jacob Lustig-Yaeger}
\affil{Department~of~Astronomy, University~of~Washington, Seattle, WA}
\affil{Virtual~Planetary~Laboratory, University~of~Washington, Seattle, WA}
\author{Russell Deitrick}
\affil{Center~for~Space~and~Habitability, University~of~Bern, Bern, Switzerland}
\affil{Virtual~Planetary~Laboratory, University~of~Washington, Seattle, WA}


\begin{abstract}
We derive analytic, closed form, numerically stable solutions for the total flux
received from a spherical planet, moon or star during an occultation
if the specific intensity map of the body is expressed as a
sum of spherical harmonics. Our expressions are valid to arbitrary degree
and may be computed recursively for speed. The formalism we develop
here applies to the computation of stellar transit light curves,
planetary secondary eclipse light curves, and planet-planet/planet-moon
occultation light curves, as well as thermal (rotational) phase curves.
In this paper we also introduce \starry, an open-source package written in \cpp
and wrapped in \Python that computes these light curves.
The algorithm in \starry is six orders of magnitude faster than direct
numerical integration and several orders of magnitude more precise.
\starry also computes analytic derivatives of the light curves with respect to all input
parameters for use in gradient-based optimization and inference, such as
Hamiltonian Monte Carlo (HMC), allowing users to quickly and efficiently
fit observed light curves to infer properties of a celestial body's
surface map.
\href{https://github.com/rodluger/starry}{\color{linkcolor}\faGithub}
\href{https://rodluger.github.io/starry}{\color{linkcolor}\faBook}
\href{https://doi.org/10.5281/zenodo.1312286}{\color{linkcolor}\faTags}
\end{abstract}

%
\section{Introduction}
\label{sec:intro}

Our understanding of the surface of Earth and the other planets in our solar
system starts with the creation of maps.  Mapping the colors, compositions, and
surface features gives us an understanding of the geological, hydrological,
and meteorological processes at play that are the basis of planetary science,
including comparative planetology.
With the discovery of planets orbiting other
stars, cartography becomes a formidable task: these planets are too distant to
resolve their surfaces into maps as we do for our own planetary suite.  One way
to overcome this drawback is to utilize the time-dependence of unresolved,
disk-integrated light from planetary bodies:  both rotational variability
\citep{Russell1906,Lacis1972,Cowan2008,OakleyCash2009}
and occultations \citep{Williams2006,Rauscher2007} yield the opportunity to constrain
the presence of static variations in the surface features of exoplanets.

The first application of time-dependent mapping to exoplanets was carried out in the
infrared with the hot Jupiter HD 189733b using both phase variations and
secondary eclipses of the exoplanet \citep{Knutson2007,Majeau2012,deWit2012}.
These yielded crude constraints on the monopole and dipole components of the thermal
emission from the thick, windy atmosphere of this giant planet.
Since then, phase curve and/or secondary eclipse measurements have been made for
hundreds of other exoplanets \citep[e.g.,][]{Shabram2016, Jansen2017, Adams2018} and have allowed for
the measurements of their average albedos and, in some cases, higher order spatial features
such as hotspot offsets. Given its unprecedented photometric precision in the thermal infrared,
the upcoming James Webb Space Telescope (JWST) is expected to dramatically push the boundaries of
what can be inferred from these observations, potentially leading to the construction of de facto
surface maps of planets in short orbital periods \citep{Beichman2014,Schlawin2018}.
Future mission concepts such as the Large UV-Optical-InfraRed telescope (LUVOIR) and the
Origins Space Telescope (OST) will likewise open doors for the mapping technique,
extending it to the study of exoplanets with solid or even liquid surfaces
\citep[e.g.,][]{KawaharaFujii2010,KawaharaFujii2011,FujiiKawahara2012,Cowan2012,CowanFuentesHaggard2013,CowanFujii2017,Fujii2017,LugerLustigYaegerAgol2017,BerdyuginaKuhn2017}.
Future direct imaging telescopes should also enable eclipse mapping from
mutual transits of binary planets or planet-moon systems \citep{Cabrera2007},
in analogy with mutual events viewed in the Solar System
\citep{Brinkmann1973,Vermilion1974,Herzog1975,Brinkmann1976,Reinsch1994,Young1999,
Young2001,Livengood2011}.

As we prepare to perform these observations, it is essential that we have
robust models of exoplanet light curves so
that we may reliably infer the surface maps that generated them. Because the
features that we seek will likely be close to the limit of detectability,
exoplanet mapping is necessarily a probabilistic problem, requiring
a careful statistical approach capable of characterizing the uncertainty
on the inferred map. Recently, \citet{Farr2018} introduced \exocartographer,
a Bayesian model for inferring surface maps and rotation states
of exoplanets directly imaged in reflected light. In a similar but
complementary vein, \citet{Louden2018} presented \spiderman, a fast
code to model phase curves and secondary eclipses of exoplanets, which
the authors show is fast enough to be used in Markov Chain Monte Carlo
(MCMC) runs for general mapping problems. However, both algorithms, along with all
others in the literature to date, rely on numerical integration methods to compute
the flux received from the planet during occultation. In addition to
the potential loss of precision due to the approximations they employ,
numerical algorithms are typically much slower than an analytic
approach, should it exist. During the writing of this paper, \citet{Haggard2018} derived
analytic solutions to the phase curve problem, demonstrating that an
exoplanet's phase curve can be computed exactly in both thermal and
reflected light if its map is expressed as a sum of spherical harmonics.

Here we present an algorithm to compute analytic occultation light curves of stars,
planets, or moons of arbitrary complexity if the surface map of the occulted body is expressed in the
spherical harmonic basis. Our algorithm generalizes the \citet{MandelAgol2002}, \citet{Gimenez2006},
and \citet{Pal2012} analytic transit formulae to model eclipses and occultations of bodies with arbitrary, non-radially
symmetric surface maps or stars with limb darkening of arbitrary order.
For radially symmetric, second-degree maps, our expressions reduce to the \citet{MandelAgol2002}
quadratic limb darkening transit model; in the limit of zero occultor size or large
impact parameter, they
reduce to the expressions of \citet{Haggard2018} for thermal phase curves.


This paper is organized as follows: in \S\ref{sec:surfacemaps} we discuss the
real spherical harmonics and introduce our mathematical formalism for
dealing with spherical harmonic surface maps. In \S\ref{sec:lightcurves} we
discuss how to compute analytic thermal phase curves and occultation light curves
for these surface maps. In \S\ref{sec:starrycode} we introduce our
light curve code, \starry, and discuss how to use it to compute full
light curves for systems of exoplanets and other celestial bodies. We
present important caveats in \S\ref{sec:caveats} and conclude in \S\ref{sec:conclusions}.
Most of the math, including the derivations of the analytic expressions for the
light curves, is folded into the Appendix. For convenience, throughout
the paper we provide links
to \Python code (\,\codeicon\,) to reproduce all of the
figures, as well as links to \Jupyter notebooks
(\,\prooficon\,) containing proofs and derivations
of the principal
equations. Finally, Table~\ref{tab:symbols} at the end lists
all the symbols used
in the paper, with references to the equations defining them.

\section{Surface Maps}
\label{sec:surfacemaps}

\edited{
In this section we discuss the mathematical framework we use to
express, manipulate, and rotate spherical harmonic surface maps. We also
introduce two bases, along with corresponding transformations, that
will come in handy when computing
light curves in \S\ref{sec:lightcurves}: the \emph{polynomial basis} and
the \emph{Green's basis}. While it is convenient to express a surface map
as a set of spherical harmonic coefficients, we will see that it is much
easier to integrate the map if we first transform to the appropriate
basis.
}

\subsection{Spherical harmonics}
\label{sec:spharm}

The orthonormal real spherical harmonics $Y_{lm}(\uptheta,\upphi)$ of degree $l \ge 0$
and order $m \in [-l,\, l]$ with the Condon-Shortley phase factor \citep[e.g.][]{Varshalovich1988}
are defined in spherical coordinates as \vspace{1em}
\begin{align}
    \label{eq:ylmtp}
    Y_{lm}(\uptheta, \upphi) =
    \begin{cases}
        \bar{P}_{lm}(\cos\uptheta)\cos(m\upphi) & \qquad m \geq 0 \\
        \bar{P}_{l|m|}(\cos\uptheta)\sin(|m|\upphi) & \qquad m < 0 \quad,
    \end{cases}\\ \nonumber
\end{align}
where $\bar{P}_{lm}$ are the normalized associated Legendre functions
(Equation~\ref{eq:plm}). On the
surface of the unit sphere, we have
\begin{align}
    \label{eq:xyz}
    \x &= \sin\uptheta \cos\upphi \nonumber \\
    \y &= \sin\uptheta \sin\upphi \nonumber \\
    \z &= \cos\uptheta \quad,
\end{align}
where $\uptheta$ is the inclination angle and $\upphi$ is the azimuthal angle
(ISO convention).
The observer is located along the $z$-axis at $z = \infty$ such
that the projected disk of the body sits at the origin on the $xy$-plane with $\xhat$ to
the right and $\yhat$ up.
\begin{figure}[t!]
    \begin{centering}
    \includegraphics[width=\linewidth]{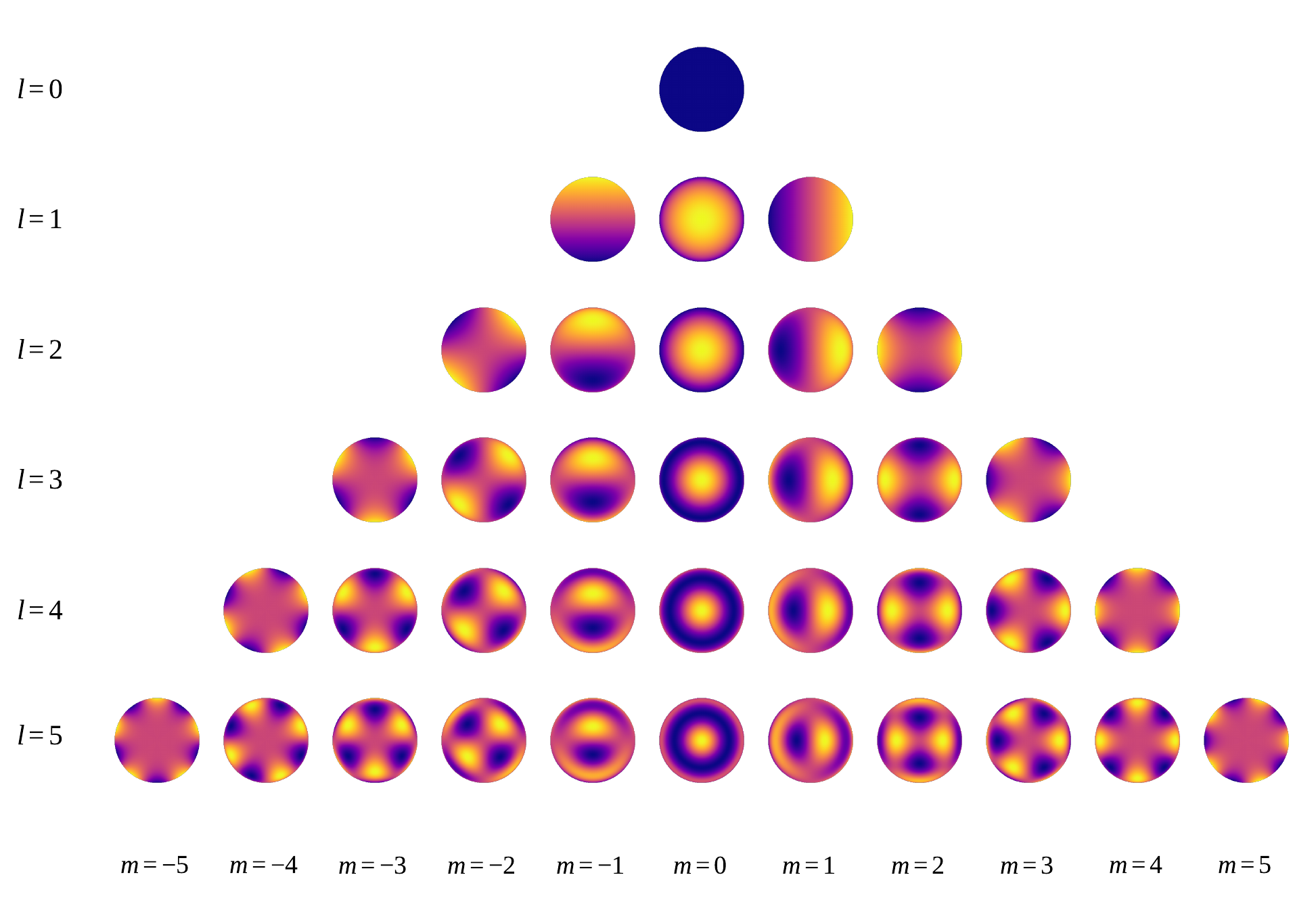}
    \caption{\label{fig:ylms}
             The real spherical harmonics up to degree $l = 5$ computed from
             Equation~(\ref{eq:ylmtp}). In these plots, the $x$-axis points
             to the right,
             the $y$-axis points up, and the $z$-axis points
             out of the page.
             \codelink{ylms}\animlink{ylms}\vspace*{3em}
             }
    \end{centering}
\end{figure}
Re-writing Equation~(\ref{eq:ylmtp}) in terms of $\x$, $\y$, and $\z$ leads
to expressions that are simply polynomials of these variables, a fact we will
heavily exploit below when computing their integrals.
We derive the polynomial representation of the spherical harmonics
in Appendix~\ref{app:spharm}. The spherical harmonics up to
degree $l = 5$ are shown in Figure~\ref{fig:ylms}.

\vspace{3em}

\subsection{Surface map vectors}
\label{sec:vectors}

Any physical surface map of a celestial body can be expanded in terms of
the real spherical harmonics defined in the previous section. For
convenience, in this paper we represent a surface map as a
vector $\bvec{y}$ of spherical harmonic
coefficients such that the specific intensity at the point
$(\x, \y)$ may be written
\begin{align}
    \label{eq:I}
    I(\x, \y) = \ybasis^\mathsf{T} (\x, \y) \, \bvec{y}
    \quad,
\end{align}
where $\ybasis$ is the \emph{spherical harmonic basis},
arranged in increasing degree and order:
\begin{align}
    \label{eq:by}
    \ybasis =
    \begin{pmatrix}
        Y_{0, 0} &
        Y_{1, -1} & Y_{1, 0} & Y_{1, 1} &
        Y_{2, -2} & Y_{2, -1} & Y_{2, 0} & Y_{2, 1} & Y_{2, 2} &
        \cdot\cdot\cdot
    \end{pmatrix}^\mathsf{T}
    \quad,
\end{align}
where $Y_{l, m} = Y_{l, m}(\x, \y)$ are given by \eq{ylmxy}.
For reference, in this basis the coefficient of the spherical harmonic
$Y_{l, m}$ is located at the index
\begin{align}
    \label{eq:n}
    n = l^2 + l + m
\end{align}
of the vector $\bvec{y}$. Conversely, the coefficient at index $n$
of $\bvec{y}$ corresponds
to the spherical harmonic of degree and order given by
\begin{align}
    \label{eq:lm}
    l &= \floor*{\sqrt{n}} \nonumber \\
    m &= n - \floor*{\sqrt{n}}^2 - \floor*{\sqrt{n}}
    \quad,
\end{align}
where $\floor*{\bigdot}$ is the floor function.

\subsection{Change of basis}
\label{sec:basis}

In order to compute the occultation light curve for a body with a given surface
map $\bvec{y}$, it is convenient to first find its polynomial representation
$\bvec{p}$, which we express as a vector of coefficients in the
\emph{polynomial basis} $\pbasis$: \vspace{2em}
\begin{proof}{bp}
    \pbasisn &=
    \begin{dcases}
        \x^\frac{\mu}{2} \y^\frac{\nu}{2} & \qquad \nu \, \mathrm{even}
        \\
        \x^\frac{\mu-1}{2} \y^\frac{\nu-1}{2} \z & \qquad \nu \, \mathrm{odd}
    \end{dcases}
    \nonumber\\[0.5em]
    \pbasis &=
    \begin{pmatrix}
        1 &
        \x & \z & \y &
        \x^2 & \x\z & \x\y & \y\z & \y^2 &
        \cdot\cdot\cdot
    \end{pmatrix}^\mathsf{T}
    \quad,
    \label{eq:bp} \\ \nonumber
\end{proof}
where
\begin{align}
    \label{eq:munu}
    \mu &= l - m \nonumber \\
    \nu &= l + m
    \quad
\end{align}
with $l$ and $m$ given by \eq{lm}.
To find $\bvec{p}$ given $\bvec{y}$, we
introduce the change of basis matrix $\AOne$,
which transforms
a vector in the spherical harmonic basis $\ybasis$ to the
polynomial basis $\pbasis$:
\begin{align}
    \bvec{p} = \AOne \, \bvec{y}
\end{align}
The columns of $\AOne$ are simply the polynomial vectors
corresponding to each of the spherical harmonics in \eq{by}; see
Appendix~\ref{app:basis} for details.
As before, the specific intensity at the point $(\x, \y)$
may be computed as \vspace{1em}
\begin{align}
    I(\x, \y) &= \pbasis^\mathsf{T} \bvec{p} \nonumber \\
              &= \pbasis^\mathsf{T} \AOne \, \bvec{y}
    \quad. \\ \nonumber
\end{align}

As we will see in the next section, integrating the surface map over the disk of
the body is easier if we apply one final transformation to our input vector,
rotating it into what we will refer to as the \emph{Green's basis}, $\gbasis$:
\begingroup\makeatletter\def\f@size{10}\check@mathfonts
\def\maketag@@@#1{\hbox{\m@th\normalsize#1}}%
\begin{proof}{bg}
    \gbasisn &=
    \begin{dcases}
        \frac{\mu+2}{2}\x^\frac{\mu}{2} \y^\frac{\nu}{2}
            & \qquad \nu \, \mathrm{even}
        \\[1em]
        \z
            & \qquad l = 1, \, m = 0
        \\[1em]
        3\x^{l-2}\y\z
            & \qquad \nu \, \mathrm{odd}, \,
                     \mu = 1, \,
                     l \, \mathrm{even}
        \\[1em]
        \z
        \bigg(
         -\x^{l-3} + \x^{l-1} + 4\x^{l-3}\y^2
        \bigg)
         & \qquad \nu \, \mathrm{odd}, \,
                  \mu = 1, \,
                  l \, \mathrm{odd}
        \\[1em]
        \z
        \bigg(
            \frac{\mu-3}{2} \x^\frac{\mu-5}{2} \y^\frac{\nu-1}{2}
            -
            \frac{\mu-3}{2} \x^\frac{\mu-5}{2} \y^\frac{\nu+3}{2}
            -
            \frac{\mu+3}{2} \x^\frac{\mu-1}{2} \y^\frac{\nu-1}{2}
        \bigg)
            & \qquad \mathrm{otherwise}
    \end{dcases}
    \nonumber\\[1.5em]
    \gbasis &=
    \begin{pmatrix}
        1 &
        2\x & \z & \y &
        3\x^2 & -3\x\z & 2\x\y & 3\y\z & \y^2 &
        \cdot\cdot\cdot
    \end{pmatrix}^\mathsf{T}
    \quad,
    \label{eq:bg}
\end{proof}
\endgroup
where the values of $l$, $m$, $\mu$, and $\nu$ are given by
Equations~(\ref{eq:lm}) and (\ref{eq:munu}). Given
a polynomial vector $\bvec{p}$, the corresponding vector in
the Green's basis, $\bvec{g}$, can be found by performing another
change of basis operation:
\begin{align}
    \bvec{g} = \bvec{\ATwo} \, \bvec{p}
\end{align}
where the columns of the matrix $\ATwo$ are the Green's vectors
corresponding to each of the polynomial terms in \eq{bp};
see Appendix~\ref{app:basis} for details.

Note that we may also transform directly from the spherical harmonic basis
to the Green's basis:
\begin{align}
    \bvec{g} &= \ATwo \, \AOne \, \bvec{y} \nonumber \\
             &= \bvec{A} \, \bvec{y}
\end{align}
where
\begin{align}
    \label{eq:A}
    \bvec{A} \equiv \ATwo \, \AOne
\end{align}
is the full change of basis matrix.
For completeness,
we again note that the specific intensity at a point on a map
described by the spherical harmonic vector $\bvec{y}$ may be written
\begin{align}
    \label{eq:fluxpoint}
    I(\x, \y) &= \gbasis^\mathsf{T}(\x, \y) \bvec{g} \nonumber \\
              &= \gbasis^\mathsf{T}(\x, \y) \bvec{A} \, \bvec{y}
    \quad.
\end{align}
%

\subsection{Rotation of surface maps}
\label{sec:rotation}

Defining a map as a vector of spherical harmonic coefficients makes it
straightforward to compute the projection of the map under arbitrary rotations
of the body via a rotation matrix $\bvec{R}$:
\begin{align}
    \label{eq:rotation}
    \bvec{y'} = \bvec{R} \, \bvec{y}
\end{align}
where $\bvec{y'}$ are the spherical harmonic coefficients of the rotated map.
In Appendix~\ref{app:rotation} we derive expressions for $\bvec{R}$ in terms
of the Euler angles $\alpha$, $\beta$, and $\gamma$, as well as in terms of
an angle $\theta$ and an arbitrary axis of rotation $\bvec{u}$. Follow the link
next to Figure~\ref{fig:ylms} to view an animation of the spherical harmonics
rotating about the $y$-axis, computed from \eq{rotation}.

\section{Computing light curves}
\label{sec:lightcurves}
%
\subsection{Rotational phase curves}
\label{sec:phasecurves}

Consider a body of unit radius centered at the origin, \edited{with an observer
located along the $z$-axis at $z = \infty$.} The body has a surface map
given by the spherical harmonic vector $\bvec{y}$ viewed at an orientation
specified by the rotation matrix $\bvec{R}$, such that
the specific intensity at a point $(\x, \y)$ on the surface is
\begin{align}
    I(\x, \y) &= \ybasis^\mathsf{T} (\x, \y) \bvec{R} \, \bvec{y}
    \nonumber \\
              &= \pbasis^\mathsf{T} (\x, \y) \AOne \, \bvec{R} \, \bvec{y}
    \quad
\end{align}
where $\pbasis$ is the polynomial basis and $\AOne$ is the corresponding
change-of-basis matrix (\S\ref{sec:basis}).
The total flux radiated
in the direction of the observer is obtained by integrating the specific
intensity over a region $S$ of the projected disk of the body:
\begin{align}
    \label{eq:phaseint}
    F &=
    \oiint I(\x, \y) \, \dd S
    \nonumber \\
    &=
    \oiint \pbasis^\mathsf{T} (\x, \y) \AOne \, \bvec{R} \, \bvec{y} \, \dd S
    \nonumber \\
    &=
    \bvec{r}^\mathsf{T} \AOne \, \bvec{R} \, \bvec{y}
    \quad,
\end{align}
where $\AOne$, $\bvec{R}$, and $\bvec{y}$ are constant and
$\bvec{r}$ is a column vector whose $n^\mathrm{th}$ component is given by
\begin{align}
    \label{eq:rn}
    r_n &\equiv
      \oiint \pbasisn (\x, \y)  \, \dd S
    \quad.
\end{align}
When the entire disk of the body is visible (i.e., when no occultation is
occurring), this may be written
\begin{proof}{rnsoln}
    r_n &=
              \int_{-1}^{1}
              \int_{-\sqrt{1-\x^2}}^{\sqrt{1+\x^2}}
              \tilde{p}_n (\x, \y)
              \,
              \dd \y \, \dd \x
        \nonumber \\[1em]
        &=
        \begin{dcases}
            \frac{
                    \Gamma\left(\frac{\mu}{4} + \frac{1}{2}\right)
                    \Gamma\left(\frac{\nu}{4} + \frac{1}{2}\right)
                }{
                    \Gamma\left(\frac{\mu + \nu}{4} + 2\right)
                }
            & \qquad \frac{\mu}{2} \, \mathrm{even}, \, \frac{\nu}{2} \, \mathrm{even}
            \\[1em]
            \frac{\sqrt{\pi}}{2}
            \frac{
                    \Gamma\left(\frac{\mu}{4} + \frac{1}{4}\right)
                    \Gamma\left(\frac{\nu}{4} + \frac{1}{4}\right)
                }{
                    \Gamma\left(\frac{\mu + \nu}{4} + 2\right)
                }
            & \qquad \frac{\mu-1}{2} \, \mathrm{even}, \, \frac{\nu-1}{2} \, \mathrm{even}
            \\[1em]
            0
            & \qquad \mathrm{otherwise.}
        \end{dcases}
    \label{eq:rnsoln}
\end{proof}
where $\Gamma(\bigdot)$ is the gamma function.
\eq{phaseint} may be used to analytically compute the rotational (thermal) phase curve of a body
with an arbitrary surface map. Since $\bvec{r}$ and $\bvec{A_1}$ are independent
of the map coefficients or its orientation, these may be pre-computed for
computational efficiency.

We note, finally, that a form of this solution was very recently found by \citet{Haggard2018};
a special case of their equations for phase curves in reflected light yields
analytic expressions for thermal phase curves of spherical harmonics.

\subsection{Occultation light curves}
\label{sec:occultationflux}

\begin{figure}[p!]
    \begin{centering}
    \includegraphics[width=\linewidth]{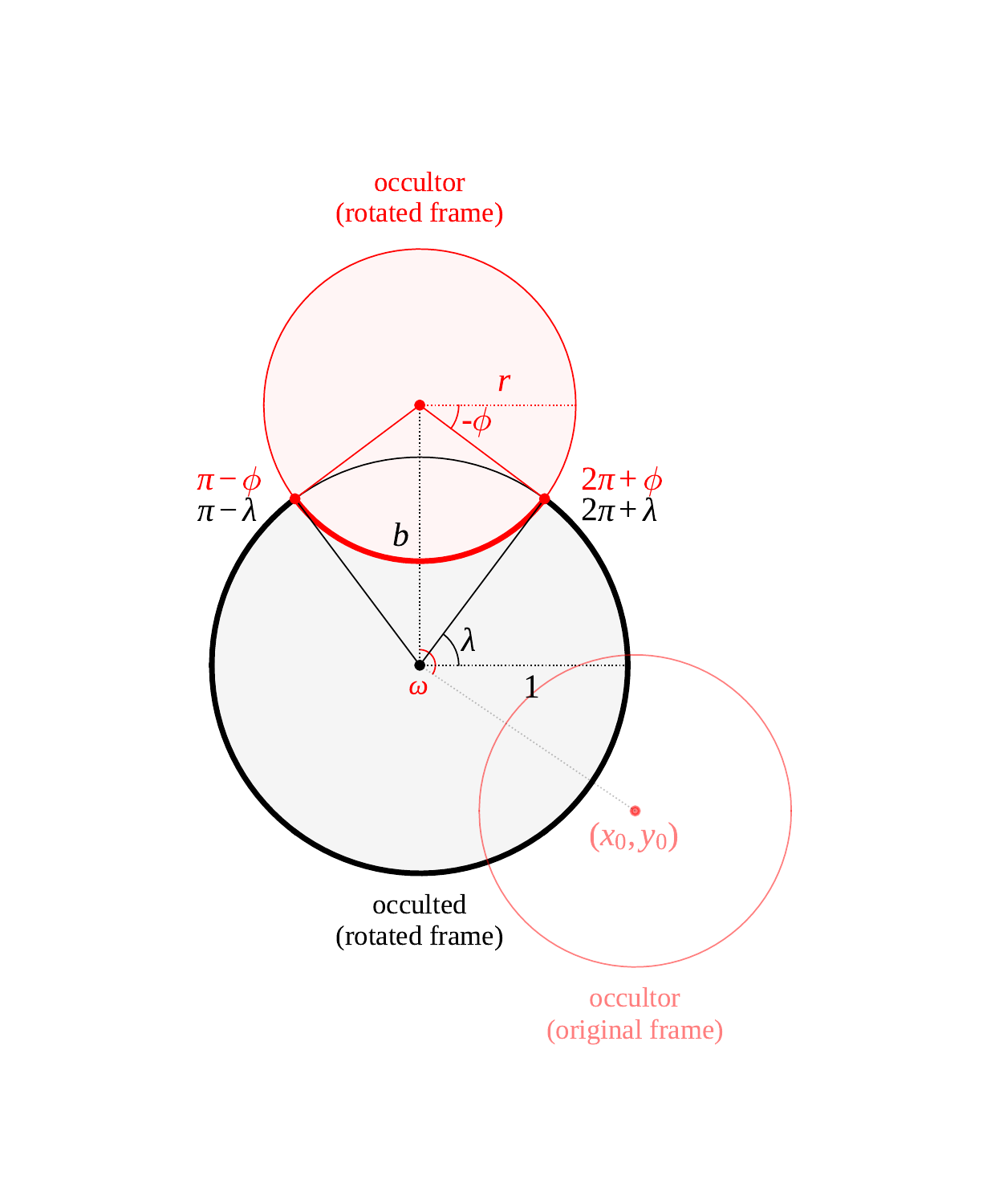}
    \caption{\label{fig:geometry}
             Geometry of the occultation problem.
             The occulted body is centered
             at the origin and has unit radius, while the occultor
             is centered at $(x_o, y_o)$ and has radius $r$.
             \edited{The observer is located at $z = \infty$.}
             We first rotate
             the two bodies about the origin through an angle
             $\omega = \nicefrac{\pi}{2} - \mathrm{arctan2}(y_o, x_o)$
             so the problem is symmetric about the $y$-axis. In this frame,
             the occultor is located at $(0, b)$, where
             $b = \sqrt{x_o^2 + y_o^2}$ is the impact parameter.
             The arc of the occultor
             that overlaps the occulted body (thick red curve) now extends from
             $\pi - \phi$ to $2\pi + \phi$, measured from the center of the
             occultor.
             The arc of the occulted body that is visible during
             the occultation (thick black curve) extends from
             $\pi - \lambda$ to $2\pi + \lambda$, measured from the origin.
             These are the curves along which the primitive integrals
             (Equations~\ref{eq:primitiveP} and \ref{eq:primitiveQ}) are evaluated.
             The angles $\phi$ and $\lambda$ are given by
             Equations~(\ref{eq:phi}) and (\ref{eq:lambda})
             and extend from $-\nicefrac{\pi}{2}$ to $\nicefrac{\pi}{2}$. When
             the occultor is completely within the disk of the occulted body,
             we define $\phi = \lambda = \nicefrac{\pi}{2}$.
             \codelink{geometry}
             }
    \end{centering}
\end{figure}

As we showed earlier, the specific intensity
at a point $(\x, \y)$ on the surface of a body described by the map $\bvec{y}$
and the rotation matrix $\vec{R}$ may also be written as
\begin{align}
    I(\x, \y) &= \ybasis^\mathsf{T} (\x, \y) \bvec{R} \, \bvec{y}
    \nonumber \\
              &= \gbasis^\mathsf{T} (\x, \y) \bvec{A} \, \bvec{R} \, \bvec{y}
    \quad,
\end{align}
where $\gbasis$ is the Green's basis and $\bvec{A}$ is the full change of basis
matrix (\S\ref{sec:basis}).
As before, the total flux radiated
in the direction of the observer is obtained by integrating the specific
intensity over a region $S$ of the projected disk of the body:
\begin{align}
    \label{eq:occint}
    F &=
    \oiint I(\x, \y) \, \dd S
    \nonumber \\
    &=
    \oiint \gbasis^\mathsf{T} (\x, \y) \, \dd S \, \bvec{A} \, \bvec{R} \, \bvec{y}\quad.
\end{align}
This time, suppose the body is
occulted by another body of radius $r$ centered at the point $(x_o, y_o)$,
so that the surface $S$ over which the integral is taken
is a function of $r$, $x_o$, and $y_o$.
In general, the integral in \eq{occint} is
difficult (and often impossible) to compute directly.
One way to simplify the problem is to first perform a rotation through an angle
\begin{align}
    \label{eq:zrot}
    \omega = \frac{\pi}{2} - \mathrm{arctan2}(y_o, x_o)
\end{align}
about the $z$-axis ($\bvec{u} = \left[0, 0, 1\right]$)
so that the occultor lies along the
$+y$-axis, with its center located a distance $b = \sqrt{x_o^2 + y_o^2}$
from the origin (see Figure~\ref{fig:geometry}).
In this rotated frame, the limits of integration (the two points of intersection
between the occultor and the occulted body, should they exist)
are symmetric about the $y$-axis.
If we define $\phi \in [-\nicefrac{\pi}{2}, \, \nicefrac{\pi}{2}]$
as the angular position of the right hand side intersection point
relative to the occultor center, measured counter-clockwise
from the $+x$ direction, the arc of the occultor that overlaps the occulted
body extends from $\pi - \phi$ to $2\pi + \phi$ (see the Figure).
Similarly, defining $\lambda \in [-\nicefrac{\pi}{2}, \, \nicefrac{\pi}{2}]$
as the angular position of the same point relative to the origin, the
arc of the portion of the occulted body that is visible during the occultation
extends from $\pi - \lambda$ to $2\pi + \lambda$ (see the Figure).
For future reference, it can be shown that
\begin{proof}{lambda}
    \label{eq:phi}
    \phi &=
    \begin{dcases}
        \arcsin\left({\frac{1 - r^2 - b^2}{2br}}\right)
                                                & \qquad |1 - r| < b < 1 + r \\
        \frac{\pi}{2}                           & \qquad b \le 1 - r
    \end{dcases} \\
\intertext{and}
    \lambda &=
    \begin{dcases}
        \arcsin\left(\frac{1 - r^2 + b^2}{2b}\right)
                                                & \qquad |1 - r| < b < 1 + r \\
        \frac{\pi}{2}                           & \qquad b \le 1 - r
        \quad.
    \end{dcases}
    \label{eq:lambda}
\end{proof}
The case $b \le 1 - r$ corresponds to an occultation during which the occultor
is fully within the planet disk, so no points of intersection exist.
In this case,
we define $\phi$ such that the arc from $\pi - \phi$ to $2\pi + \phi$ spans the
entire circumference of the occultor, and $\lambda$ such that the arc
from $\pi - \lambda$ to $2\pi + \lambda$ spans the
entire circumference of the occulted body.
Note that if $b \ge 1 + r$, no occultation occurs and the flux may
be computed as in \S\ref{sec:phasecurves}, while
if $b \le r - 1$, the entire disk of the body is occulted and the total flux
is zero.

The second trick we employ to solve \eq{sn} is to use
Green's theorem to express the surface integral of $\gbasis_n$ as the
line integral of a vector function $\bvec{G}_n$ along the boundary of
the same surface \citep{Pal2012}. Defining the ``solution'' column vector
\begin{align}
    \label{eq:sndef}
    \bvec{s}^\mathsf{T} &\equiv
      \oiint \gbasis^\mathsf{T} (\x, \y)  \, \dd S
    \quad,
\end{align}
we may write its $n^\mathrm{th}$ component as
\begin{align}
    \label{eq:greens}
    s_n &=
    \oiint \gbasisn (\x, \y) \, \dd S
    =
    \oint \bvec{G}_n (\x, \y) \cdot \dd \bvec{r}
    \quad,
\end{align}
where $\bvec{G}_n (\x, \y) = {G_n}_x (\x, \y) \, \xhat + {G_n}_y (\x, \y) \, \yhat$ is
chosen such that
\begin{align}
    \label{eq:DGg}
    \bvec{D} \wedge \bvec{G}_n = \gbasisn(\x, \y)
    \quad.
\end{align}
The operation $\bvec{D} \wedge \bvec{G}_n$ denotes the
\emph{exterior derivative} of $\bvec{G}_n$. In two-dimensional Cartesian
coordinates, it is given by
\begin{align}
    \label{eq:extderiv}
    \bvec{D} \wedge \bvec{G}_n &\equiv \frac{\dd {G_n}_y}{\dd \x}
                                   - \frac{\dd {G_n}_x}{\dd \y} \quad.
\end{align}
Thus, in order to compute $s_n$ in \eq{greens}, we must (1) apply a rotation
to our map $\bvec{y}$ to align the occultor with the $+y$-axis;
(2) find a vector function
$\bvec{G}_n$ whose exterior derivative is the $n^\mathrm{th}$ component of the
vector basis $\gbasis$ (Equation~\ref{eq:bg}); and
(3) integrate it along the boundary of the visible portion of the occulted
body's surface. In general, for an occultation involving two bodies,
this boundary consists of two arcs: a segment of the circle bounding the
occultor (thick red curve in Figure~\ref{fig:geometry}),
and a segment of the circle bounding the occulted body (thick black curve
in Figure~\ref{fig:geometry}).
If we happen to know $\bvec{G}_n$, the integral in \eq{greens} is just
\begin{align}
    \label{eq:sn}
    s_n &= \mathcal{Q}(\bvec{G}_n) - \mathcal{P}(\bvec{G}_n)
    \quad,
\end{align}
where, as in \citet{Pal2012}, we define the \emph{primitive integrals}
\begin{align}
    \label{eq:primitiveP}
    \mathcal{P}(\bvec{G}_n) &=
    \int\displaylimits_{\pi-\phi}^{2\pi + \phi}
        \big[ {G_n}_y(r c_\varphi, b + r s_\varphi) c_\varphi -
              {G_n}_x(r c_\varphi, b + r s_\varphi) s_\varphi \big] r \dd \varphi
    \\
\intertext{and}
    \label{eq:primitiveQ}
    \mathcal{Q}(\bvec{G}_n) &=
    \int\displaylimits_{\pi-\lambda}^{2\pi + \lambda}
        \big[ {G_n}_y(c_\varphi, s_\varphi) c_\varphi -
              {G_n}_x(c_\varphi, s_\varphi) s_\varphi \big] \dd \varphi
    \quad,
\end{align}
where we defined
$c_\varphi \equiv \cos \varphi$
and
$s_\varphi \equiv \sin \varphi$
and we used the fact that along the arc of a circle,
\begin{align}
    \label{eq:dr}
    \dd \bvec{r} &= -r s_\varphi \, \dd \varphi \, \xhat +
                     r c_\varphi \, \dd \varphi \, \yhat
    \quad.
\end{align}
In Equations~(\ref{eq:primitiveP}) and (\ref{eq:primitiveQ}), $\mathcal{P}(\bvec{G}_n)$
is the line integral along the arc of the occultor of radius $r$,
and $\mathcal{Q}(\bvec{G}_n)$ is the line integral along the arc of the occulted
body of radius one.

As cumbersome as the Green's basis (Equation~\ref{eq:bg}) may appear, the reason
we introduced it is that its anti-exterior derivatives are conveniently simple.
It can be easily shown that one possible solution to \eq{DGg} is
\begin{proof}{Gn}
    \bvec{G}_n (\x, \y) &=
    \begin{dcases}
        \x^{\frac{\mu + 2}{2}}
        \y^{\frac{\nu}{2}}
        \,\yhat
            & \qquad \nu \, \mathrm{even}
        \\[1em]
        \frac{1-z^3}{3(1-z^2)}(-\y \, \xhat + \x \, \yhat)
            & \qquad l = 1, \, m = 0
        \\[1em]
        \x^{l-2}
        \z^3
        \,\xhat
            & \qquad \nu \, \mathrm{odd}, \,
                     \mu = 1, \,
                     l \, \mathrm{even}
        \\[1em]
        \x^{l-3}
        \y
        \z^3
        \,\xhat
         & \qquad \nu \, \mathrm{odd}, \,
                  \mu = 1, \,
                  l \, \mathrm{odd}
        \\[1em]
        \x^{\frac{\mu-3}{2}}
        \y^{\frac{\nu-1}{2}}
        \z^3
        \,\yhat
            & \qquad \mathrm{otherwise,}
    \end{dcases}
    \label{eq:Gn}
\end{proof}
where $l$ and $m$ are given by \eq{lm} and $\mu$ and $\nu$ are given by
\eq{munu}.
\footnote{
\edited{
It is important to note that our definition of the Green's basis
(Equation~\ref{eq:bg}) is by no means unique. Rather, we imposed solutions of
the form
$\bvec{G}_n = \x^{i}\y^{j}\z^{k} \, \xhat$ and
$\bvec{G}_n = \x^{i}\y^{j}\z^{k} \, \yhat$ and
used \eq{DGg} to find each of the terms in the basis, choosing $i$, $j$,
and $k$ to ensure the basis was complete.
}
}
Solving the occultation problem is therefore a matter of
evaluating the primitive integrals of $\bvec{G}_n$
(Equations~\ref{eq:primitiveP} and \ref{eq:primitiveQ}).
The solutions are in general tedious, but
they are all analytic, involving sines, cosines, and complete elliptic integrals.
In Appendix~\ref{app:solutionvector} we derive recurrence
relations to quickly compute these. We note, in particular, that the
solutions all involve complete elliptic integrals of the \emph{same} argument,
so that the elliptic integrals need only be evaluated once for a map
of arbitrary degree, greatly improving the evaluation speed and the
scalability of the problem to high order. In practice we find that
the stability in evaluation of these expressions is improved by using a
rapidly converging series expansion for occultors of large and small radius.

\subsection{Summary}
\label{sec:summary}

Here we briefly summarize how to analytically compute the flux during
an occultation of a body whose specific intensity profile is described
by a sum of spherical harmonics.
\edited{
The first step is to
compute the change-of-basis matrix $\bvec{A}$ (\S\ref{sec:basis}) to
convert our vector of spherical harmonic coefficients to a vector
of polynomial coefficients in the Green's basis
(Equation~\ref{eq:bg}). Since $\bvec{A}$ is constant,
this matrix may be pre-computed for speed.
}

Then, given a body of unit radius with a
surface map described by the vector of spherical harmonic coefficients
$\mathbf{y}$ (Equation~\ref{eq:by}),
occulted by another body of radius $r$
centered at the point $(x_o, y_o)$,
\edited{and viewed by an observer located at $z = \infty$,}
we must:
\begin{enumerate}
    \item Compute the rotation matrix $\bvec{R}$ to rotate the map to the correct
          viewing orientation, which may be
          specified by the Euler angles $\alpha$, $\beta$, and $\gamma$
          (Appendix~\ref{app:euler}) or by an axis $\bvec{u}$ and an angle $\theta$
          (Appendix~\ref{app:axisangle}).
    \item Compute the rotation matrix $\bvec{R'}$ to rotate the map
          by an angle $\omega$ about the $+z$-axis
          (Equation~\ref{eq:zrot}) so the center of the occultor is a
          distance $b = \sqrt{x_o^2 + y_o^2}$ along the $+y$-axis
          from the center of the occulted body.
    \item Compute the solution vector $\bvec{s}$ (Equation~\ref{eq:sn}), with
          $\mathcal{P}(\bvec{G}_n)$ and $\mathcal{Q}(\bvec{G}_n)$ given by
          the equations in Appendix~\ref{app:generalterm}.
          Note that $s_2$ is special and must be computed separately
          (Equation~\ref{eq:s2}).
\end{enumerate}
Given these quantities, the total flux $f$ during an occultation is then just
\begin{equation}
    \label{eq:starry}
    \boxed{
        f = \bvec{s}^{\boldsymbol{\mathsf{T}}} \bvec{A} \, \bvec{R'} \, \bvec{R} \, \bvec{y}
        }
    \quad.
\end{equation}

\section{The \textbf{STARRY} code package}
\label{sec:starrycode}

\color{editcolor}

The \starry code package provides code to analytically
compute light curves for celestial bodies using the formalism developed
in this paper. \starry is coded entirely in \cpp for speed and wrapped
in \Python using \pybind \citep{pybind11} for quick and easy
light curve calculations. The code may be installed three different ways:
using \textsf{conda} (recommended),
\begin{lstlisting}[language=bash]
conda install -c conda-forge starry
\end{lstlisting}
via \textsf{pip},
\begin{lstlisting}[language=bash]
pip install starry
\end{lstlisting}
or from source by cloning the
\href{https://github.com/rodluger/starry}{\textsf{GitHub}} repository,
\begin{lstlisting}[language=bash]
git clone https://github.com/rodluger/starry.git
cd starry
python setup.py develop
\end{lstlisting}
There are two primary ways of interfacing with \starry: via the surface
map class \Map and via the celestial body system class
\System. The former gives users the most flexibility to
create and manipulate surface maps and compute their fluxes for a variety
of applications, while the latter provides an easy
way to generate light curves for simple Keplerian systems. Let us
discuss the map class first.

\begin{figure}[t!]
    \begin{centering}
    \includegraphics[width=0.925\linewidth]{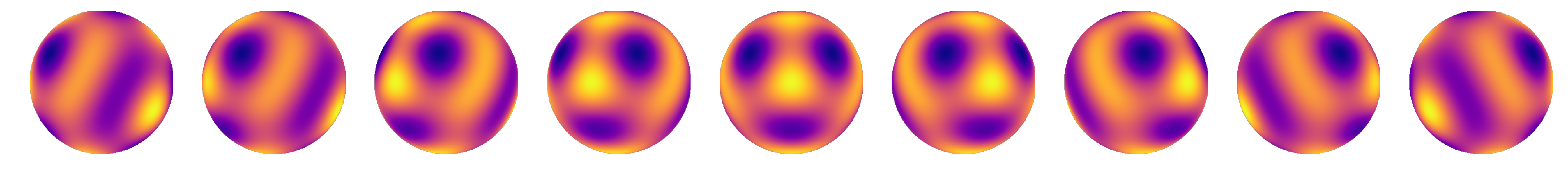}
    \caption{\label{fig:smiley}
             Rotation of the map given by Equation~(\ref{eq:ylm_code_example})
             about $\yhat$.
             \codelink{smiley}
             }
    \end{centering}
\end{figure}
%
%

\subsection{Creating a map}
\label{sec:starrymap}

To begin using \starry, execute the following in a \Python environment:
\begin{lstlisting}[language=Python]
from starry import Map
\end{lstlisting}
A \starry \Map is a vector of spherical harmonic coefficients, indexed by
increasing degree and order, as in \eq{by}. As an example, we can create a map of
spherical harmonics up to degree $l_\mathrm{max} = 5$ by typing
\begin{lstlisting}[language=Python,firstnumber=last]
map = Map(lmax=5)
\end{lstlisting}
By default, the first coefficient ($y_0$, the coefficient multiplying the
$Y_{0,0}$ harmonic) is set to unity and
all other coefficients are set to zero. Importantly, maps in \starry are
normalized
such that the \textbf{average disk-integrated intensity is equal to the
coefficient of the $Y_{0,0}$ harmonic}. By default, the average amount of
flux visible from an unocculted map is therefore unity.

Say our surface map is given by the function
\begin{align}
    \label{eq:ylm_code_example}
    I(\x, \y) &= Y_{0,0} - 2 Y_{5,-3}(\x, \y) + 2 Y_{5,0}(\x, \y) + Y_{5,4}(\x, \y)
    \quad.
\end{align}
To create this map, we set the corresponding coefficients by direct
assignment to the \textsf{(l, m)} indices of the \Map instance:
\begin{lstlisting}[language=Python,firstnumber=last]
map[5, -3] = -2
map[5, 0] = 2
map[5, 4] = 1
\end{lstlisting}
Users can also directly access the spherical harmonic vector $\bvec{y}$,
polynomial vector $\bvec{p}$, and Green's polynomial vector $\bvec{g}$
via the read-only attributes \Mapy, \Mapp, and \Mapg, respectively.
Once a map is instantiated, users may quickly visualize it by calling
\begin{lstlisting}[language=Python,firstnumber=last]
map.show()
\end{lstlisting}
or
\begin{lstlisting}[language=Python,firstnumber=last]
map.animate()
\end{lstlisting}
where the editable attribute \Mapaxis defines the axis
of rotation for the animation.
Rotation of this map about $\yhat$ yields the sequence shown in
Figure~\ref{fig:smiley}.

Alternatively, users may provide a two-dimensional \textsf{numpy} array
of intensities on a latitude-longitude grid or the path to an image
file of the surface map on a latitude-longitude grid:
\begin{lstlisting}[language=Python,firstnumber=last]
map.load_image(array)
\end{lstlisting}
or
\begin{lstlisting}[language=Python,firstnumber=last]
map.load_image("/path/to/image.jpg")
\end{lstlisting}
In both cases, \starry uses the \textsf{map2alm()} function
of the \textsf{healpy} package to find the expansion of the map in
terms of spherical harmonics. Keep in mind that if the image contains very dark
pixels (with \textsf{RGB} values close to zero), its spherical harmonic
expansion may lead to regions with \emph{negative} specific intensity, which
is of course unphysical.

\begin{figure}[t!]
    \begin{centering}
    \includegraphics[width=0.75\linewidth]{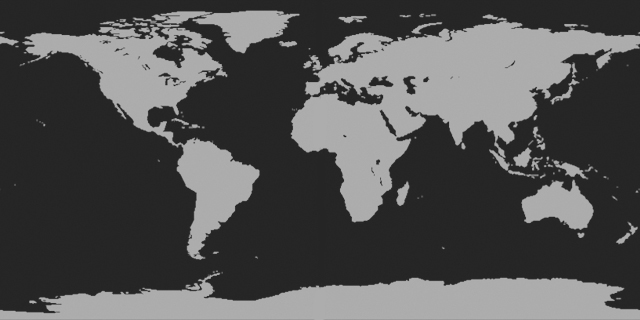}
    \\[1em]
    \includegraphics[width=0.8\linewidth]{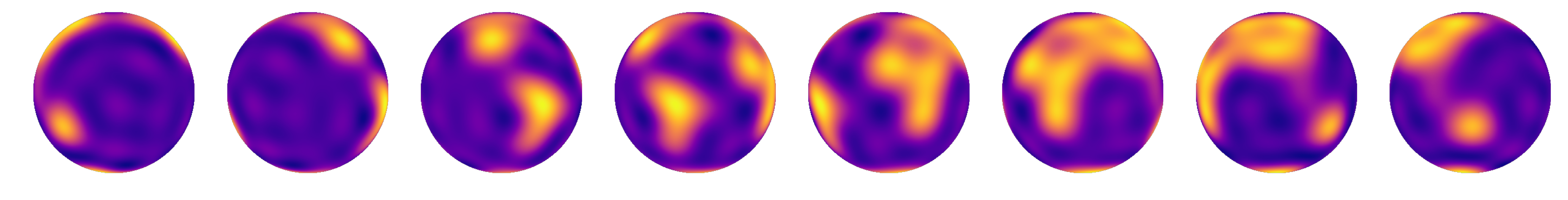}
    \caption{\label{fig:earth}
             A simplified two-color map of the cloudless Earth (top) and the
             corresponding tenth-degree spherical harmonic expansion,
             rotated about $\yhat$ (bottom).
             \codelink{earth}
             }
    \end{centering}
\end{figure}

In Figure~\ref{fig:earth} we show a
simplified two-color map of the cloudless Earth and its corresponding
\starry instance for
$l_\mathrm{max} = 10$, rotated successively about $\yhat$.
%

\subsection{Computing rotational phase curves}
\label{sec:starryphasecurves}
%
\begin{figure}[p!]
    \begin{centering}
    \includegraphics[width=\linewidth]{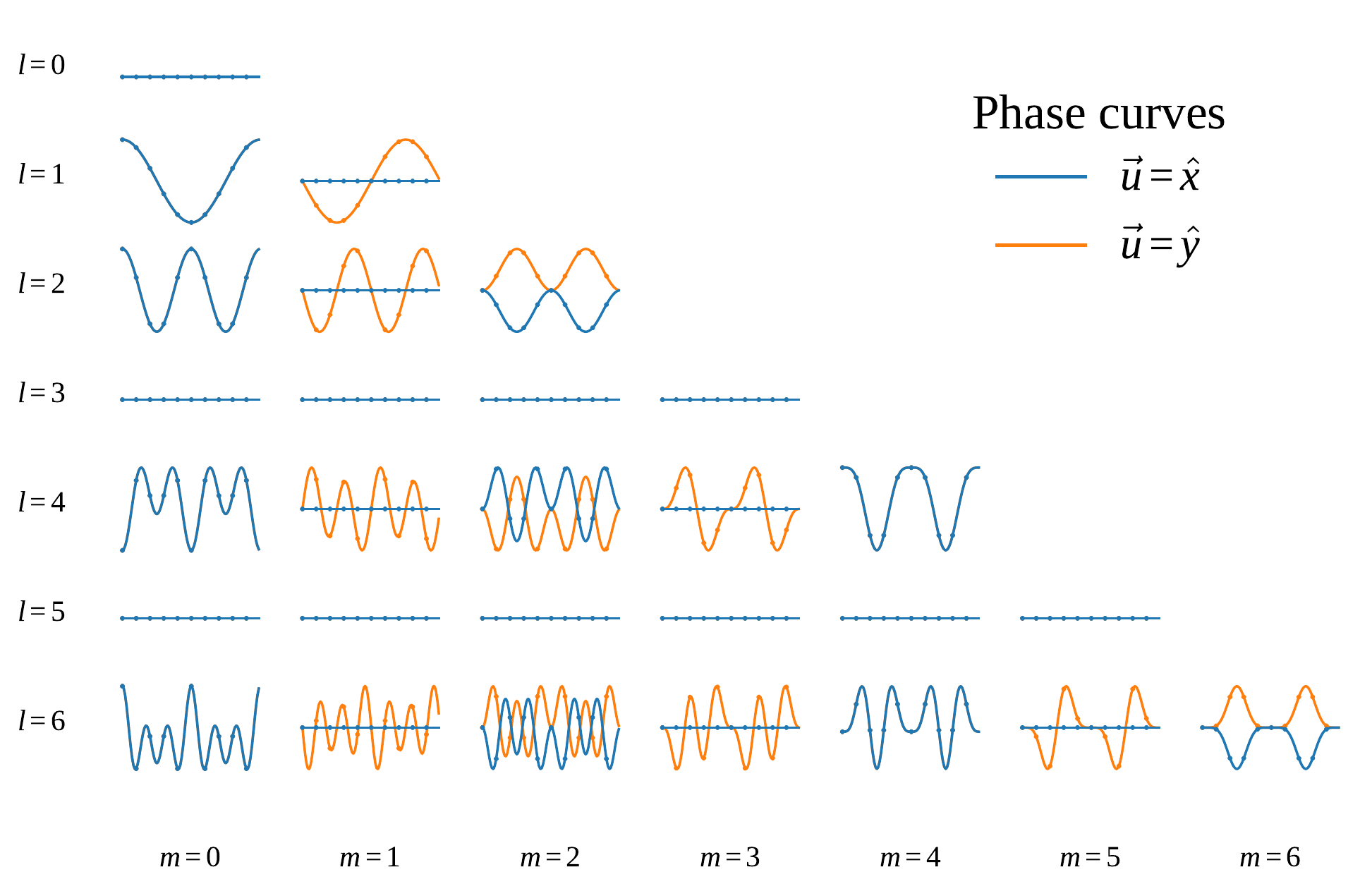}
    \\[1em]
    \includegraphics[width=\linewidth]{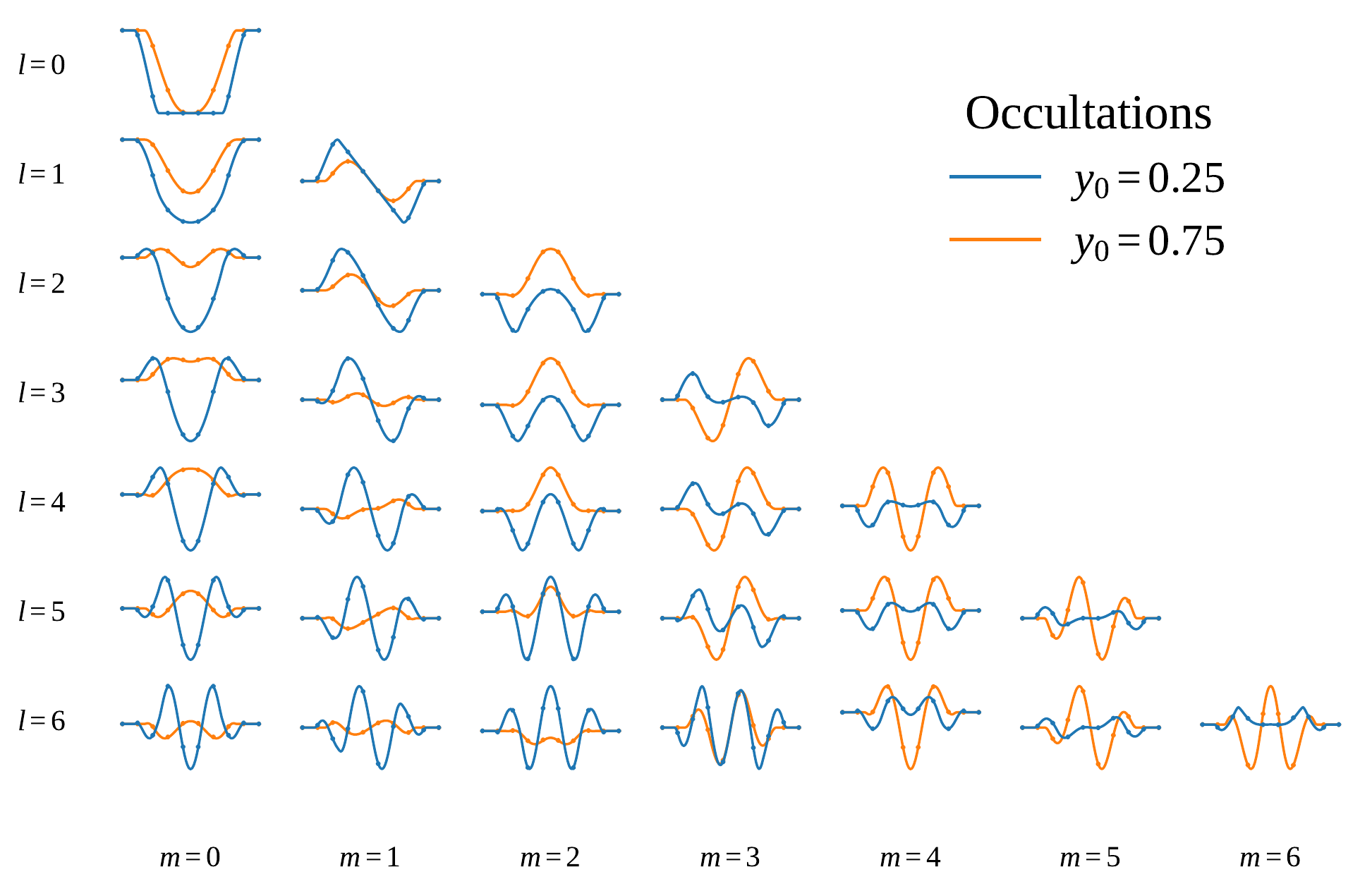}
    \caption{\label{fig:ylmlightcurves}
             \emph{Top:} Phase curves for the first several spherical
             harmonics with order $m \ge 0$ rotated about the $x$-axis
             (blue) and about the $y$-axis (orange).
             Odd harmonics with $l > 1$ and harmonics with $m < 0$ are
             in the phase curve null space \citep{CowanFuentesHaggard2013}.
             \emph{Bottom:} Occultation light curves for the same
             set of harmonics. An occultor of radius $r=0.3$
             transits the body along the $+\x$ direction at $y_o = 0.25$
             (blue) and $y_o = 0.75$ (orange).
             \codelink{ylmphasecurves}\codelink{ylmlightcurves}
             }
    \end{centering}
\end{figure}
Once a map is instantiated, it is easy to compute its rotational
phase curve, \textsf{F}:
\begin{lstlisting}[language=Python,firstnumber=last]
F = map.flux(theta=theta)
\end{lstlisting}
where \textsf{theta} is an array of angles (in degrees) for which to
compute the flux. Note that rotations performed
by \Mapflux are not cumulative; instead, all angles should be specified
relative to the original, unrotated map frame. As before, the axis of
rotation can be set via the \Mapaxis attribute.
In the top panel of Figure~\ref{fig:ylmlightcurves} we plot rotational phase curves
for all spherical harmonics
up to $l_\mathrm{max} = 6$ for rotation about $\xhat$ (blue curves) and $\yhat$
(orange curves). The small dots correspond to phase curves computed by numerical
evaluation of the flux on an adaptive radial mesh (see \S\ref{sec:starrybenchmarks}).
As discussed by \citet{CowanFuentesHaggard2013}, harmonics with
odd $l > 1$ and those with $m < 0$ (not plotted) are in the null space and
therefore do not exhibit rotational phase variations
\edited{when rotated about \xhat or \yhat.}

\begin{figure}[t!]
    \begin{centering}
    \includegraphics[width=\linewidth]{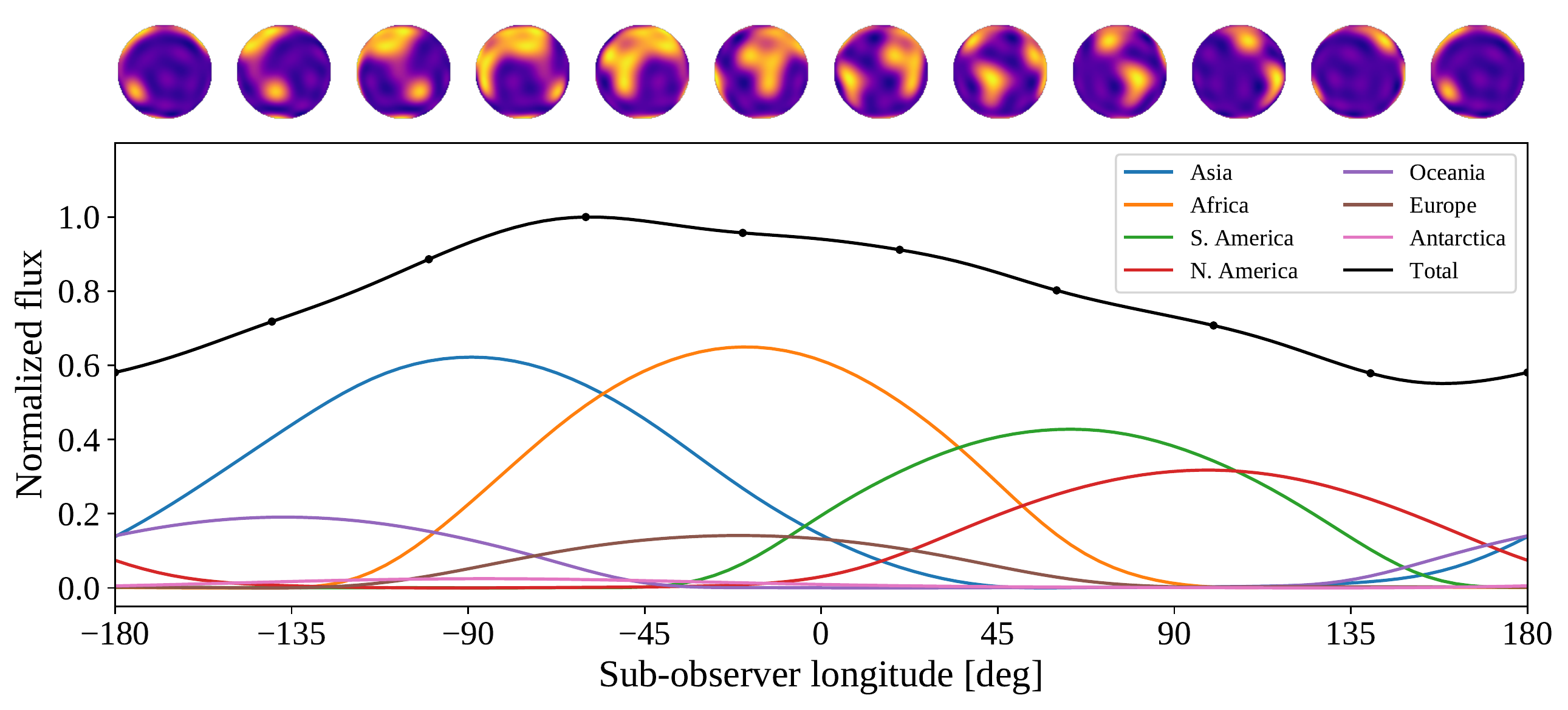}
    \caption{\label{fig:earthphasecurve}
             Phase curve for the Earth rotating about its axis, computed
             from the $l_\mathrm{max} = 10$ expansion from
             Figure~\ref{fig:earth}. The full rotational phase curve is shown in black,
             and the flux due to each of the seven continents is shown as
             the colored curves (see legend). The black dots correspond to the
             numerical solution (see \S\ref{sec:starrybenchmarks}).
             \codelink{earthphasecurve}
             }
    \end{centering}
\end{figure}

As a second example, we can compute the rotational phase curve of the simplified Earth model
(Figure~\ref{fig:earth}) for rotation about $\yhat$ (its actual spin axis)
by executing
\begin{lstlisting}[language=Python,firstnumber=last]
theta = np.linspace(0, 360, 100)
F = map.flux(theta=theta)
\end{lstlisting}
The variable $\textsf{F}$ is
an array of flux values computed from \eq{phaseint}; we plot this in
Figure~\ref{fig:earthphasecurve}, alongside the rotational phase curves due to each of
the seven individual continents.
For more complex phase curves, such as those of planets on inclined orbits,
see \S\ref{sec:starryphotodynamics}.
%

\subsection{Computing occultation light curves}
\label{sec:starryoccultation}

Occultation light curves are similarly easy to compute:
\begin{lstlisting}[language=Python,firstnumber=last]
F = map.flux(theta=theta, xo=xo, yo=yo, ro=ro)
\end{lstlisting}
where \textsf{theta} is the same as above, and
\textsf{xo}, \textsf{yo}, and \textsf{ro} are the occultor parameters
($\x$ position, $\y$ position, and radius, all in units of the
occulted body's radius), which may be either scalars
or arrays.

In the bottom panel of Figure~\ref{fig:ylmlightcurves} we plot
occultation light curves for the spherical harmonics with $m \ge 0$
up to $l_\mathrm{max} = 6$. The occultor has radius $r = 0.3$ and
moves at a constant speed along the $\x$ direction at $y_o = 0.25$
(blue curves) and $y_o = 0.75$ (orange curves). The light curve of
any body undergoing such an occultation can be expressed as a weighted
sum of these light curves. Note that because the value of individual
spherical harmonics can be negative, an increase in the flux is visible
at certain points during the occultation; however, this would of course not
occur for any physical map constructed from a linear combination of
the spherical harmonics. Note also that unlike in the case of rotational phase curves,
there is no null space for occultations, as all spherical harmonics (including
those with $m < 0$, which are not shown) produce a flux signal during
occultation. As before, the numerical solutions are shown as the small dots.

\begin{figure}[t!]
    \begin{centering}
    \includegraphics[width=\linewidth]{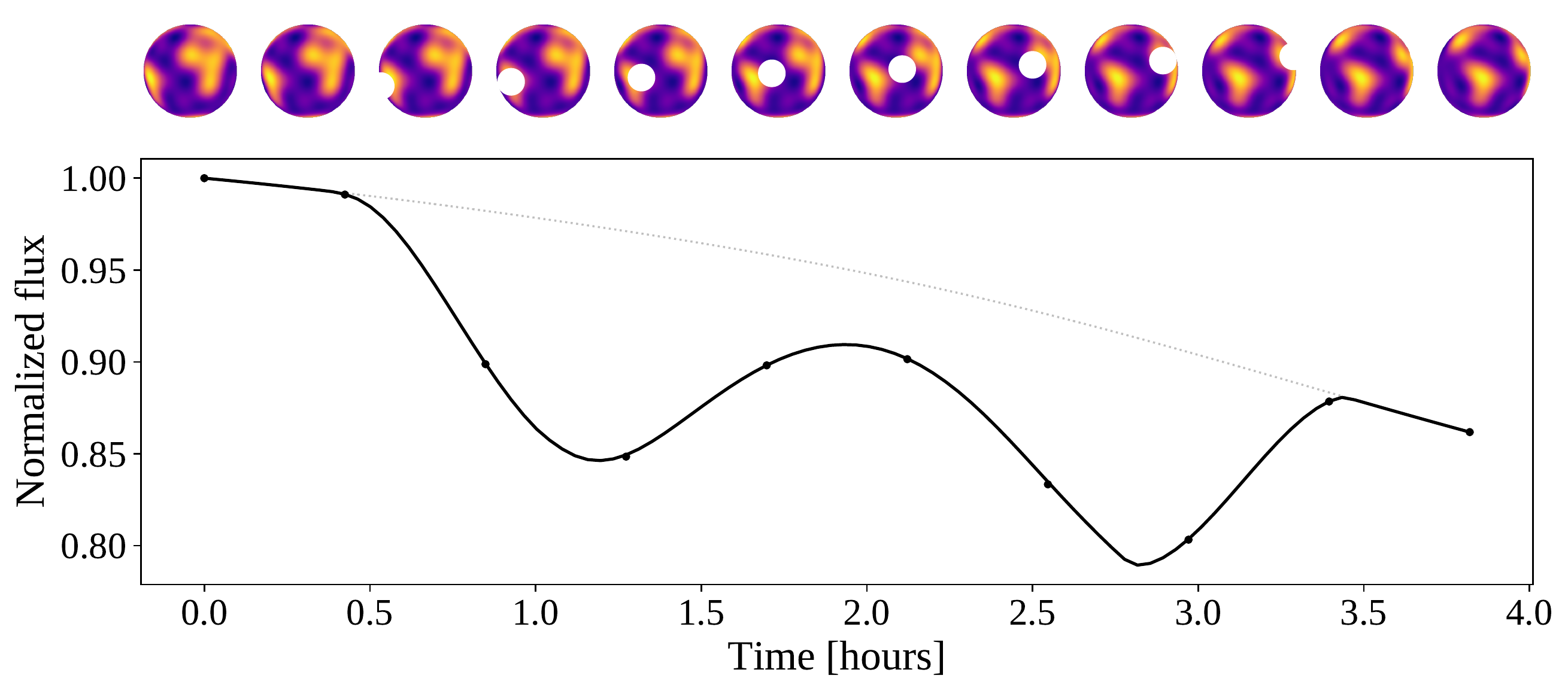}
    \caption{\label{fig:earthoccultation}
             Occultation light curve for the Moon transiting the
             rotating Earth,
             computed from the $l_\mathrm{max} = 10$ expansion from
             Figure~\ref{fig:earth}. The two largest dips are due
             to the occultations of South America (left) and Africa
             (right). Once again, the black dots correspond to the
             numerical solution \edited{(see \S\ref{sec:starrybenchmarks}).
             For reference, the light
             grey dots correspond to the rotational light curve in
             the absence of the occultor.}
             \codelink{earthoccultation}
             }
    \end{centering}
\end{figure}

To further illustrate the code, we return to our spherical harmonic
expansion of the Earth.
Figure~\ref{fig:earthoccultation} shows an occultation light curve
computed for a hypothetical transit of the Earth by the Moon. The
occultation lasts about four hours, during which time
the sub-observer point rotates from Africa to South America, causing a
steady flux decrease as the Pacific Ocean rotates into view. The
occultation is double-dipped: one dip due to the occultation of South
America, and one dip due to the occultation of Africa.

\subsection{Computing light curves of limb-darkened bodies}
\label{sec:starrytransits}

The formalism developed in this paper can easily be extended to the case
of occultations of limb-darkened maps (such as transits of planets
across stars) by noting that any
radially symmetric specific intensity profile can be expressed as a sum
over the $m = 0$ spherical harmonics (see Figure~\ref{fig:ylms}).
In particular, \citet{limbdark} show how a limb darkening profile
that is an order $l$ polynomial function of the radial coordinate,
$\upmu = \z = \sqrt{1 - \x^2 - \y^2}$, can be exactly
expressed in terms of the $m = 0$ spherical harmonics up to order $l$.

All \Map instances in \starry have an additional read-only attribute, \Mapu,
which stores the limb darkening coefficients $\{u_1, u_2, u_3, ...\}$
of the map. These are all zero by default, and can be changed by direct
assignment to index \textsf{l} of the map instance:%
\footnote{
Remember: the \textsf{(l, m)} index of a map instance corresponds to the
coefficient of the $Y_{l,m}$ spherical harmonic, while the (single) \textsf{l}
index corresponds to the coefficient of the $l^\mathrm{th}$ order limb
darkening term. Note, importantly, that
the $l = 0$ limb darkening coefficient cannot be set, as it is
automatically computed to enforce the correct normalization.}
\begin{lstlisting}[language=Python,firstnumber=last]
map[1] = u1
map[2] = u2
...
map[lmax] = ulmax
\end{lstlisting}
In the case of quadratic limb darkening ($l_\mathrm{max} = 2$),
this sets the map's limb darkening profile to
\begin{align}
    \label{eq:quadraticld}
    \frac{I(\upmu)}{I(1)} &= 1 - u_1 (1 - \upmu) - u_2 (1 - \upmu)^2
    \quad,
\end{align}
with $\upmu$ given above.
It is straightforward to show that this corresponds to the spherical
harmonic sum
\begin{proof}{qldylm}
    \label{eq:qldylm}
    \frac{I(\x, \y)}{I(1)} =
            \frac{2\sqrt{\pi}}{3} (3 - 3u_1 - 4u_2) \, Y_{0,0}
          + \frac{2\sqrt{\pi}}{\sqrt{3}} (u_1 + 2u_2) \, Y_{1,0}
          - \frac{4\sqrt{\pi}}{3\sqrt{5}} u_2 \, Y_{2,0}
      \quad,
\end{proof}
where we set
\begin{align}
    \label{eq:I1}
    I(1) = \frac{1}{\pi(1 - \frac{1}{3}u_1 - \frac{1}{6}u_2)}
\end{align}
to enforce the integral of the specific intensity over the visible disk
is unity. Limb darkening profiles of arbitrary degree are supported in \starry,
and in all cases the corresponding light curve is computed analytically.

Note, importantly, that the limb darkening coefficients are treated separately
from the spherical harmonic coefficients in \starry. In particular, the limb darkening
profile does not rotate along with the rest of the map when a rotation is applied.
Moreover, while most users will find it sufficient to specify either the spherical harmonic
coefficients \emph{or} the limb darkening coefficients of a surface map, it is
also possible to specify \emph{both}. This may be convenient in the case of a
limb-darkened star with rotating starspots or other surface inhomogeneities.
In this case, the limb darkening coefficients
are applied to the map as a multiplicative filter following any requested rotation
operations. Since products of spherical harmonics are spherical harmonics, applying
limb darkening to a spherical harmonic map simply raises its degree by an amount
equal to the degree of the limb darkening profile. Hence users must be careful
not to exceed the maximum degree of the map when setting limb darkening
coefficients. For instance, a map instantiated with \textsf{lmax=10} having
nonzero spherical harmonic coefficients up to degree $l = 5$ can have at most
fifth order ($l = 5$) limb darkening.

In principle, one could also model limb-darkened planetary atmospheres in this
fashion, but we do not in general recommend this. Inhomogeneities in the
planetary atmosphere could lead to asymmetries in the limb darkening, which
should probably be treated using a radiative transfer model.

\begin{figure}[t!]
    \begin{centering}
    \includegraphics[width=0.95\linewidth]{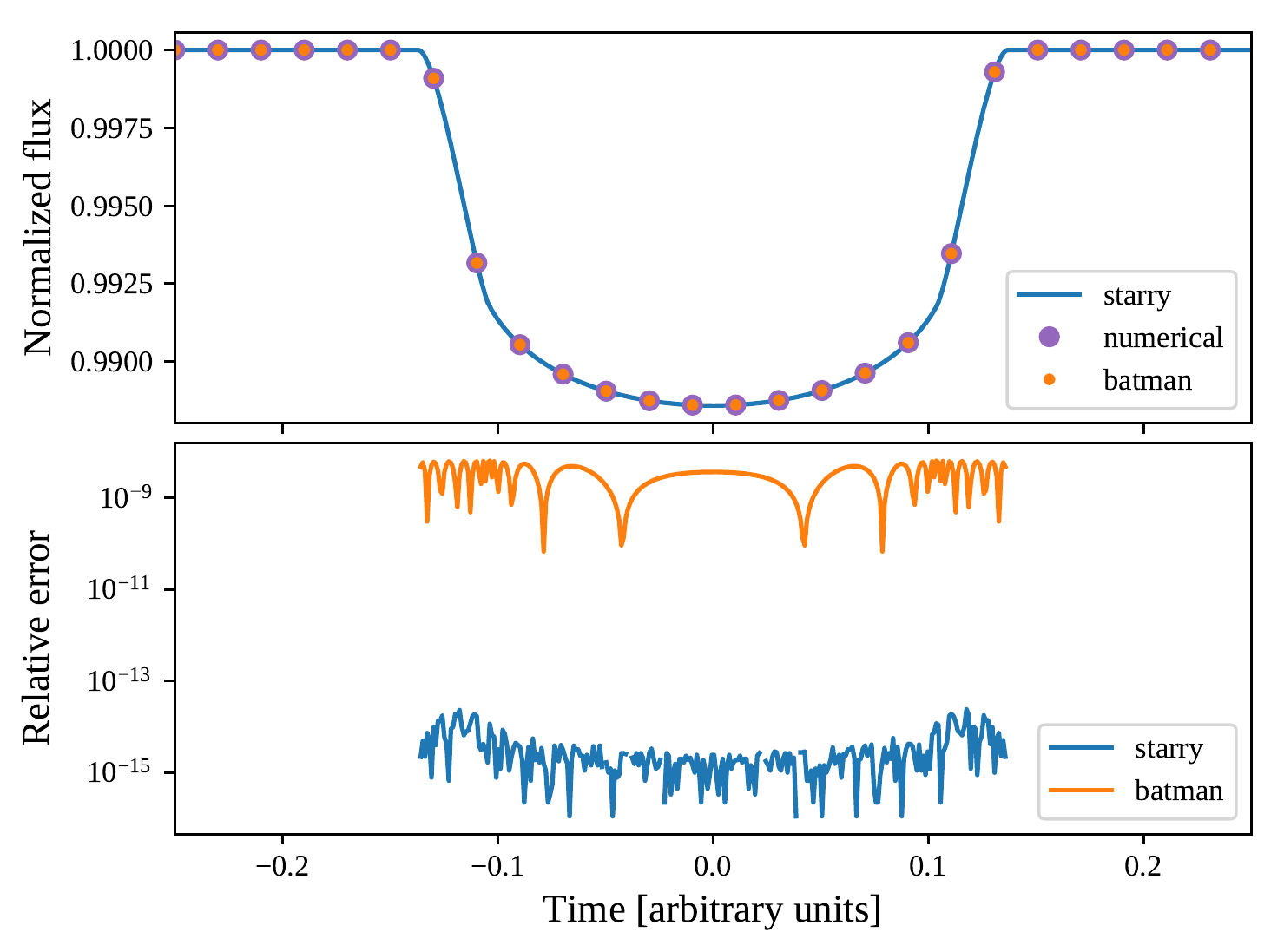}
    \caption{\label{fig:transit}
             Sample transit light curve for a planet ($r = 0.1$) transiting a
             quadratically limb-darkened star ($u_1 = 0.4, u_2 = 0.26$). The
             top panel shows the \starry (blue curve) and \batman
             (orange dots) light curves, as well as a light curve generated
             by a high precision direct numerical integration of the surface
             integral (purple dots). The bottom panel shows the relative
             \edited{error on the flux} compared to the high
             precision numerical solution for \starry (blue) and \batman (orange).
             \codelink{transit}}
    \end{centering}
\end{figure}

Figure~\ref{fig:transit} shows the light curve of a planet transit across
a quadratically limb-darkened star
with $u_1 = 0.4, u_2 = 0.26$ computed using \starry.
The planet/star radius ratio is $r = 0.1$ and
the planet transits at impact parameter $b = 0.5$. For comparison, we also
compute the flux with \batman \citep{Kreidberg2015} and with a
high precision numerical integration of the surface integral of
\eq{quadraticld}
using
\edited{the \textsf{scipy.integrate.dblquad} \citep{scipy} routine
with a tolerance of $10^{-14}$}. The relative error
on the flux for \starry flux is less
than $10^{-7}$ parts per million everywhere in the light curve.
%

\subsection{Photodynamics}
\label{sec:starryphotodynamics}

The \Map class discussed above is convenient when the rotational
state of the body in question and/or the position of the occultor is known,
or when these can easily be computed by some other means. For convenience,
\starry implements a Keplerian solver to compute
light curves of simple star-planet, star-star, or planet-moon systems
given the orbital parameters as input. Users can access this functionality
by instantiating a
\Primary and any number of \Secondary objects, then passing them to a \System
instance. As an example, let us create a central star:
\begin{lstlisting}[language=Python,firstnumber=last]
from starry.kepler import Primary
star = Primary(lmax=2)
\end{lstlisting}
A \Primary instance has unit radius and unit luminosity; the secondary bodies' radii,
semi-major axes, and luminosities are all defined relative to these values.
All \Primary and \Secondary instances derive from the \Map class, so we can
limb-darken the star in the same way as before:
\begin{lstlisting}[language=Python,firstnumber=last,]
star[1] = 0.40
star[2] = 0.26
\end{lstlisting}
where we arbitrarily set $u_1 = 0.40$ and $u_2 = 0.26$.
Next, we will instantiate a planet by typing
\begin{lstlisting}[language=Python,firstnumber=last]
from starry.kepler import Secondary
planet = Secondary(lmax=1)
\end{lstlisting}
Let us set its orbital parameters as follows:
\begin{lstlisting}[language=Python,firstnumber=last]
planet.r = 0.1       # Radius in units of primary radius
planet.L = 5.e-3     # Luminosity in units of primary
planet.a = 50        # Semi-major axis in units of primary
planet.inc = 90      # Inclination in degrees
planet.ecc = 0       # Eccentricity
planet.w = 90        # Longitude of pericenter in degrees
planet.Omega = 0     # Longitude of ascending node in degrees
planet.lambda0 = 90  # Mean longitude in deg. at the ref. time
planet.tref = 0      # Reference time in days
planet.porb = 4.3    # Orbital period in days
planet.prot = 4.3    # Rotational period in days
\end{lstlisting}
These properties, along with their default values, are detailed in full
in the
\href{https://rodluger.github.io/starry/api.html\#starry.kepler.Secondary}{documentation}.

Suppose we wish to give the planet a simple
dipole map ($Y_{1,0}$) with peak brightness at the sub-stellar point.
\starry expects the planet map to be instantiated at an eclipsing
configuration (full phase), so we want to set the coefficient for the $Y_{1,0}$
harmonic (see Figure~\ref{fig:ylms}):
\begin{lstlisting}[language=Python,firstnumber=last]
planet[1, 0] = 0.5
\end{lstlisting}
Finally, care should be taken to ensure the map is positive everywhere. By default,
the coefficient of the $Y_{0,0}$ term of a map is fixed at unity, since
changing this term would change the total luminosity (this should instead be
modified via the \SecondaryL property). In the case of a simple dipole map of
the form $Y_{0,0} \, + \, y_1 Y_{1, -1} \, + \, y_2 Y_{1, 0} \, + \, y_3 Y_{1, 1}$,
it can be shown that as long as we enforce
\begin{proof}{l1positive}
    \label{l1positive}
    y_1^2 + y_2^2 + y_3^2 \le \frac{1}{3}
\end{proof}
the map will be non-negative everywhere along the unit sphere.
Since $0.5^2 < \nicefrac{1}{3}$,
our map is in fact positive semi-definite.
For more details on ensuring surface maps are positive everywhere, see
\S\ref{sec:nonnegative}.

\begin{figure}[t!]
    \begin{centering}
    \includegraphics[width=\linewidth]{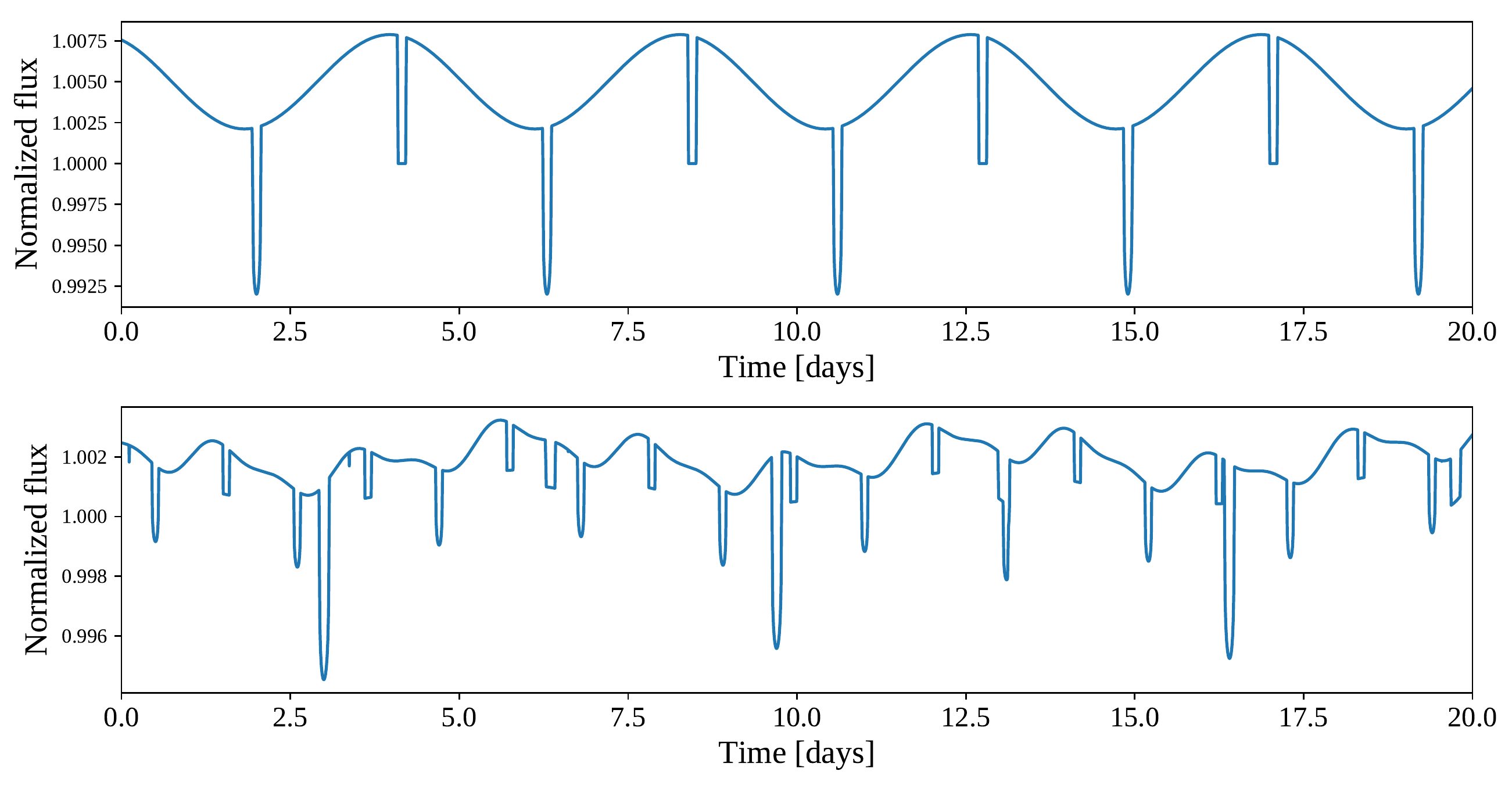}
    \caption{\label{fig:system}
             Sample analytic exoplanet system light curves computed with \starry.
             \emph{Top:} a hot Jupiter
             transiting a Sun-like star. The planet's map is a simple dipole,
             with the hotspot offset 15$^\circ$ from stellar noon; the offset
             in the secondary eclipse from the peak of the phase curve is
             apparent. \emph{Bottom:} a two-planet system with more complex
             surface maps. In addition to transits and secondary eclipses,
             a few planet-planet occultations are visible (e.g., the very short
             events at $t=0.1$ and $t=3.4$ days).
             \codelink{system}}
    \end{centering}
\end{figure}

We are now ready to instantiate the planetary system:
\begin{lstlisting}[language=Python,firstnumber=last]
from starry.kepler import System
system = System(star, planet)
\end{lstlisting}
(note that the primary body must always be listed first).
We can now compute the full light curve:
\begin{lstlisting}[language=Python,firstnumber=last]
system.compute(time)
\end{lstlisting}
where \textsf{time} is the array of times (in days) at which to compute the
light curve. This command internally calls the \Mapflux method of each
of the surface maps, populating the \textsf{flux} attribute of each body
with its respective light curve. The total light curve (the sum of the
light curves of each of the bodies in the system, including the star) is
stored in \Systemlightcurve. The top panel of Figure~\ref{fig:system} shows
the light curve for the system we instantiated above: both the transits
and secondary eclipses of the planet are clearly visible. For flair, we added
a hotspot offset of $15^\circ$ to simulate advection of heat by \edited{an
eastward} wind, causing the peak of the planet's phase curve to occur slightly
before secondary eclipse (refer to the \Python script for details).

The bottom panel of the figure shows a two-planet system with more elaborate
surface maps. In addition to the transits, eclipses, and complex phase curve
morphology, several planet-planet occultations are also visible in the
light curve.

\subsection{Gradients of the light curves}
\label{sec:gradients}
Since all expressions derived in this paper are analytic, so too are their
derivatives. The ability to compute derivatives of a light curve model with
respect to the model parameters can be extremely useful in both optimization
and inference problems. When fitting a model to data with an optimization
algorithm, knowledge of the gradient of the objective function can greatly
speed up convergence, as the optimizer always ``knows'' which direction to
take a step in to improve the fit. Gradients can also be used in Hamiltonian
Monte Carlo (HMC) simulations, in which the gradient of the likelihood is used
to improve the efficiency of the sampler and greatly speed up convergence of
the chains \citep[e.g.,][]{Betancourt2017}.

In principle, one could differentiate the recurrence relations in
the Appendix and arrive at expressions for the derivatives of a light curve with
respect to any of the input parameters. \citet{Pal2008} derived gradients in
this fashion for the case of transits across a quadratically limb-darkened star.
However, for the complex surface maps we consider here, differentiating all our
equations would be an extremely tedious
task. Instead, we can take advantage of the analytic nature of our expressions
and compute all derivatives using automatic differentiation \citep[autodiff; e.g.,][]{Wengert1964}.
Despite the complexity of the expressions we derive here, each of the individual
steps involved in computing a light curve is either a basic arithmetic operation
or the evaluation of an elementary function and is therefore trivially differentiable.
Autodiff algorithms exploit this fact by repeatedly applying the chain rule to
compute the derivatives of any function during its evaluation, returning derivatives
that can be accurate to high precision and at a speed that can be significantly greater
than that of numeric (or symbolic) differentiation.

We employ the autodiff algorithm of the Eigen \citep{eigen} C++ library to compute
derivatives of the flux with respect to all input parameters. Although fast,
evaluation of the derivatives introduces overhead to the computation
and is therefore disabled by default. To enable it, users should pass the
\textsf{gradient=True} keyword argument to \Mapflux or \Systemcompute. In the former
case, the gradient is returned in a tuple alongside the flux; in the latter,
the gradient of the total light curve is stored in the \textsf{gradient}
property of the system, and the gradient of each body's light curve is stored
in that body's own \textsf{gradient} property.
As an example, let us instantiate a \Map
class and compute the flux at one point during an occultation:
\begin{lstlisting}[language=Python,firstnumber=last]
map = Map(lmax=1)
map[1, 0] = 0.5
flux, gradient = map.flux(theta=30, xo=0.1, yo=0.1, ro=0.1,
                          gradient=True)
\end{lstlisting}
Running the code above returns the value of the flux, \textsf{1.48216...},
as well as a dictionary of derivatives of the flux with respect to all
input parameters:
\begin{lstlisting}[language=Python,firstnumber=last]
{'theta': array([-0.0049768]),
 'xo': array([-0.00356856]),
 'yo': array([0.00076157]),
 'ro': array([-0.35638527]),
 'y': array([[ 0.99      ],
             [-0.00173205],
             [ 0.98432307],
             [-0.57029919]]),
 'u': array([])}
\end{lstlisting}
These are the derivatives with respect to
the rotational phase, the position and radius of the occultor,
and each of the spherical harmonic and limb darkening coefficients.
\begin{figure}[p!]
    \begin{centering}
    \includegraphics[width=0.9\linewidth]{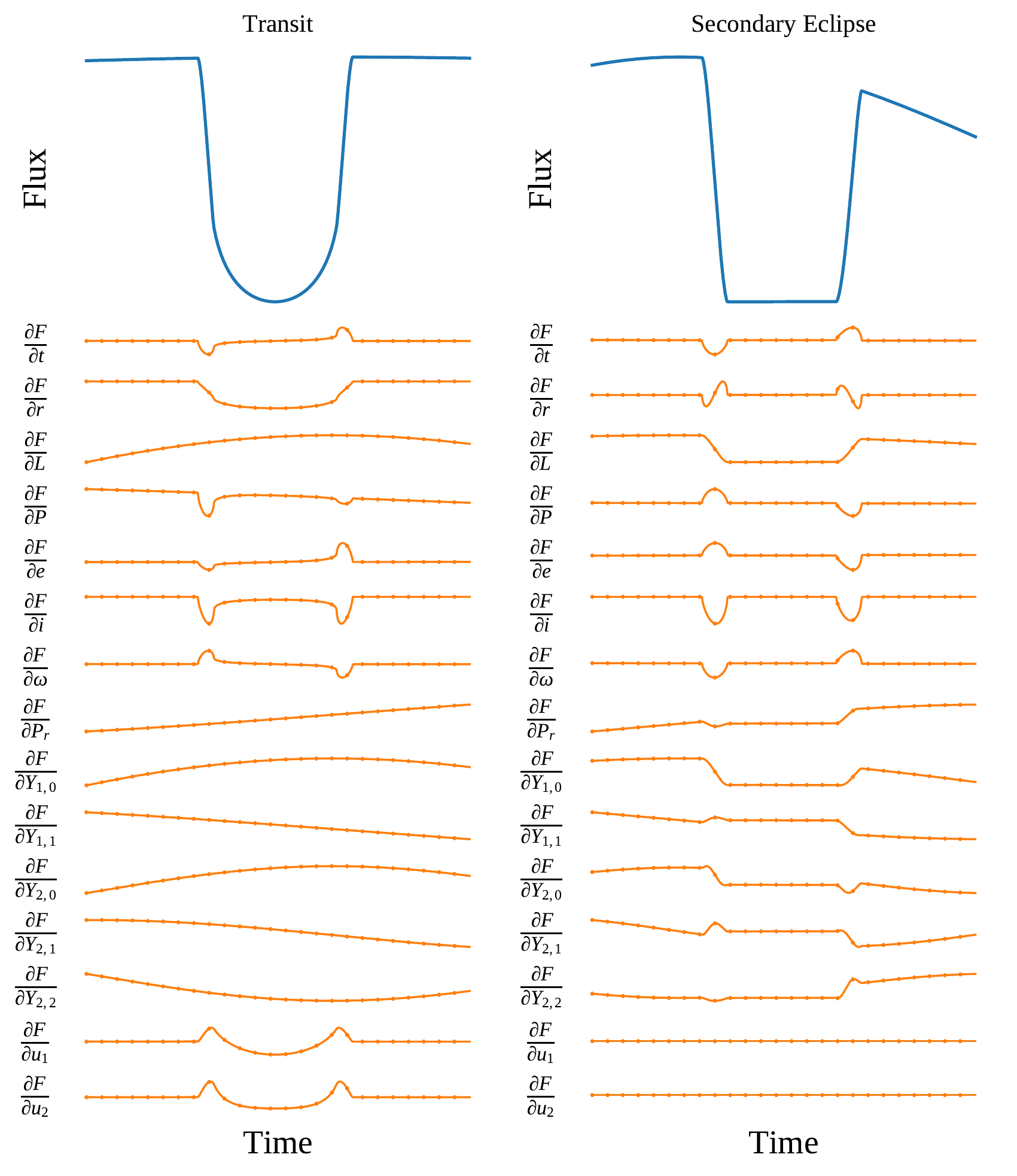}
    \caption{\label{fig:autodiff}
             Transit (top left) and secondary eclipse (top right) of a mildly
             eccentric, slightly inclined, quickly-rotating hot Jupiter with a
             dipole map, computed with \starry. Derivatives
             as a function of time for several of these parameters are plotted in
             orange below each light curve. Solid lines correspond to the analytic
             derivatives and dots correspond to derivatives evaluated numerically
             using finite differences. From top to bottom, the curves correspond
             to derivatives with respect to time, planet radius, planet luminosity,
             orbital period, eccentricity, inclination, longitude of pericenter,
             rotational period, five of the planet surface map coefficients, and
             the linear and quadratic stellar limb darkening coefficients.
             \codelink{autodiff}}
    \end{centering}
\end{figure}
Figure~\ref{fig:autodiff} shows an example of the autodiff capabilities of
\starry for a transit and a secondary eclipse of a hot Jupiter.

\color{black}

\subsection{Benchmarks}
\label{sec:starrybenchmarks}

We validate all our calculations of rotational phase curves and occultation
light curves by comparing them to numerical solutions of the corresponding
surface integrals. We integrate the specific intensity of the body by
discretely summing over its surface map on an adaptive radial mesh whose
resolution is iteratively increased wherever the spatial gradient of the
%
%
specific intensity is large and in the vicinity of the limb of the occultor.

%
%

All light curves in Figure~\ref{fig:ylmlightcurves} show the flux computed
in this way as the small points along each of the curves. We find that our
analytic light curves agree with the numerical solutions to within the error
of the latter, \edited{which is on the order of $10^{-5}$}.

\begin{figure}[p!]
    \begin{centering}
    \includegraphics[width=0.925\linewidth]{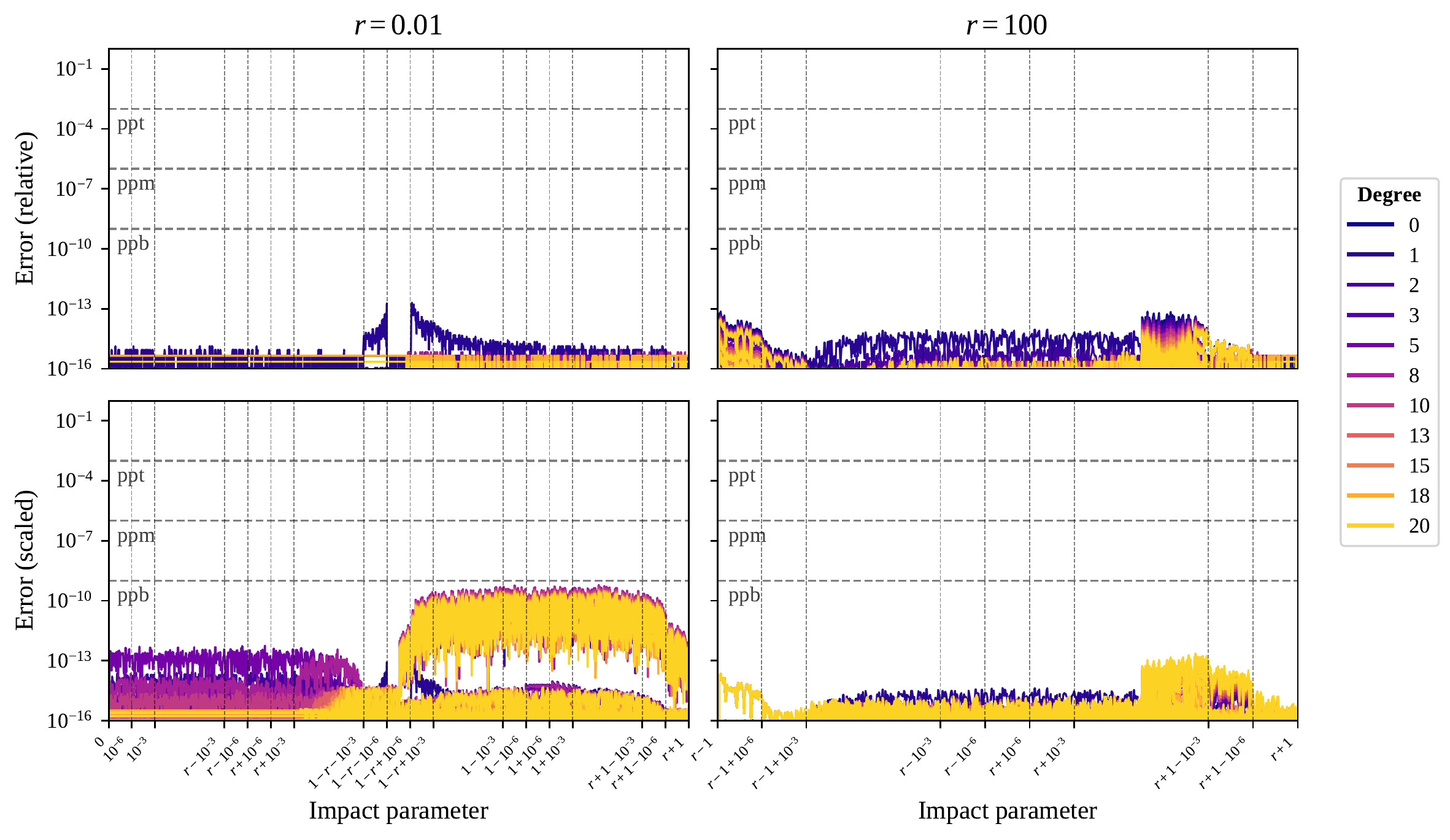}
    \caption{\label{fig:stability}
             Error in the terms of the solution vector $\mathbf{s}$ for a small
             occultor ($r = 0.01$, left) and a large occultor ($r = 100$, right),
             computed relative to calculations using quadruple
             floating point precision. The error is plotted as a function of impact parameter
             for terms with $\mu$ even (left)
             and $\mu$ odd (right). The horizontal axis extends from
             $0$ to $r + 1$ \edited{(left panel)}
             $r - 1$ to $r + 1$ \edited{(right panel)}
             and covers
             the entire range of $b$ during an occultation, with extra resolution near
             potentially unstable regions. The top panel shows the relative error
             and the bottom panel shows the fractional error, scaled to the largest value
             of $\mathbf{s}$ over the course of the occultation.
             \codelink{stability}}
    \end{centering}
\end{figure}

\begin{figure}[p!]
    \begin{centering}
    \includegraphics[width=0.925\linewidth]{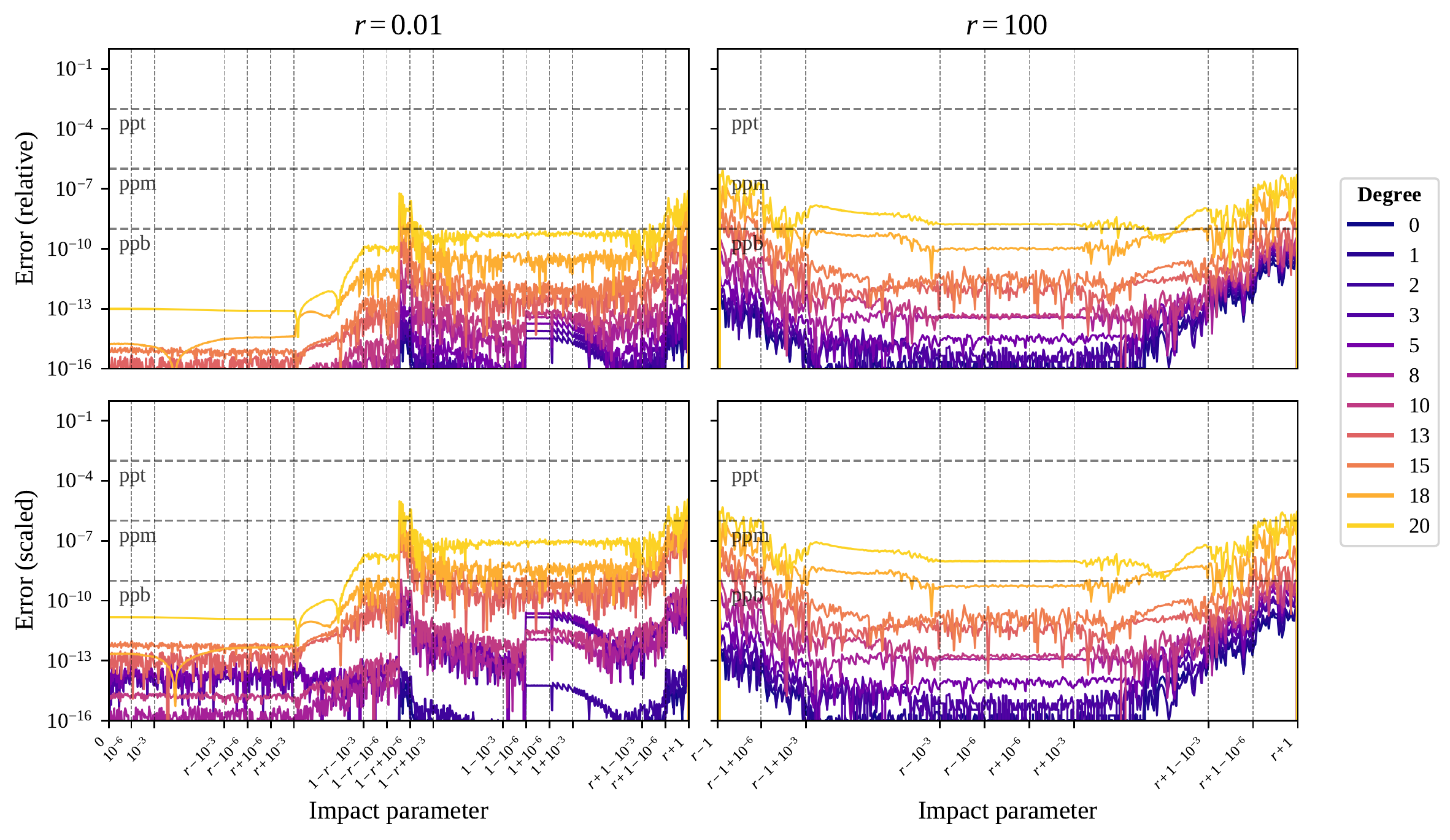}
    \caption{\label{fig:stability_grad}
             Similar to Figure~\ref{fig:stability}, but showing instead the
             error on the \emph{derivative} of the flux with respect to the impact
             parameter computed analytically with autodifferentiation. The error is computed
             relative to a numerical derivative computed at 128 bit precision.
             \codelink{stability_grad}}
    \end{centering}
\end{figure}

To test for numerical stability, we also compare our calculations to
the same calculations performed at quadruple (128-bit) floating-point precision.
Figure~\ref{fig:stability} shows the relative error on the computation of each of the
terms in the solution vector $\mathbf{s}$ up to $l = 20$ for a small occultor (left)
and a large occultor (right). The horizontal axis corresponds to the impact
parameter, spanning all possible values of $b$ during an occultation.
In both cases, the maximum relative error (top panel) is less than one part per
trillion. The bottom panel shows the fractional error, equal to the relative
error scaled to the largest value of the function during the occultation. In cases
where the largest value of the flux is less than $10^{-9}$, we scale the relative
error to this value to avoid division by a very small number. In all cases, the
fractional error is less than 1 ppb.

In Figure~\ref{fig:stability_grad} we show similar curves for the numerical error
on the \emph{derivative} of the flux with respect to the impact parameter. While
the derivatives are in general more prone to numerical instabilities, particularly in the
limits $b \rightarrow |1 - r|$ and $b \rightarrow 1 + r$, we find that the error
is less than 1 ppb over most of the domain for both small and large occultors. For
large values of $l$ near the unstable regions, the error approaches 1 ppm.

\begin{figure}[t!]
    \begin{centering}
    \includegraphics[width=0.85\linewidth]{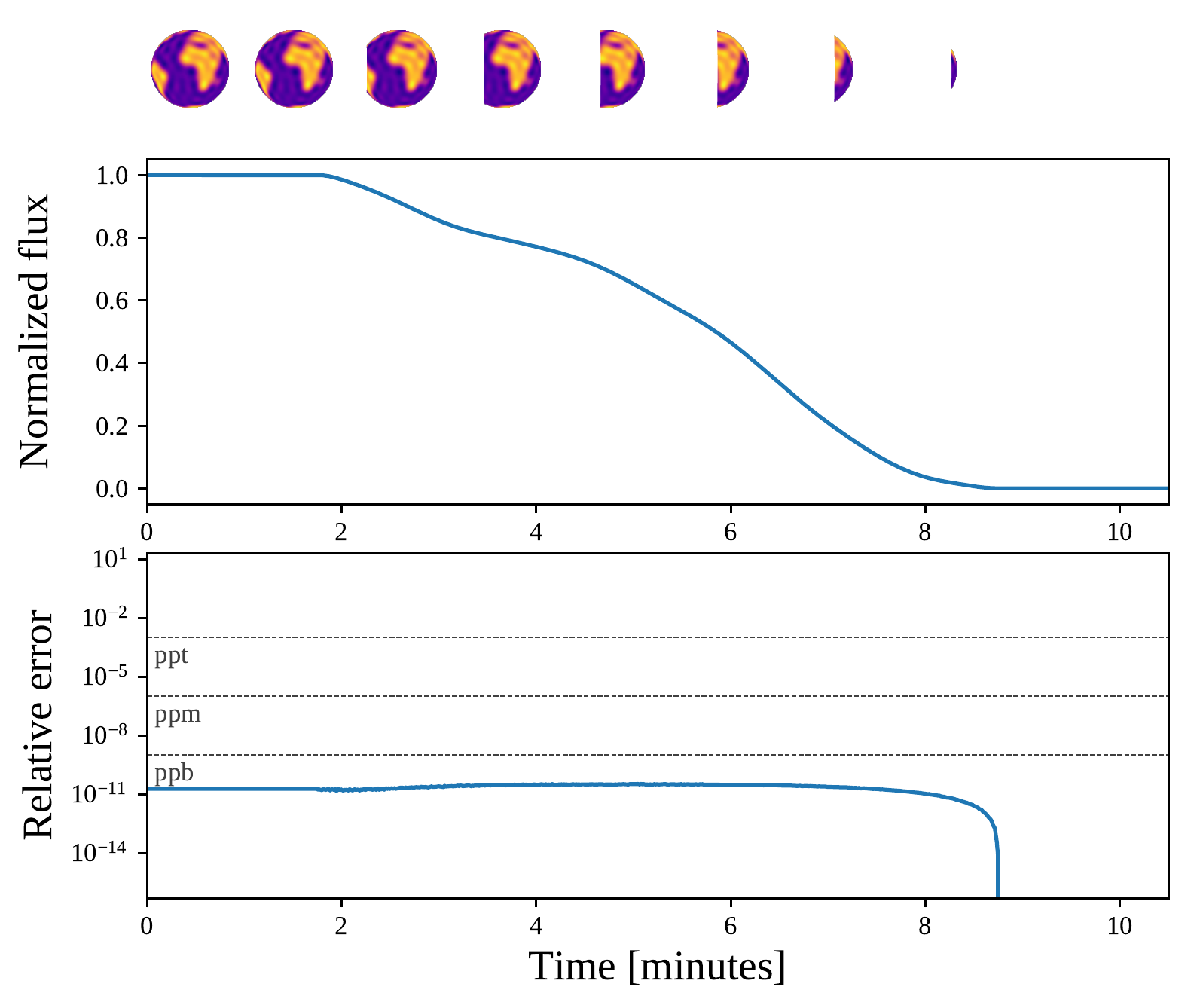}
    \caption{\label{fig:stability_earth}
             Secondary eclipse ingress for the Earth being occulted by the Sun
             ($r = 110$), computed for a $l = 20$ expansion of the planet's
             surface map. The relative error due to floating point precision loss
             is shown at the bottom and is
             less than one part per billion everywhere.
             \codelink{stability}}
    \end{centering}
\end{figure}

Figure~\ref{fig:stability_earth} shows the error on a secondary eclipse
light curve for an $l = 20$ expansion of the Earth being occulted by the Sun.
As expected, the relative error (relative to the Earth's flux) is much less than 1 part
per billion everywhere.

\begin{figure}[t!]
    \begin{centering}
    \includegraphics[width=0.75\linewidth]{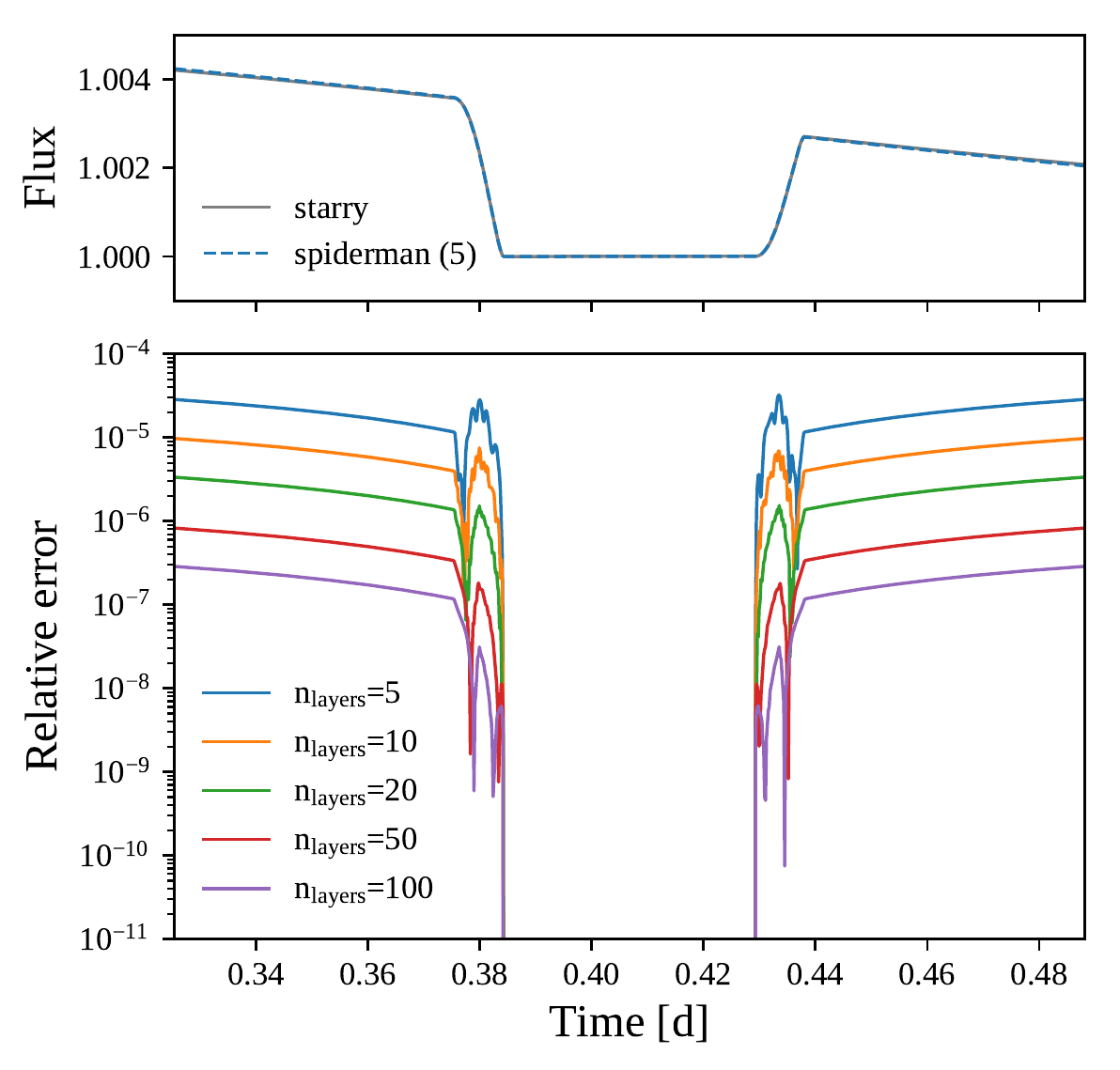}
    \caption{\label{fig:spidercomp_flux}
             Comparison to a light curve generated using the \spiderman code
             \citep{Louden2018}. The top panel shows a secondary eclipse of a
             hot Jupiter with an offset dipole computed with \starry (solid grey)
             and \spiderman (dashed blue), and the bottom panel shows the absolute
             value of the difference between the two light curves as the number
             of layers in the \spiderman grid is increased. The \spiderman solution
             \edited{slowly approaches} the \starry solution as the number of layers
             is increased.
             \codelink{spidercomp}}
    \end{centering}
\end{figure}

Finally, we compare our secondary eclipse and phase curve computations to
light curves generated using the \spiderman package \citep{Louden2018}.
The top panel of Figure~\ref{fig:spidercomp_flux} shows a secondary eclipse light curve
for a hot Jupiter with an offset dipole map ($l = 1$) computed with
\starry (solid blue) and \spiderman (dashed orange) using the default number
of layers ($n_\mathrm{layers} = 5$) in their discretized surface intensity grid.
The bottom panel shows the relative difference between the \spiderman flux and
the \starry flux for different values of $n_\mathrm{layers}$. For the default
number of layers, the maximum relative error in the \spiderman flux is on the
order of 30 ppm (relative to the \emph{stellar} flux)
during ingress and egress and is somewhat higher at the peak
of the phase curve. Relative to the \emph{planet} flux, the error is more significant: $\sim 30 / 0.004 = 7500$ ppm.
This error decreases linearly as the number of layers increases, and the \spiderman solution
appears to \edited{approach} the \starry solution in the limit $n_\mathrm{layers} \rightarrow \infty$.

\subsection{Speed tests}
\label{sec:starryspeed}

\begin{figure}[t!]
    \begin{centering}
    \includegraphics[width=0.9\linewidth]{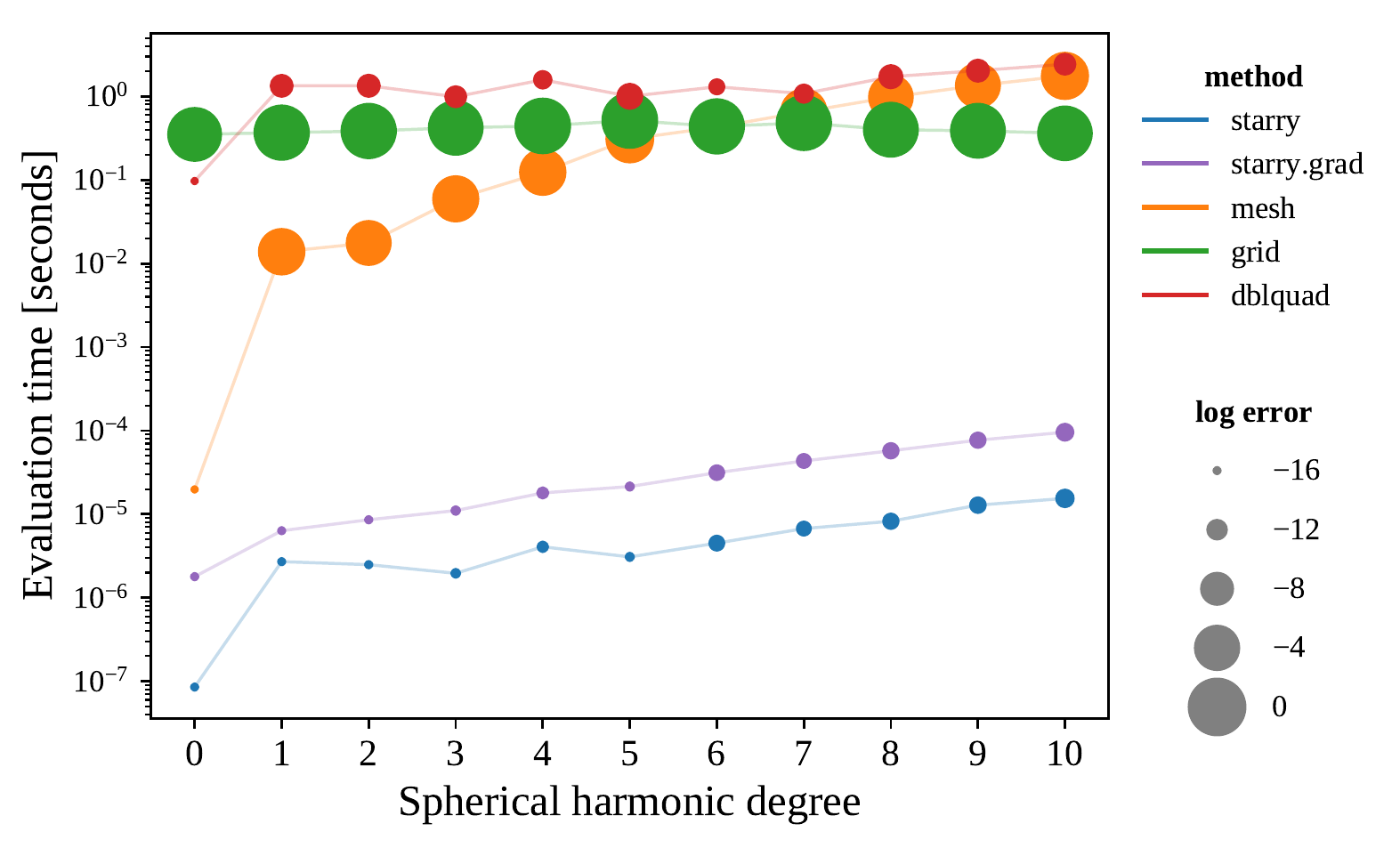}
    \caption{\label{fig:compare_to_numerical}
             Evaluation time for a single occultation calculation as a function
             of the spherical harmonic degree of the map using
             \starry (blue), \starry with gradients (purple), the adaptive mesh technique (\S\ref{sec:starrybenchmarks}, orange),
             brute force integration on a Cartesian grid (green), and
             \textsf{scipy}'s \textsf{dblquad} two-dimensional numerical
             integration routine (red).
             The size of each point is proportional to the log of the error relative
             to the \starry quadruple-precision solution.
             \codelink{speed}}
    \end{centering}
\end{figure}

Figure~\ref{fig:compare_to_numerical} shows the evaluation time for
occultation calculations as a function of the spherical
harmonic degree $l$ of the map. Analytic solutions computed with \starry
are shown as the blue dots (\edited{purple dots for solutions with gradients enabled}).
Also shown are calculations using the adaptive
mesh technique described in the previous section (orange dots),
brute-force integration on a 300$\times$300 Cartesian grid (green dots),
and numerical evaluation of the double integral using the
\textsf{scipy.integrate.dblquad} \citep{scipy} routine (red). The size of each point
is proportional to the log of the fractional error relative to the
\starry \edited{quadruple floating-point precision} solution. Light curve computation using \starry is several orders of
magnitude faster and more accurate than any other evaluation technique.

\begin{figure}[p!]
    \begin{centering}
    \includegraphics[width=\linewidth]{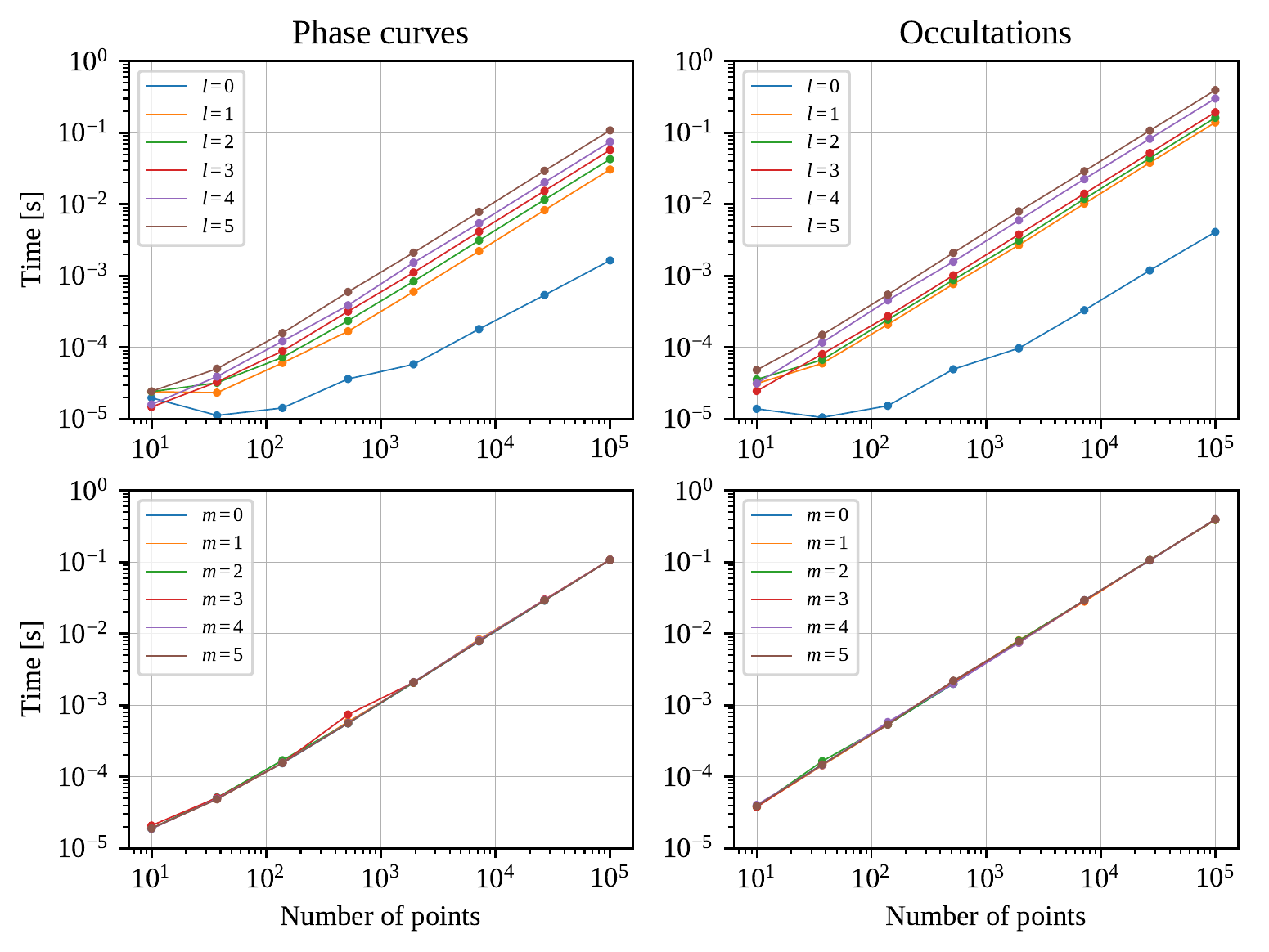}
    \caption{\label{fig:speed}
             Speed tests for \starry, showing the light curve evaluation time
             as a function of number of light curve points for rotational
             phase curves (left) and occultation light curves (right) of individual
             spherical harmonics. The top
             panels show the evaluation time for spherical harmonics of different
             degrees $l$, averaged over all orders $m$. The bottom panels
             show the time for each of the non-negative orders ($m \ge 0$)
             of the $l = 5$ harmonics.
             \codelink{speed}}
    \end{centering}
\end{figure}

\begin{figure}[p!]
    \begin{centering}
    \includegraphics[width=0.65\linewidth]{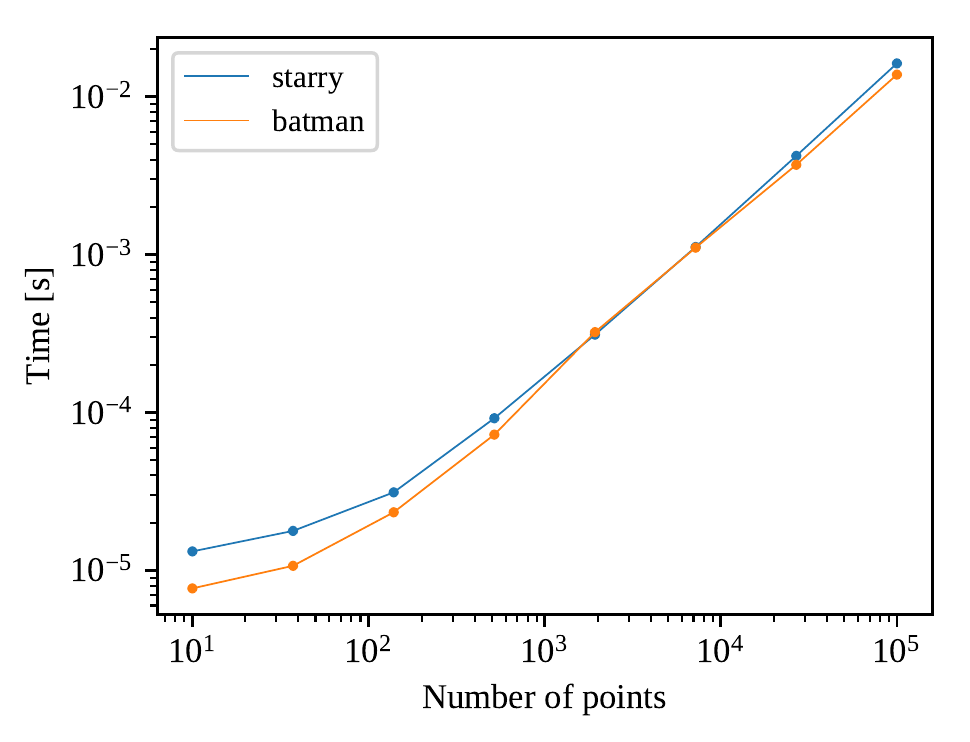}
    \caption{\label{fig:speed_batman}
             Speed comparison to the \batman transit modeling package
             \citep{Kreidberg2015} for a hot Jupiter transit across a
             quadratically limb-darkened star.
             \codelink{speed_batman}}
    \end{centering}
\end{figure}

Figure~\ref{fig:speed} shows the evaluation time for \starry
as a function of the number of points in the light curve for phase curves
(left) and occultation light curves (right). The top panel shows curves for
maps of different degree $l$, and the bottom panel shows curves for single-order
maps of degree $l = 5$. Evaluation time scales exponentially with increasing
degree, but \starry can compute full occultation light curves for $l = 5$ maps
with $10^5$ points in under one second. Evaluation time is roughly constant
across the different orders at fixed degree.
In Figure~\ref{fig:speed_batman} we show a speed comparison
to the \batman transit package \citep{Kreidberg2015} for transits across a
quadratically limb-darkened star. \starry is as efficient as \batman at
computing transit light curves.

\begin{figure}[t!]
    \begin{centering}
    \includegraphics[width=\linewidth]{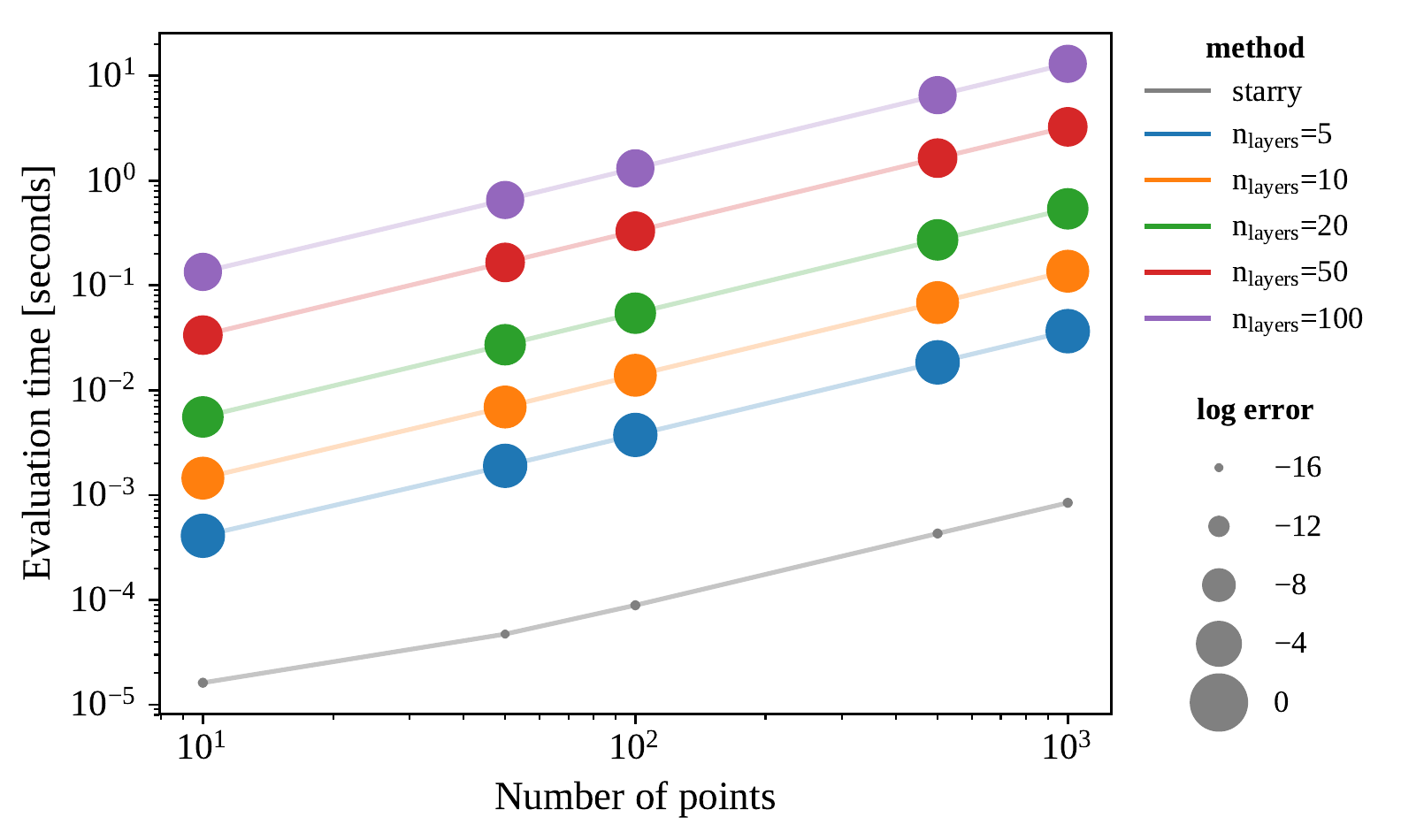}
    \caption{\label{fig:spidercomp}
             Speed comparison to the \spiderman code package \citep{Louden2018}.
             For a dipole ($l = 1$) map, \starry computes secondary eclipse
             light curves and phase curves in about an order of magnitude less
             time compared to \spiderman with 5 layers in their surface intensity
             grid. The size of the markers is proportional to the log of the
             error in the solutions, compared to evaluation of the equations
             presented in this paper using quadruple floating-point precision.
             As the number of layers is increased in \spiderman, the error
             decreases linearly, but the evaluation time increases proportionally.
             \codelink{spidercomp}}
    \end{centering}
\end{figure}

Finally, Figure~\ref{fig:spidercomp} shows the evaluation time for \starry
compared to that for \spiderman \citep{Louden2018} for a secondary eclipse
of a simple $l = 1$ map and varying values of the number of layers in the
\spiderman grid. The size of the points is proportional to the log of the
relative error on the solution (see Figure~\ref{fig:spidercomp_flux}).
The evaluation time for \starry is about one order of magnitude less than
\spiderman for the default $n_\mathrm{layers}=5$, for which the error
is about 30 ppm (relative to the stellar flux). Computation of the \spiderman
light curves with larger values of $n_\mathrm{layers}$ improves the precision
but leads to proportionally longer evaluation times.

\subsection{Application to real data: HD 189733b}
\label{sec:hd189}

As a brief example of the application of \starry to a real dataset, we analyzed the
well-studied \textit{Spitzer}/IRAC 8 $\mu$m secondary eclipse light curve of the
hot jupiter HD 189733b from \citet{Knutson2007}.
Our analysis is very similar to that of \citet{Majeau2012}, who also fit a spherical
harmonic map to the secondary eclipse data, but evaluated their model numerically.
We fit for the $l = 1$ map coefficients
and the planet luminosity, holding the orbital parameters constant for simplicity.
Unlike \citet{Majeau2012}, we fit each of the nearly 128,000 observations in the
timeseries without binning. We first find the maximum likelihood fit to the
data using \edited{gradient-descent optimization}, then initialize an MCMC sampler in a
Gaussian ball about this solution and run a chain of 40 walkers for 10,000 steps
using the \textsf{emcee} package \citep{Foreman-Mackey2013}.
\edited{At a rate of about one million flux evaluations per second,
the full calculation took on the order of 10 CPU hours. Note that this is almost certainly
overkill; given the extremely low signal-to-noise ratio of each measurement,
binning the dataset in time could allow for runtimes of less than one hour that would yield
virtually the same results.}

\begin{figure}[t!]
    \begin{centering}
    \includegraphics[width=\linewidth]{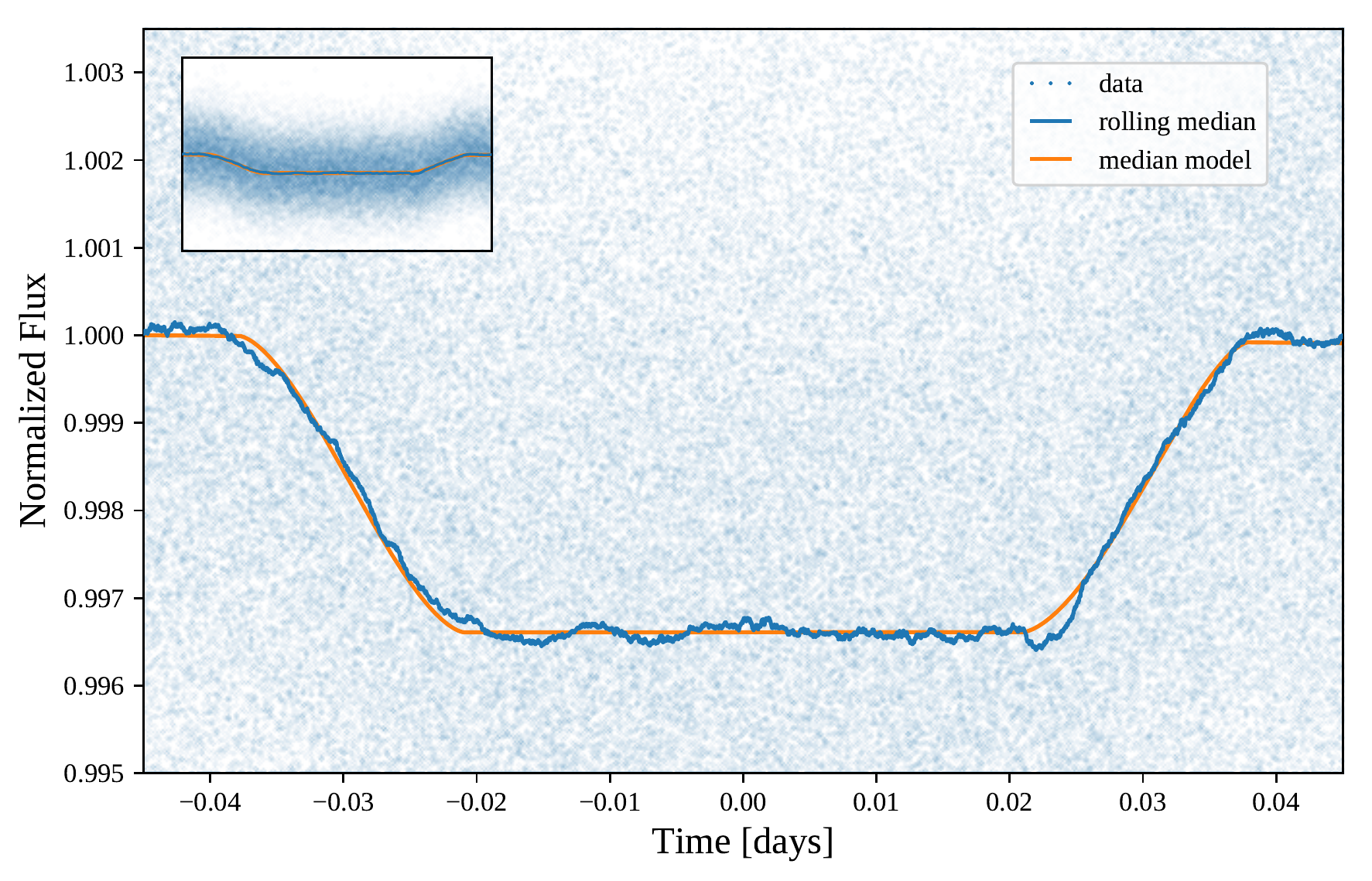}
    \caption{\label{fig:hd189_mcmc_fit} Secondary eclipse light curve of
             HD 189733b observed with \textit{Spitzer}/IRAC at 8 $\mu$m from
             \citet{Knutson2007} along with our fit to the data. The inset
             at the top left shows a zoomed-out version of the timeseries.
             \codelink{hd189733b}}
    \end{centering}
\end{figure}

\begin{figure}[t!]
    \begin{centering}
    \includegraphics[width=\linewidth]{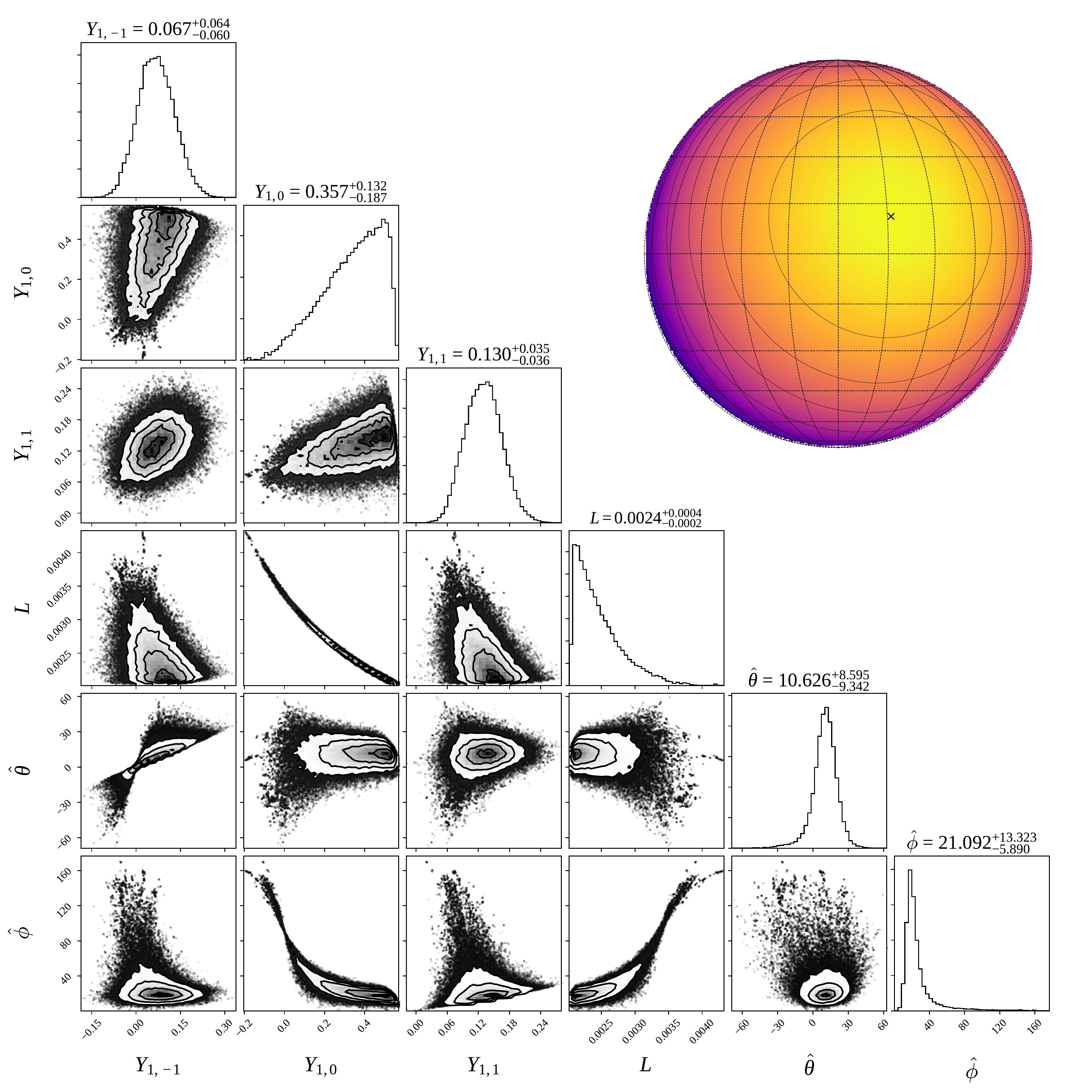}
    \caption{\label{fig:hd189_mcmc_corner} Posterior distributions
             of our model parameters from the fit to the HD189733b
             secondary eclipse, plotted using the \textsf{corner} package
             \citep{Foreman-Mackey2016}. We fit for the spherical harmonic
             coefficients $Y_{1,-1}$,
             $Y_{1,0}$, and $Y_{1,1}$, as well as the planet luminosity $L$. The
             latitude $\hat{\phi}$ and longitude $\hat{\theta}$
             of the hotspot were calculated from these coefficients. The median
             value of each parameter was used to generate the map of HD189733b
             shown in the upper right. \edited{An \textsf{X} marks the inferred location
             of the hotspot and contour levels indicate 10\% drops in specific
             intensity. For reference, a latitude/longitude grid is
             superimposed with cells measuring $15^\circ$ on a side.}
             \codelink{hd189733b}}
    \end{centering}
\end{figure}

Figure \ref{fig:hd189_mcmc_fit} shows the secondary eclipse light curve and our
median model fit to the data. Figure \ref{fig:hd189_mcmc_corner}
shows the marginalized posteriors and covariances of our four model parameters, as
well as the latitude ($\hat{\phi}$) and longitude ($\hat{\theta}$) of the hotspot
relative to the substellar
point. A map corresponding to \edited{the maximum likelihood model} is plotted at the top right,
showing a statistically significant eastward
offset in the location of the hotspot in agreement with previous studies
\citep{Knutson2007, Majeau2012, deWit2012}. There is also evidence for a slight northward offset,
although it is less statistically significant and consistent with zero.
Note that because we did not simultaneously
fit phasecurve data, there is a strong degeneracy between the planet's
total luminosity and the $Y_{1,0}$ spherical harmonic coefficient. Moreover,
since we did not account for the uncertainty in the planet's orbital parameters,
we are likely underestimating the uncertainty on the map coefficients.

%

\section{Caveats}
\label{sec:caveats}

\subsection{Wavelength dependence}
In our formalism thus far we have avoided mention of wavelength dependence
of a body's surface map. In our derivations we treated the specific intensity at a point
on the body's map as a scalar: a single number corresponding to the total power
emitted to space by an infinitesimal area element on the body's surface.
When applying \starry to actual data, this intensity can either
be the power integrated over a range of wavelengths, in which case the light curve
has units of flux proper,
corresponding to (say) the quantity measured by an instrument performing filter
photometry; or the power at a \emph{specific} wavelength, in which case the light curve
computed by \starry has units of \emph{spectral} flux, corresponding to (say) the
flux measured in a tiny wavelength bin by a spectrometer. Note, importantly,
that in the former case the ``surface map'' is in reality the integral of the body's
wavelength-dependent specific intensity convolved with the instrument's spectral
response function over a given wavelength range.

\edited{
Alternatively, inspection of \eq{starry} suggests that the methods outlined
above for computing light curves can be trivially extended to
wavelength-dependent maps. Since neither the solution vector, the
change of basis matrix, nor the rotation matrices depend on the values of the
map coefficients, one can compute a wavelength-dependent light curve as
\begin{equation}
    \label{eq:starryspectral}
        \bvec{f} = \bvec{s}^{\boldsymbol{\mathsf{T}}} \bvec{A} \, \bvec{R'} \, \bvec{R} \, \bvec{Y}
    \quad,
\end{equation}
where $\bvec{f}$ is the vector of fluxes, one per wavelength bin, and
$\bvec{Y}$ is now a \emph{matrix} of spherical harmonic coefficients, where
each column is the usual $\bvec{y}$ vector corresponding to a specific
wavelength bin. This method makes it extremely fast to compute
wavelength-dependent light curves, since the solution vector and rotation
matrices need only be computed \emph{once} for all wavelength bins.

In \starry, users can set the \textsf{nwav} keyword argument when instantiating
maps to indicate the number of wavelength bins (the default is 1). For
multi-wavelength maps, the coefficient at a given value of \textsf{(l, m)} is
a \emph{vector}, corresponding to the value of the spherical harmonic coefficient
in each wavelength bin. All intensities, fluxes, and gradients computed by
\starry gain an extra dimension in this case.
}

\subsection{Reflectance light curves}
At present, \starry can only compute thermal phase curves and occultation
light curves for planets and moons. Reflectance light curves are significantly
more difficult to compute analytically because of the sharp discontinuity in the
illumination gradient at the terminator. In principle, the stellar illumination
pattern could be modeled with a high order spherical harmonic expansion, but
this approach cannot accurately capture the sharp day/night transition at the
terminator and typically leads to spurious ringing on the night side. A
better approach is to treat the terminator as one of the boundaries of the
surface integral and use Green's theorem to compute the line integral about
this elliptical curve. This will be the topic of a future paper and will
be implemented in future versions of the code.

\subsection{Anisotropic emission}
\edited{
It is important to note that the formalism developed here for computing
light curves (excepting our treatment of limb darkening)
implicitly assumes isotropic emission from the body's surface.
While this is usually an appropriate assumption for emission, it could
break down due to scattering by, say, clouds or hazes in a planet's
atmosphere.
}

\subsection{Physical surface maps}
\label{sec:nonnegative}
While spherical harmonics are a convenient way to approximate surface
maps of celestial bodies, it is not trivial to ensure
that a given spherical harmonic expansion $\bvec{y}$ evaluates to
non-negative values everywhere on the unit sphere. This is because
there is no analytic way to compute the extrema of a function of
spherical harmonics of arbitrary degree.
This fact makes it difficult to enforce the physical prior that the
specific intensity of a celestial body cannot be negative, which
could be desirable when fitting a model to real data. The minimum
can, of course, be found numerically. In \starry, users can check
whether a map is positive semi-definite (P.S.D.) by evaluating
\begin{lstlisting}[language=Python,firstnumber=last]
map.is_physical()
\end{lstlisting}
which returns either \textsf{True} (the map is non-negative
everywhere) or \textsf{False} (at least one region on the map has
a negative specific intensity).
This method evaluates the surface map on a coarse grid in
$\uptheta$ and $\upphi$, locates the
approximate location of the minimum, and performs a gradient-descent
optimization to locate the global minimum of the map.

\edited{
Note that if any limb darkening coefficients are set, this method
will \emph{separately} determine whether the limb darkening profile
is physical by ensuring that it is positive everywhere \emph{and}
monotonically decreasing toward the limb.
}

\subsection{Maps of very large degree}
For \emph{very} large values of the spherical harmonic degree $l$, the
equations presented here may become numerically unstable. The evaluation
of the spherical harmonics depends on ratios of several factorials, whose
precision can degrade for large $l$ and $m$. Similarly, the coefficients
of the occultation solution vector $\mathbf{s}$ can drop below machine
precision at large $l$, leading to further numerical issues. While
\starry is specifically coded up to minimize numerical instabilities,
we find that for $l_\mathrm{max} \geq 30$ numerical issues may occur.
Fortunately, situations in which maps of such high degree are necessary
are not likely for exoplanet science in the foreseeable future.
Nevertheless, if users wish to perform calculations for very large $l$,
they can avoid these numerical issues by instantiating a multi-precision
map:
\begin{lstlisting}[language=Python,firstnumber=last]
map = Map(multi=True)
\end{lstlisting}
By default, this will perform all calculations using quadruple (128-bit)
floating point precision. We caution, however, that this will increase
computation time by at least an order of magnitude.

\subsection{Three-body events}
The occultation formalism developed in this paper applies specifically to the
case of a single occultor, so \starry cannot at present handle mutual occultations
involving more than two bodies; \edited{if a three-body event occurs, the computed
flux will be incorrect}. However, even for an arbitrary number of bodies
the problem is still analytic, since Green's theorem may be employed in the same
way, but instead evaluating the line integrals along the more complex network of
arcs defining the edges of the visible portion of the body's surface. This was
first noted by \cite{Pal2012}, whose \textsf{mttr} code computes analytic
transit light curves for mutually overlapping bodies such as a transiting
planet with a moon. Future versions of \starry will extend the calculations
to this general case.

\section{Conclusions}
\label{sec:conclusions}

In this paper, we derived a formalism to compute analytic thermal light curves of celestial
bodies in occultation, provided their specific intensity maps can be expressed as a sum
of spherical harmonics. Our expressions extend the analytic results of the
\citet{MandelAgol2002} transit model for limb-darkened stars to transits and
occultations of celestial bodies whose surface maps are not radially symmetric
and/or possess higher order features, and are thus generally applicable to
stars, planets, and moons. We derived recurrence relations to quickly compute
occultation light curves for surface maps expressed at arbitrary spherical harmonic degree.
We showed, in particular, that the flux contribution from higher degree terms
depends on the same elliptic integrals as the linear limb darkening term,
so these need only be evaluated once per light curve cadence. This results in
evaluation times for higher degree maps that are extremely fast, and
only marginally slower than in the quadratic limb darkening case.
In the limit of zero occultor size, our expressions trivially reduce to equations
for thermal phase curves of celestial bodies.

We introduced \starry, a \Python-wrapped model coded in \cpp
that can be used to compute phase curves and occultation light curves for
individual celestial bodies or entire exoplanet systems. \starry computes
transits, secondary eclipses, phase curves, and planet-planet occultations
analytically and is comparable in speed to other transit-modeling packages such
as \batman \citep{Kreidberg2015}. Because the light curves are all analytic,
\starry can also easily compute analytic gradients of the light curves with respect
to all input parameters via autodifferentiation, facilitating its interface with
gradient-based inference schemes such as Hamiltonian Monte Carlo (HMC)
or gradient-descent optimization methods.

Although we have in mind the application of this \starry to exoplanets, it could
in principle be applied to eclipsing binaries as well.  If the deformation of
the body is small and reflection is negligible, as is the case for long orbital periods,
then the surface brightness of each star can be decomposed into spherical harmonics,
and the \starry formalism may be used to integrate their phase curves and eclipses.
One could imagine applying \starry, for instance, to secondary eclipses of white
dwarfs to search for non-uniform surface brightness.

At present, \starry supports only
monochromatic surface maps, making it ideally suited for the modeling of
light curves collected via filter photometry, but future work will extend
it to spectrophotometry. \starry is also limited to thermal light curves of
planets and moons, since the discontinuity in the gradient of the
illumination pattern at the terminator makes it more challenging to analytically
solve the surface integrals in reflected light. However, an analytic solution
is likely to exist, and future work aims to extend \starry to this case.

The upcoming James Webb Space Telescope (JWST) and eventual
next-generation telescopes such as the Origins Space Telescope (OST) will
measure exoplanet secondary eclipses and
phase curves in the thermal infrared to unprecedented precision. \starry
can compute extremely fast and high-precision models for these
light curves, enabling the reconstruction of two-dimensional maps of these
alien worlds.

\acknowledgments
The \starry code is open source under the GNU General Public License
and is available at \url{https://github.com/rodluger/starry}, with
documentation and tutorials hosted at \url{https://rodluger.github.io/starry}.
A permanent version of the code used to generate the figures and results in this
paper is archived at \url{https://doi.org/10.5281/zenodo.1312286}.
This work was supported by the NASA Astrobiology
Institute's Virtual Planetary Laboratory under Cooperative
Agreement number NNA13AA93A.  EA is supported by NSF grant 1615315.
Some of the results in this paper have been derived using the HEALPix
\citep{Gorski2005} package.

\software{
starry v0.1.2 (\href{https://doi.org/10.5281/zenodo.1312286}{Luger et al. 2018}),
HEALPix (Gorski et al. 2005),
emcee (\href{http://adsabs.harvard.edu/abs/2013ascl.soft03002F}{Foreman-Mackey et al. 2013}),
corner.py (\href{https://doi.org/10.21105/joss.00024}{Foreman-Mackey 2016}),
batman (\href{http://adsabs.harvard.edu/abs/2015ascl.soft10002K}{Kreidberg 2015}),
spiderman (\href{http://adsabs.harvard.edu/abs/2017ascl.soft11019L}{Louden and Kreidberg 2017}),
pybind11 (Jakob et al. 2017),
Eigen v3 (Guennebaud et al. 2010),
scipy (Jones et al. 2001).
}

\bibliography{starry}
%

\pagebreak
\appendix

\section{Spherical harmonics}
\label{app:spharm}

In spherical coordinates, the spherical harmonics may be compactly represented
as in Equation~(\ref{eq:ylmtp}). The formalism in this paper requires us to express
them in Cartesian form, which is somewhat more cumbersome but still tractable.
Using Equation~(\ref{eq:xyz}) and expanding Equation~(\ref{eq:ylmtp})
via the multiple angle formula, we obtain
\begin{align}
    \label{eq:ylm0}
    Y_{lm}(\x, \y , \z) =
    \left(\frac{1}{\sqrt{1 - \z^2}}\right)^{|m|}
    \begin{dcases}
        \bar{P}_{lm}(\z)
        \sum_{j\, \mathrm{even}}^{m}
        \left(-1\right)^\frac{j}{2}
        \binom{m}{j}
        \x^{m - j}
        \y^j
         & \qquad m \geq 0
         \\[1em]
        \bar{P}_{l|m|}(\z)
        \sum_{j\, \mathrm{odd}}^{|m|}
        \left(-1\right)^\frac{j-1}{2}
        \binom{|m|}{j}
        \x^{|m| - j}
        \y^j
        & \qquad m < 0 \quad ,
    \end{dcases}
\end{align}
where $\binom{\bigdot}{\bigdot}$ is the binomial
coefficient. The normalized associated Legendre functions are defined as
\begin{align}
    \label{eq:plm}
    \bar{P}_{lm}(\z) &= A_{lm} \left(\sqrt{1-\z^2}\right)^m
                       \frac{\dd^m}{\dd \z^m}
                       \left[
                       \frac{1}{2^l l!}
                       \frac{\dd^l}{\dd \z^l}
                       \left(
                       \z^2 - 1
                       \right)^l
                       \right] \quad,
\end{align}
where
\begin{align}
    \label{eq:alm}
    A_{lm} = \sqrt{\frac{(2 - \delta_{m0})(2l + 1)(l - m)!}{4\pi(l + m)!}}
             \quad.
\end{align}
Expanding out the $\z$ derivatives, we obtain
\begin{align}
    \label{eq:plm_exp}
    \bar{P}_{lm}(\z) &= A_{lm} \left(\sqrt{1-\z^2}\right)^m\sum_{k=0}^{l-m}
                       \frac{2^l \left(\frac{l + m + k - 1}{2}\right)!}
                            {k!(l-m-k)!
                             \left(\frac{-l + m + k - 1}{2}\right)!}
                       \z^k
                       \quad,
\end{align}
which we combine with the previous results to write
\begin{proof}{ylmxyz}
    \label{eq:ylmxyz}
    Y_{lm}(\x, \y , \z) &=
    \begin{dcases}
        \sum_{j\, \mathrm{even}}^m\sum_{k=0}^{l-m}
        \left(-1\right)^{\frac{j}{2}}
        A_{lm}
        B_{lm}^{jk}
        \x^{m - j}
        \y^j
        \z^k
        \qquad & m \ge 0 \\
        \sum_{j\, \mathrm{odd}}^{|m|}\sum_{k=0}^{l-|m|}
        \left(-1\right)^{\frac{j-1}{2}}
        A_{l|m|}
        B_{l|m|}^{jk}
        \x^{|m| - j}
        \y^j
        \z^k
        \qquad & m < 0
    \end{dcases}
\end{proof}
where
\begin{align}
    \label{eq:blmjk}
    B_{lm}^{jk} =
    \frac{2^l m! \left(\frac{l + m + k - 1}{2}\right)!}
         {j! k! (m - j)! (l - m - k)!
          \left(\frac{-l + m + k - 1}{2}\right)!} \quad.
\end{align}
Since we are confined to the surface of the unit sphere, we have
$\z = \sqrt{1 - \x^2 - \y^2}$ and we may expand $\z^k$ using
the binomial theorem:
\begin{align}
    \z^{k} &= (1 - \x^2 - \y^2)^\frac{k}{2} \nonumber \\[0.5em]
          &=
          \begin{dcases}
              \sum_{p\,\mathrm{even}}^{k}
              \sum_{q\,\mathrm{even}}^p
              (-1)^\frac{p}{2}
              C_{pq}^{k}
              \x^{p-q} \y^{q}
              \qquad & k\,\mathrm{even} \\
              \sum_{p\,\mathrm{even}}^{k - 1}
              \sum_{q\,\mathrm{even}}^p
              (-1)^\frac{p}{2}
              C_{pq}^{k-1}
              \x^{p-q} \y^{q} \sqrt{1 - \x^2 - \y^2}
              \qquad & k\,\mathrm{odd} \quad,
          \end{dcases}
          \label{eq:zk}
\end{align}
where
\begin{align}
    \label{eq:ckpq}
    C_{pq}^{k} =
    \frac{\left(\frac{k}{2}\right)!}{\left(\frac{q}{2}\right)!
    \left(\frac{k-p}{2}\right)! \left(\frac{p-q}{2}\right)!} \quad.
\end{align}
This gives us an expression for the spherical harmonics $Y_{lm}$
as a function of $\x$ and $\y$ only:
\begin{proof}{ylmxyz}
    \label{eq:ylmxy}
    Y_{lm}(\x, \y) &=
    \begin{dcases}
        \!\begin{aligned}
            &
                \sum_{j\, \mathrm{even}}^m
                \sum_{k\, \mathrm{even}}^{l-m}
                \sum_{p\,\mathrm{even}}^{k}
                \sum_{q\,\mathrm{even}}^p
                \left(-1\right)^{\frac{j+p}{2}}
                A_{lm}
                B_{lm}^{jk}
                C_{pq}^{k}
                \x^{m - j + p - q}
                \y^{j + q}
            \, + \\
            &
                \sum_{j\, \mathrm{even}}^m
                \sum_{k\, \mathrm{odd}}^{l-m}
                \sum_{p\,\mathrm{even}}^{k - 1}
                \sum_{q\,\mathrm{even}}^p
                \left(-1\right)^{\frac{j+p}{2}}
                A_{lm}
                B_{lm}^{jk}
                C_{pq}^{k - 1}
                \x^{m - j + p - q}
                \y^{j + q}
                \z
       \end{aligned}
       &
       \quad m \ge 0 \\
       \\
       \!\begin{aligned}
           &
               \sum_{j\, \mathrm{odd}}^{|m|}
               \sum_{k\, \mathrm{even}}^{l-|m|}
               \sum_{p\,\mathrm{even}}^{k}
               \sum_{q\,\mathrm{even}}^p
               \left(-1\right)^{\frac{j+p-1}{2}}
               A_{l|m|}
               B_{l|m|}^{jk}
               C_{pq}^{k}
               \x^{|m| - j + p - q}
               \y^{j + q}
           \, + \\
           &
               \sum_{j\, \mathrm{odd}}^{|m|}
               \sum_{k\, \mathrm{odd}}^{l-|m|}
               \sum_{p\,\mathrm{even}}^{k - 1}
               \sum_{q\,\mathrm{even}}^p
               \left(-1\right)^{\frac{j+p-1}{2}}
               A_{l|m|}
               B_{l|m|}^{jk}
               C_{pq}^{k - 1}
               \x^{|m| - j + p - q}
               \y^{j + q}
               \z
      \end{aligned}
      &
      \quad m < 0
   \end{dcases}
\end{proof}
where $\z = \z(\x, \y) = \sqrt{1 - \x^2 - \y^2}$. Evaluating the nested sums
may be computationally slow, but these operations need only be performed a
single time to construct our change of basis matrix (following section).

\section{Change of Basis}
\label{app:basis}

In this section we discuss how to compute the change of basis matrices
$\AOne$ and $\ATwo$ from \S\ref{sec:basis} and provide links to
\Jupyter scripts to compute them.
Recall that the columns of the change of basis matrix from spherical
harmonics to polynomials, $\AOne$, are just the polynomial vectors
corresponding to each of the spherical harmonics in \eq{by}. From
Equations (\ref{eq:bp}) and (\ref{eq:ylmxy}), we can calculate
the first few spherical harmonics and their corresponding polynomial vectors:
\begin{equation}
\def\arraystretch{1.3}
\begin{array}{@{}lcccccccl@{}}
    \phantom{..}
    Y_{0,0} = \frac{1}{2\sqrt{\pi}}
    & & & & & & & &
    \bvec{p} = \frac{1}{2\sqrt{\pi}}
                  \begin{pmatrix}
                        1 & 0 & 0 & 0 & \cdot\cdot\cdot
                  \end{pmatrix}^\mathsf{T}
    \\
    Y_{1,-1} = \frac{\sqrt{3}}{2\sqrt{\pi}}\y
    & & & & & & & &
    \bvec{p} = \frac{1}{2\sqrt{\pi}}
                  \begin{pmatrix}
                        0 & 0 & 0 & \sqrt{3} & \cdot\cdot\cdot
                  \end{pmatrix}^\mathsf{T}
    \\
    \phantom{..}
    Y_{1,0} = \frac{\sqrt{3}}{2\sqrt{\pi}}\z
    & & & & & & & &
    \bvec{p} = \frac{1}{2\sqrt{\pi}}
                \begin{pmatrix}
                      0 & 0 & \sqrt{3} & 0 & \cdot\cdot\cdot
                \end{pmatrix}^\mathsf{T}
    \\
    \phantom{..}
    Y_{1,1} = \frac{\sqrt{3}}{2\sqrt{\pi}}\x
    & & & & & & & &
    \bvec{p} = \frac{1}{2\sqrt{\pi}}
                  \begin{pmatrix}
                        0 & \sqrt{3} & 0 & 0 & \cdot\cdot\cdot
                  \end{pmatrix}^\mathsf{T}
    \\
    Y_{2,-2} = \cdot\cdot\cdot
    & & & & & & & &
    \bvec{p} = \cdot\cdot\cdot
\end{array}
\end{equation}
From these we can construct $\AOne$. As an example, for spherical
harmonics up to degree $l_\mathrm{max} = 2$, this is
\begin{proof*}{A}
    \label{eq:AOne}
    \setstackgap{L}{1.25\baselineskip}
    \fixTABwidth{T}
    \AOne =
    \frac{1}{2 \sqrt{\pi }}
        \parenMatrixstack{
            1 & 0 & 0 & 0 & 0 & 0 & \sqrt{5} & 0 & 0 \\
            0 & 0 & 0 & \sqrt{3} & 0 & 0 & 0 & 0 & 0 \\
            0 & 0 & \sqrt{3} & 0 & 0 & 0 & 0 & 0 & 0 \\
            0 & \sqrt{3} & 0 & 0 & 0 & 0 & 0 & 0 & 0 \\
            0 & 0 & 0 & 0 & 0 & 0 & -\frac{3 \sqrt{5}}{2} & 0 & \frac{\sqrt{15}}{2} \\
            0 & 0 & 0 & 0 & 0 & 0 & 0 & \sqrt{15} & 0 \\
            0 & 0 & 0 & 0 & \sqrt{15} & 0 & 0 & 0 & 0 \\
            0 & 0 & 0 & 0 & 0 & \sqrt{15} & 0 & 0 & 0 \\
            0 & 0 & 0 & 0 & 0 & 0 & -\frac{3 \sqrt{5}}{2} & 0 & -\frac{\sqrt{15}}{2}
        }\quad.
\end{proof*}

We compute the change of basis matrix from polynomials to Green's
polynomials, $\ATwo$, in a similar manner. In practice,
it is easier to express the elements of the Green's basis $\gbasis$ in terms
of the elements of the polynomial basis $\pbasis$ and use those to populate the
columns of the matrix $\ATwo^{-1}$. Continuing our example
for $l_\mathrm{max} = 2$, our second change of basis matrix
is
\begin{proof*}{A}
    \setstackgap{L}{1.1\baselineskip}
    \fixTABwidth{T}
    \ATwo =
        \parenMatrixstack{
            \quad\quad\, 1\, \quad\quad\quad\quad & 0 & 0 & 0 & 0 & 0 & 0 & 0 & 0 \\
            0 & \frac{1}{2} & 0 & 0 & 0 & 0 & 0 & 0 & 0 \\
            0 & 0 & 1 & 0 & 0 & 0 & 0 & 0 & 0 \\
            0 & 0 & 0 & 1 & 0 & 0 & 0 & 0 & 0 \\
            0 & 0 & 0 & 0 & \frac{1}{3} & 0 & 0 & 0 & 0 \\
            0 & 0 & 0 & 0 & 0 & -\frac{1}{3} & 0 & 0 & 0 \\
            0 & 0 & 0 & 0 & 0 & 0 & \frac{1}{2} & 0 & 0 \\
            0 & 0 & 0 & 0 & 0 & 0 & 0 & \frac{1}{3} & 0 \\
            0 & 0 & 0 & 0 & 0 & 0 & 0 & 0 & 1
        }\quad.
\end{proof*}

Finally, recall that the complete change of basis matrix from spherical
harmonics to Green's polynomials, $\bvec{A}$, is just the matrix product
of $\ATwo$ and $\AOne$. For $l_\mathrm{max} = 2$, we have
\begin{proof*}{A}
    \setstackgap{L}{1.25\baselineskip}
    \fixTABwidth{T}
    \bvec{A} =
        \frac{1}{2\sqrt{\pi}}
        \parenMatrixstack{
        1 & 0 & 0 & 0 & 0 & 0 & \sqrt{5} & 0 & 0 \\
        0 & 0 & 0 & \frac{\sqrt{3}}{2} & 0 & 0 & 0 & 0 & 0 \\
        0 & 0 & \sqrt{3} & 0 & 0 & 0 & 0 & 0 & 0 \\
        0 & \sqrt{3} & 0 & 0 & 0 & 0 & 0 & 0 & 0 \\
        0 & 0 & 0 & 0 & 0 & 0 & -\frac{\sqrt{5}}{2} & 0 & \frac{\sqrt{\frac{5}{3}}}{2} \\
        0 & 0 & 0 & 0 & 0 & 0 & 0 & -\sqrt{\frac{5}{3}} & 0 \\
        0 & 0 & 0 & 0 & \frac{\sqrt{15}}{2} & 0 & 0 & 0 & 0 \\
        0 & 0 & 0 & 0 & 0 & \sqrt{\frac{5}{3}} & 0 & 0 & 0 \\
        0 & 0 & 0 & 0 & 0 & 0 & -\frac{3 \sqrt{5}}{2} & 0 & -\frac{\sqrt{15}}{2}
        }\quad.
\end{proof*}
%

\vspace*{4em}
\section{Rotation of spherical harmonics}
\label{app:rotation}

\subsection{Euler angles}
\label{app:euler}

\citet{AlvarezCollado1989} derived expressions for the rotation matrices for
the real spherical harmonics of a given degree $l$ from the corresponding
complex rotation matrices \citep{Steinborn1973}:
\begin{align}
    \label{eq:rl}
    \bvec{R}^l = \bvec{U}^{-1} \bvec{D}^l \bvec{U}
\end{align}
where
\begin{proof}{R}
    \label{eq:dl}
    \bvec{D}^l_{m,m'} &= \mathrm{e}^{-\ii (\alpha m' + \gamma m)}
                       (-1)^{m' + m}
                       \sqrt{
                            (l - m)! (l + m)! (l - m')! (l + m')!
                       }
                       \nonumber \\
                       & \phantom{=}
                       \times
                       \sum_k (-1)^k
                              \frac{
                                \cos\left(\frac{\beta}{2}\right)^{2l + m - m' - 2k}
                                \sin\left(\frac{\beta}{2}\right)^{-m + m' + 2k}
                              }{
                                k! (l + m - k)! (l - m' - k)! (m' - m + k)!
                              }
\end{proof}
is the $(m, m')$ \edited{index of the}
rotation matrix for the complex spherical harmonics of degree $l$ and
\begin{proof*}{R}
    \label{eq:U}
    \setstackgap{L}{1.25\baselineskip}
    \fixTABwidth{T}
    \bvec{U} =
    \frac{1}{\sqrt{2}}
        \parenMatrixstack{
            \quad\quad\, \ddots \, \quad\quad\quad\quad
                   &     &     &     &          &     &     &     & \Ddots \\
                   & \ii &     &     &          &     &     &  1  &        \\
                   &     & \ii &     &          &     &  1  &     &        \\
                   &     &     & \ii &          &  1  &     &     &        \\
                   &     &     &     & \sqrt{2} &     &     &     &        \\
                   &     &     & \ii &          & -1  &     &     &        \\
                   &     & -\ii&     &          &     &  1  &     &        \\
                   & \ii &     &     &          &     &     & -1  &        \\
            \Ddots &     &     &     &          &     &     &     & \ddots
        }\quad.
\end{proof*}
describes the transformation from complex to real spherical harmonics. In
\eq{dl} above, $\alpha$, $\beta$, and $\gamma$ are the (proper) Euler angles
for rotation in the $z{-}y{-}z$ convention.
To obtain a rotation matrix for an arbitrary vector $\bvec{y}$ with spherical
harmonics of different orders up to $l = l_\mathrm{max}$, we define the
block-diagonal matrix $\bvec{R}$:
\begin{proof*}{R}
    \label{eq:rblockdiag}
    \setstackgap{L}{1.25\baselineskip}
    \fixTABwidth{T}
    \bvec{R} =
        \parenMatrixstack{
            \quad\quad \, \bvec{R}^0 \, \quad\quad
                       &            &            &            &  \\
                       & \bvec{R}^1 &            &            &  \\
                       &            & \bvec{R}^2 &            &  \\
                       &            &            & \bvec{R}^3 &  \\
                       &            &            &            & \ddots
        }\quad.
\end{proof*}
Rotation of $\bvec{y}$ by the Euler angles $\alpha$, $\beta$, and $\gamma$
is performed via \eq{rotation} with $\bvec{R}$ given by \eq{rblockdiag}.

\subsection{Axis-angle}
\label{app:axisangle}

It is often more convenient to define a rotation by an axis $\bvec{u}$
and an angle $\theta$ of rotation about that axis. Given a unit vector
$\bvec{u}$ and an angle $\theta$, we can find the corresponding Euler
angles by comparing the 3-dimensional Cartesian rotation matrices for
both systems,
\begin{equation}
    \label{eq:rotP}
    \setstackgap{L}{1.25\baselineskip}
    \fixTABwidth{T}
    \mathbf{P} =
        \parenMatrixstack{
        c_\theta + u_x^2 \left(1 - c_\theta\right)
        &
        u_x u_y \left(1 - c_\theta\right) - u_z s_\theta
        &
        u_x u_z \left(1 - c_\theta\right) + u_y s_\theta
        \\
        u_y u_x \left(1 - c_\theta\right) + u_z s_\theta
        &
        c_\theta + u_y^2\left(1 - c_\theta\right)
        &
        u_y u_z \left(1 - c_\theta\right) - u_x s_\theta
        \\
        \quad\,\, u_z u_x \left(1 - c_\theta\right) - u_y s_\theta \,\,\quad\,
        &
        u_z u_y \left(1 - c_\theta\right) + u_x s_\theta
        &
        c_\theta + u_z^2\left(1 - c_\theta\right)
        }
\end{equation}
for axis-angle rotations and
\begin{equation}
    \label{eq:rotQ}
    \setstackgap{L}{1.25\baselineskip}
    \fixTABwidth{T}
    \mathbf{Q} =
        \parenMatrixstack{
        c_\alpha c_\beta c_\gamma - s_\alpha s_\gamma
        &
        -c_\gamma s_\alpha - c_\alpha c_\beta s_\gamma
        &
        c_\alpha s_\beta
        \\
        c_\alpha s_\gamma + c_\beta c_\gamma s_\alpha
        &
        c_\alpha c_\gamma - c_\beta s_\alpha s_\gamma
        &
        s_\alpha s_\beta
        \\
        -c_\gamma s_\beta
        &
        s_\beta s_\gamma
        &
        c_\beta
        }\quad,
\end{equation}
for Euler rotations,
where $c_{\bigdot} \equiv \cos(\cdot)$
and $s_{\bigdot} \equiv \sin(\cdot)$.
\edited{Equating the} two matrices gives us expressions for the Euler
angles in terms of $\bvec{u}$ and $\theta$:
\begin{proof}{R}
    \label{eq:eulerangles}
    \begin{matrix}
        \cos\alpha = \frac{P_{0,2}}{\sqrt{P_{0,2}^2 + P_{1,2}^2}}
        & & & &
        \cos\beta = P_{2,2}
        & & & &
        \cos\gamma = -\frac{P_{2,0}}{\sqrt{P_{2,0}^2 + P_{2,1}^2}}
        \\
        \sin\alpha = \frac{P_{1,2}}{\sqrt{P_{0,2}^2 + P_{1,2}^2}}
        & & & &
        \sin\beta = \sqrt{1 - P_{2,2}^2}
        & & & &
        \sin\gamma = \frac{P_{2,1}}{\sqrt{P_{2,0}^2 + P_{2,1}^2}}
    \end{matrix}
    \quad.
\end{proof}
Thus, given a spherical harmonic vector $\bvec{y}$, we can calculate
how it transforms under rotation by an angle $\theta$ about an axis $\bvec{u}$
by first computing the Euler angles (Equation~\ref{eq:eulerangles}) and using
those to construct the spherical harmonic rotation matrix
(Equation~\ref{eq:rblockdiag}).

\section{Computing the solution vector $\MakeLowercase{s_n}$}
\label{app:solutionvector}

Here we seek a solution to \eq{sn}, which gives the total flux
during an occultation of the $n^\mathrm{th}$ term in the Green's basis
(Equation~\ref{eq:bg}). The primitive integrals $\mathcal{P}$ and
$\mathcal{G}$ in that equation are given by Equations~(\ref{eq:primitiveP})
and (\ref{eq:primitiveQ}), with
$\bvec{G}_n$ defined in \eq{Gn}. Note that all of the terms in \eq{Gn},
with the exception of the $l = 1, m = 0$ case, are simple polynomials
in $\x$, $\y$, and $\z$, which facilitates their integration.
The $l = 1, m = 0$ term (corresponding to the $n = 2$ term in the Green's basis)
is more difficult to integrate, but an analytic
solution exists \citep{Pal2012}. It is, however, more convenient
to note that this term corresponds to a surface map given by the polynomial
$I(x, y) = \tilde{g}_2(x, y) = \sqrt{1 - x^2 - y^2}$, which is the same function used
to model linear limb darkening in stars \citep{MandelAgol2002}. We therefore
evaluate this term separately in Appendix~\ref{app:linearld} below, followed by
the general term in Appendix~\ref{app:generalterm}.

\subsection{Linear limb darkening ($n = 2$, $l=1$, $m=0$)}
\label{app:linearld}

From \citet{MandelAgol2002}, the total flux visible during the occultation of a
body whose surface map is given by $I(x, y) = \sqrt{1 - \x^2 -\y^2}$ may be computed
as
\begin{align}
    \label{eq:s2}
    s_2 = \frac{2\pi}{3} \left(1 - \frac{3\Lambda}{2} - \Theta(r - b) \right)
\end{align}
where $\Theta(\bigdot)$ is the Heaviside step function and
\begingroup\makeatletter\def\f@size{9}\check@mathfonts
\def\maketag@@@#1{\hbox{\m@th\normalsize#1}}%
\begin{proof}{biglam}
    \label{eq:biglam}
    \mbox{\normalsize$\Lambda$} &=
    \begin{dcases}
          %
          %
          \frac{1}{9 \pi \sqrt{b r}} \Bigg[
                \frac{(r + b)^2 - 1}{r + b}
                \Big(
                    -2r \,
                    \big(
                        2 (r + b)^2 + (r + b)(r - b) - 3
                    \big)
                    K(k^2)
                    &\\ \phantom{XXXX}
                    + 3 (b - r) \, \Pi\big(k^2 (b + r)^2, \, k^2\big)
                \Big)
                - 4 b r (4 - 7 r^2 - b^2) E(k^2)
          \Bigg]
          & \qquad k^2 < 1
          \\[1.5em]
          \frac{2}{9 \pi} \Bigg[
                \big(1 - (r + b)^2\big)
                \Bigg(
                    \sqrt{1 - (b - r)^2} \,
                    K\left(\frac{1}{k^2}\right)
                    + 3 \left(\frac{b-r}{(b+r)\sqrt{1 - (b - r)^2}}\right)
                    &\\ \phantom{XX}
                    \times \Pi\left(\frac{1}{k^2(b+r)^2}, \, \frac{1}{k^2}\right)
                \Bigg)
                - \sqrt{1 - (b - r)^2}
                (4 - 7 r^2 - b^2)
                E\left(\frac{1}{k^2}\right)
          \Bigg]
          & \qquad k^2 \ge 1
    \end{dcases}
\end{proof}
\endgroup
with
\begin{align}
    \label{eq:k2}
    k^2 &= \frac{1 - r^2 - b^2 + 2 b r}{4 b r}
    \quad.
\end{align}
In the expressions above, $K(\bigdot)$, $E(\bigdot)$, and $\Pi(\bigdot, \bigdot)$
are the complete elliptic integrals of the first, second kind, and third kind,
respectively, defined as
\begin{align}
    \label{eq:elliptic}
    K(k^2) &\equiv \int_0^{\frac{\pi}{2}} \frac{\dd \varphi}{\sqrt{1 - k^2 \sin^2 \varphi}}
    \nonumber \\[0.5em]
    E(k^2) &\equiv \int_0^{\frac{\pi}{2}} \sqrt{1 - k^2 \sin^2 \varphi} \, \dd \varphi
    \nonumber \\[0.5em]
    \Pi(n, k^2) &\equiv \int_0^{\frac{\pi}{2}} \frac{\dd \varphi}{(1 - n \sin^2 \varphi)\sqrt{1 - k^2 \sin^2 \varphi}}
    \quad.
\end{align}
In some cases, the expressions above can become unstable. For $r > 1$,
$b \approx r$, $b + r \approx 1$, and $|b - r| \approx 1$, we re-parametrize
these expressions in terms of the modified elliptic integral
$\mathrm{cel}(k_c, p, a, b)$ \citep{Bulirsch1969} as described in
\citet{limbdark}.

\subsection{All other terms}
\label{app:generalterm}

\subsubsection{Setting up the equations}
\label{app:generaltermsetup}

We evaluate all other terms in $s_n$ by integrating the primitive integrals of
$\bvec{G}_n$. These are given by
\begingroup\makeatletter\def\f@size{9}\check@mathfonts
\def\maketag@@@#1{\hbox{\m@th\normalsize#1}}%
\begin{align}
    \label{eq:PGn}
    \mbox{\normalsize$\mathcal{P}(\bvec{G}_n)$} &=
    \begin{dcases}
        +\int\displaylimits_{\pi - \phi}^{2\pi + \phi}
            (r c_\varphi)^{\frac{\mu+2}{2}}
            (b + r s_\varphi)^{\frac{\nu}{2}}
            r c_\varphi
            \, \dd\varphi
            & \qquad \frac{\mu}{2} \, \mathrm{even}
        \\[1em]
        -\int\displaylimits_{\pi - \phi}^{2\pi + \phi}
            (r c_\varphi)^{l-2}
            (1 {-} r^2 {-} b^2 {-} 2 b r s_\varphi)^{\frac{3}{2}}
            r s_\varphi
            \, \dd\varphi
            & \qquad \mu = 1, \,
                     l \, \mathrm{even}
        \\[1em]
        -\int\displaylimits_{\pi - \phi}^{2\pi + \phi}
            (r c_\varphi)^{l-3}
            (b + r s_\varphi)
            (1 {-} r^2 {-} b^2 {-} 2 b r s_\varphi)^{\frac{3}{2}}
            r s_\varphi
            \, \dd\varphi
            & \qquad \mu = 1, \, l \ne 1, \,
                     l \, \mathrm{odd}
        \\[1em]
        +\int\displaylimits_{\pi - \phi}^{2\pi + \phi}
            (r c_\varphi)^{\frac{\mu-3}{2}}
            (b + r s_\varphi)^{\frac{\nu-1}{2}}
            (1 {-} r^2 {-} b^2 {-} 2 b r s_\varphi)^{\frac{3}{2}}
            r c_\varphi
            \, \dd\varphi
            & \qquad \frac{\mu - 1}{2} \, \mathrm{even}, \, l \ne 1
        \\[1em]
        \textrm{---} \ \mathrm{(c.f.\ Appendix\ \ref{app:linearld})} & \qquad \mu=1, \, l=1
        \\[1em]
        0 & \qquad \mathrm{otherwise}
    \end{dcases}
\end{align}
\endgroup
and
\begin{align}
    \label{eq:QGn}
    \mathcal{Q}(\bvec{G}_n) &=
    \begin{dcases}
        +\int\displaylimits_{\pi - \lambda}^{2\pi + \lambda}
            c_\varphi^{\frac{\mu+2}{2}}
            s_\varphi^{\frac{\nu}{2}}
            c_\varphi
            \, \dd\varphi
            & \qquad\qquad \frac{\mu}{2} \, \mathrm{even}
        \\[1em]
        \phantom{XXXXX}0
            & \qquad\qquad \mathrm{otherwise,}
    \end{dcases}
\end{align}
where we have used the fact that the line integral of any function
proportional to $\z$ taken along the limb of the occulted planet
(where $\z = \sqrt{1-\x^2-\y^2} = 0$) is zero.
%

\subsubsection{The $\mathcal{Q}$ integral}
\label{app:Q}

We begin with the expression for $\mathcal{Q}$ (Equation~\ref{eq:QGn}), as this
is the most straightforward. Defining the integral
\begin{align}
    \label{eq:Huv}
    \mathcal{H}_{u,v} &=
    \int\displaylimits_{\pi - \lambda}^{2\pi + \lambda}
            c_\varphi^u
            s_\varphi^v
            \, \dd\varphi
            \quad,
\end{align}
we may write
\begin{align}
    \nonumber \\
    \label{eq:QGnI}
    \mathcal{Q}(\bvec{G}_n) &=
    \begin{dcases}
        \mathcal{H}_{\frac{\mu+4}{2}, \frac{\nu}{2}}
        & \qquad \qquad \qquad \qquad \quad \quad \quad \quad \frac{\mu}{2} \, \mathrm{even}
        \\[1em]
        0
        & \qquad \qquad \qquad \qquad \quad \quad \quad \quad \mathrm{otherwise} \quad.
    \end{dcases}
\end{align}
\citet{Pal2012} derived simple recurrence relations for this integral:
\begin{proof}{Huvsol}
    \label{eq:Huvsol}
    \mathcal{H}_{u,v} &=
    \begin{dcases}
        0
        & \qquad u \ \mathrm{odd}
        \\[0.5em]
        2\lambda + \pi
        & \qquad u = v = 0
        \\[0.5em]
        -2\cos\lambda
        & \qquad u = 0, v = 1
        \\[0.5em]
        \frac{2}{u + v}
        (\cos\lambda)^{u - 1} (\sin\lambda)^{v + 1} +
        \frac{u - 1}{u + v} \mathcal{H}_{u-2,v}
        & \qquad u \ge 2
        \\[0.5em]
        -\frac{2}{u + v}
        (\cos\lambda)^{u + 1} (\sin\lambda)^{v - 1} +
        \frac{v - 1}{u + v} \mathcal{H}_{u,v-2}
        & \qquad v \ge 2
        \quad
    \end{dcases}
\end{proof}
%

\subsubsection{The $\mathcal{P}$ integral}
\label{app:P}

In general, the $\mathcal{P}$ integral is more difficult to evaluate because
of the term to the $\nicefrac{3}{2}$ power in several of the cases.
Moreover, the presence of terms proportional to powers of $b$
and $r$ and terms of order unity in several of the integrands in
Equation~(\ref{eq:PGn}) can lead to severe numerical instabilities when
either $b$ or $r$ are very large (which is typically the case for secondary eclipses
of small planets) or very small (which occurs for small transiting bodies).
To enforce numerical stability in all regimes, we find that is convenient to
define the parameters
\begin{equation}
    \label{eq:delta}
    \delta = \frac{b-r}{2r}
\end{equation}
and
\begin{align}
    \label{eq:kappa}
    \kappa &= \phi + \frac{\pi}{2} \nonumber \\
           &= \cos^{-1} \left(\frac{r^2+b^2-1}{2br}\right)
           \quad.
\end{align}
The latter variable can be defined more simply in terms of $\sin^2\tfrac{\kappa}{2} = k^2$,
or $\kappa = 2 \sin^{-1}k$. Note that when $r+b \le 1$, $\phi=\pi/2$, so $\kappa = \pi$.
With this transformed variable, the limits of integration of $\mathcal{P}(\bvec{G}_n)$
become $\tfrac{3\pi}{2}-\kappa$ to $\tfrac{3\pi}{2}+\kappa$.  Transforming
$\varphi$ to $\varphi^\prime = \frac{1}{2}(\varphi - \tfrac{3\pi}{2})$ yields
\begingroup\makeatletter\def\f@size{9}\check@mathfonts
\def\maketag@@@#1{\hbox{\m@th\normalsize#1}}%
\begin{proof}{PGn_reparam}
    \label{eq:PGn_reparam}
    \mbox{\normalsize$\mathcal{P}(\bvec{G}_n)$} &=
    \begin{dcases}
        2(2r)^{l+2}\int\displaylimits_{-\kappa/2}^{\kappa/2}
            (s_\varphi^2 -s_\varphi^4)^{\frac{\mu+4}{4}}
            (\delta + s_\varphi^2)^{\frac{\nu}{2}}
            \, \dd\varphi
            & \qquad \frac{\mu}{2} \, \mathrm{even}
        \\[1em]
        \mathcal{F}\int\displaylimits_{-\kappa/2}^{\kappa/2}
            (s_\varphi^2-s_\varphi^4)^{\tfrac{l-2}{2}}
            (k^2- s_\varphi^2)^{\frac{3}{2}}
            (1-2s_\varphi^2)
            \, \dd\varphi
            & \qquad \mu = 1, \,
                     l \, \mathrm{even}
        \\[1em]
        \mathcal{F}\int\displaylimits_{-\kappa/2}^{\kappa/2}
            (s_\varphi^2-s_\varphi^4)^{\tfrac{l-3}{2}}
            (\delta + s_\varphi^2)
            (k^2- s_\varphi^2)^{\frac{3}{2}}
            (1-2 s_\varphi^2)
            \, \dd\varphi
            & \qquad \mu = 1, \, l \ne 1, \,
                     l \, \mathrm{odd}
        \\[1em]
        2\mathcal{F}\int\displaylimits_{-\kappa/2}^{\kappa/2}
            (s_\varphi^2-s_\varphi^4)^{\frac{\mu-1}{4}}
            (\delta + s_\varphi^2)^{\frac{\nu-1}{2}}
            (k^2 - s_\varphi^2)^{\frac{3}{2}}
            \, \dd\varphi
            & \qquad \frac{\mu-1}{2} \, \mathrm{even}, \, l \ne 1
        \\[1em]
        \textrm{---} \ \mathrm{(c.f.\ Appendix\ \ref{app:linearld})} & \qquad \mu=1, \, l=1
        \\[1em]
        0 & \qquad \mathrm{otherwise},
    \end{dcases}
\end{proof}
\endgroup
where $\mathcal{F} = (2r)^{l-1}(4br)^{3/2}$ and we have subsequently
dropped the prime from $\varphi^\prime$ in these integrals.
Expanding the term $(1-s_\varphi^2)^u(\delta + s_\varphi^2)^v$ as a polynomial
in $s_\varphi^2$, we find
\begin{proof}{vieta}
    \label{eq:poly_expansion}
    (1-s_\varphi^2)^u(\delta + s_\varphi^2)^v = \sum_{i=0}^{u+v} \mathcal{A}_{i,u,v} s_\varphi^{2i},
\end{proof}
where
\begin{proof}{vieta}
    \label{eq:vieta}
    \mathcal{A}_{i,u,v} = \sum_{j=\mathrm{max}(0,u-i)}^{\mathrm{min}(u+v-i,u)}
                          \binom{u}{j} \binom{v}{u+v-i-j}(-1)^{u+j}\delta^{u+v-i-j}.
\end{proof}
The coefficients $\mathcal{A}_{i,u,v}$ are computed from Vieta's formulae for the
coefficients of a polynomial in terms of sums and products of its roots, and are
equal to the elementary symmetric polynomials of the roots of $(1-x)^u(x+\delta)^v$.
This expansion yields a sum over terms which are integrals over powers of $s_\varphi^{2v}$.
We use this expansion to rewrite the expressions for $\mathcal{P}(\mathbf{G}_n)$ as
\begin{proof}{PGn_reparam2}
    \label{eq:PGn_reparam2}
    \mathcal{P}(\bvec{G}_n) &=
    \begin{dcases}
        2(2r)^{l+2} \mathcal{K}_{\frac{\mu+4}{4}, \frac{\nu}{2}}
            & \qquad \frac{\mu}{2} \, \mathrm{even}
        \\[1em]
        \mathcal{F} \left(\mathcal{L}^{(0)}_{\frac{l-2}{2},0}-2\mathcal{L}^{(1)}_{\frac{l-2}{2}, 0} \right)
            & \qquad \mu = 1, \,
                     l \, \mathrm{even}
        \\[1em]
        \mathcal{F} \left(\mathcal{L}^{(0)}_{\frac{l-3}{2},1}-2\mathcal{L}^{(1)}_{\frac{l-3}{2},1}   \right)
            & \qquad \mu = 1, \, l \ne 1, \,
                     l \, \mathrm{odd}
        \\[1em]
        2\mathcal{F} \mathcal{L}^{(0)}_{\frac{\mu-1}{4}, \frac{\nu-1}{2}}
            & \qquad \frac{\mu-1}{2} \, \mathrm{even}, \, l \ne 1
        \\[1em]
        \textrm{---} \ \mathrm{(c.f.\ Appendix\ \ref{app:linearld})} & \qquad \mu=1, \, l=1
        \\[1em]
        0 & \qquad \mathrm{otherwise}
        \quad,
    \end{dcases}
\end{proof}
where
\begin{proof}{PGn_reparam2}
    \label{eq:KILJ}
    \mathcal{K}_{u,v} &= \int_{\kappa/2}^{\kappa/2} s_\varphi^{2u} (1-s_\varphi^2)^u (\delta + s_\varphi^2)^v d\varphi \nonumber \\
    &= \sum_{i=0}^{u+v} \mathcal{A}_{i,u,v} \mathcal{I}_{i+u} \quad,\\[0.5em]
    %
    %
    \mathcal{L}^{(t)}_{u,v} &= k^3 \int_{-\kappa/2}^{\kappa/2}s_\varphi^{2(u+t)}(1-s_\varphi^2)^u(\delta+s_\varphi^2)^v
    \left(1-k^{-2}s_\varphi^2\right)^{3/2}d\varphi, \nonumber \\
    &= k^3 \sum_{i=0}^{u+v} \mathcal{A}_{i,u,v} \mathcal{J}_{i+u+t} \quad,\\
    \intertext{and}
    \mathcal{I}_{v} &= \int_{-\kappa/2}^{\kappa/2} s_\varphi^{2v} d\varphi \quad,\\[0.5em]
    %
    %
    \label{eq:J}
    \mathcal{J}_v &= \int_{-\kappa/2}^{\kappa/2} d\varphi s^{2v}_\varphi\left(1-k^{-2}s^2_\varphi\right)^{3/2} \quad,
\end{proof}
recalling that $\kappa = 2\sin^{-1}(k)$ for $b+r > 1$ and $\kappa = \pi$ for $b + r \le 1$.

Given this formulation, evaluating $\mathcal{P}(\mathbf{G}_n)$ is a matter of finding
formulae for the integrals $\mathcal{I}_v$ and $\mathcal{J}_v$, which are in fact
analytic.
Using integration by reduction, $\mathcal{I}_v$ can be expressed
in terms of sums of powers of $\sin^{-1}k$, $k$ and $k_c \equiv \sqrt{1-k^2}$,
while $\mathcal{J}_v$ can be expressed
as sums of complete elliptic integrals of $k^2$ times polynomials in $k^2$. The solutions
are different depending on whether $k^2$ is less than or greater than unity.

\subsubsection{Evaluating $\mathcal{I}_v$ and $\mathcal{J}_v$ for $k^2 < 1$}
\label{app:IJklt1}

In the $k^2 < 1$ ($b+r > 1$) limit, we make the substitution $w=k^{-2}\sin^2{\varphi}$, giving
\begin{proof}{IJHypergeo}
    \label{eq:IJHypergeo}
    \mathcal{I}_v &= k^{1+2v} \int_0^1 (1-k^2w)^{-\tfrac{1}{2}} w^{\tfrac{2v-1}{2}} dw \nonumber \\
                  &= \frac{2k^{1+2v}}{1+2v} \,_2F_1\left(\tfrac{1}{2},v+\tfrac{1}{2};v+\tfrac{3}{2};k^2\right) \quad,\\[0.5em]
    \mathcal{J}_v &= k^{1+2v} \int_0^1 (1-k^2 w)^{-\tfrac{1}{2}} w^{\tfrac{2v-1}{2}} (1-w)^{3/2} dw \nonumber \\
                  &= k^{1+2v} \frac{3\pi}{4} \frac{(2v-1)!!}{2^v(2+v)!} \,_2F_1\left(\tfrac{1}{2},v+\tfrac{1}{2}; v+3; k^2\right) \quad,
\end{proof}
where $_2F_1(a,b;c;x)$ is the generalized Hypergeometric function.
These functions can alternatively be expressed as series in $k^2$
by expanding $(1-k^2w)^{-1/2}$ as a series in $k^2w$, and then integrating each term over $w$, giving
\begin{proof}{IJSeries}
    \label{eq:IJseries}
    \mathcal{I}_{v} &= 2k^{1+2v}  \sum_{j=0}^\infty \frac{(2j-1)!!}{2^j j!(2j+2v+1)}(k^2)^j, \nonumber\\[0.5em]
    \mathcal{J}_{v} &= \frac{3\pi}{4}k^{1+2v}\sum_{j=0}^\infty \frac{(2j-1)!!(2j+2v-1)!!}{2^{2j+v} j!(j+v+2)!}(k^2)^j \quad.
\end{proof}

For computational efficiency, both $\mathcal{I}_v$ and $\mathcal{J}_v$ can be evaluated
recursively via either upward or downward iteration. When iterating upward, we use
the recursion relations
\begin{proof}{IJupward}
    \label{eq:IJupward}
    \mathcal{I}_{v} &= \frac{1}{v} \left(\frac{2v - 1}{2} \mathcal{I}_{v - 1} - k^{2v - 1} k_c\right) \\
    \label{eq:IJupward:J}
    \mathcal{J}_{v} &= \frac{1}{2v + 3} \left[2 \left(v + \left(v - 1\right) k^2 + 1\right) \mathcal{J}_{v - 1} - k^2 (2v - 3) \mathcal{J}_{v - 2}\right] \quad,
\end{proof}
along with the initial values
\begin{proof}{IJupward}
    \label{eq:IJseries0}
    \mathcal{I}_0 &= \kappa = 2\sin^{-1}k \\[0.5em]
    \mathcal{J}_0 &= \frac{2}{3k^3} \left[2(2k^2-1)E(k^2)+(1-k^2)(2-3k^2)K(k^2)\right] \nonumber \\
    \mathcal{J}_1 &= \frac{2}{15k^3} \left[(-3k^4+13k^2-8)E(k^2)+(1-k^2)(8-9k^2)K(k^2)\right] \quad.
\end{proof}
When iterating downward, we can re-arrange Equation~(\ref{eq:IJupward}) to obtain the relations
\begin{proof}{IJdownward}
    \label{eq:IJdownward}
    \mathcal{I}_{v} &= \frac{2}{2v + 1} \left[\left(v + 1\right) \mathcal{I}_{v + 1} + k^{2v + 1} k_c\right] \\
    \label{eq:IJdownward:J}
    \mathcal{J}_{v} &= \frac{1}{(2v + 1)k^2} \left[2 \left(3 + v + \left(1 + v\right)k^2\right) \mathcal{J}_{v + 1} - (2v + 7) \mathcal{J}_{v + 2}\right] \quad.
\end{proof}
In this case, the starting values are obtained directly from Equation~(\ref{eq:IJHypergeo}) or Equation~(\ref{eq:IJseries}).

Because the Hypergeometric function (Equation~\ref{eq:IJHypergeo}) can be costly to
evaluate, it is in general more computationally efficient to evaluate the expressions
in Equation~(\ref{eq:IJseries0}) and iterate upward. Moreover, note that the elliptic
integrals $E$ and $K$ in those expressions are exactly the same as those used to evaluate the
linear limb darkening ($s_2$) term of the solution vector, so these need only be
computed \emph{once} to obtain solutions for spherical harmonic maps of arbitrary order,
which makes this algorithm fast.

However, in practice, the upward recursion relations can sometimes be
numerically unstable due to cancellation of low-order terms, particularly when the
occultor radius is large ($r \gg 1$). To leading order in $k$, when $k^2 <1$ ($b + r > 1$),
$\mathcal{I}_v \propto k^{2v+1}$ and $\mathcal{J}_v \propto k^{2v+1}$.
Consequently, when these equations are computed by
recursion in $v$, the lower powers of $k$ cancel out, leading to
round-off errors that grow as $v$ gets large.
In practice, we find that when $k^2 > \frac{1}{2}$, the upward recursion relations
are numerically stable, so we use Equation~(\ref{eq:IJupward}) to evaluate the
integrals. When $k^2 \le \frac{1}{2}$, we instead use Equations(\ref{eq:IJdownward})
and (\ref{eq:IJdownward:J}),
and start by computing $\mathcal{I}_{v_\mathrm{max}}$,
$\mathcal{J}_{v_\mathrm{max}}$, and $\mathcal{J}_{v_\mathrm{max}-1}$, where
$v_\mathrm{max}$ is the maximum value needed to compute $\mathcal{K}_{u,v}$ when
$l = l_\mathrm{max}$. We find that the series in Equation~(\ref{eq:IJseries})
converge rapidly, so we use those expressions to evaluate the initial conditions.


\subsubsection{Evaluating $\mathcal{I}_v$ and $\mathcal{J}_v$ for $k^2 \ge 1$}
\label{app:IJkge1}

In the $k^2 \ge 1$ ($b+r \le 1$) limit, the expression for $\mathcal{I}_v$ is
simpler:
\begin{proof}{Ilargek}
    \label{eq:Ilargek}
    \mathcal{I}_v = \pi \frac{(2v-1)!!}{2^v v!} \quad,
\end{proof}
so $\mathcal{K}_{u,v}$ (Equation~\ref{eq:KILJ}) is simply a polynomial in $\delta$.
For $\mathcal{J}_v$, we make the substitution $w = \sin^2{\varphi}$ to obtain
\begin{proof}{Jlargek}
    \label{eq:Jlargek}
\mathcal{J}_v &= \int_0^1 w^{v-\tfrac{1}{2}} (1-k^{-2}w)^{3/2} (1-w)^{-1/2} dw \nonumber \\
            &= \sqrt{\pi}\Gamma(v+\tfrac{1}{2})\left(_2\tilde{F}_1\left(-\tfrac{1}{2},v+\tfrac{1}{2},v+1,k^{-2}\right)\right. \nonumber \\
            &\phantom{MMMMMMm}
                -\left.\left(\frac{1}{2}+v\right)k^{-2}\,_2\tilde{F}_1\left(-\tfrac{1}{2},v+\tfrac{3}{2},v+2,k^{-2}\right)\right) \quad, \nonumber \\[0.5em]
            &= \pi \sum_{j=0}^\infty (-1)^j \binom{3/2}{j} \frac{(2j+2v-1)!!}{2^{j+v} (j+v)!} (k^2)^{-j} \quad,
\end{proof}
where $_2\tilde{F}_1(a,b;c;x)$ is the regularized Hypergeometric function.

As in the $k^2 < 1$ case, we can evaluate $\mathcal{J}_v$ via either upward (Equation~\ref{eq:IJupward:J})
or downward (Equation~\ref{eq:IJdownward:J}) recursion. For upward recursion, the initial
values are given by
\begin{proof}{Jseries0largek}
    \label{Jseries0largek}
    \mathcal{J}_0 &= \frac{1}{3}\left[(8-4k^{-2})E(k^{-2})-2(1-k^{-2})K(k^{-2})\right] \nonumber \\
    \mathcal{J}_1 &= \frac{1}{15}\left[(-6k^2+26-16k^{-2})E(k^{-2}) + 2(1-k^{-2})(3k^2-4)K(k^{-2})\right] \quad.
\end{proof}
which, as before, make use of the same elliptic integrals as the $s_2$ term. For downward
recursion, we evaluate $\mathcal{J}_{v_\mathrm{max}}$ and $\mathcal{J}_{v_\mathrm{max}-1}$
from the series solution (Equation~\ref{eq:Jlargek}). In practice, we find that our results
are numerically stable if we perform upward recursion for $k^2 < 2$ and downward recursion
if $k^2 \ge 2$.

\clearpage
\begin{center}
\renewcommand*{\arraystretch}{1.08}
\begin{longtable}{cll}
\caption{Symbols used in this paper} \label{tab:symbols} \\
\toprule
\multicolumn{1}{c}{\textbf{Symbol}} &
\multicolumn{1}{c}{\textbf{Definition}} &
\multicolumn{1}{c}{\textbf{Reference}} \\
\midrule
\endfirsthead
\multicolumn{3}{c}%
{{\bfseries \tablename\ \thetable{} --} continued from previous page} \\
\toprule
\multicolumn{1}{c}{\textbf{Symbol}} &
\multicolumn{1}{c}{\textbf{Definition}} &
\multicolumn{1}{c}{\textbf{Reference}} \\
\midrule
\endhead
\bottomrule
\endfoot
\bottomrule
\endlastfoot
$A_{lm}$        & Legendre function normalization       & \eq{alm} \\
$\bvec{A}$      & Change of basis matrix:
                  $Y_{lm}$s to Green's
                  polynomials                           & \eq{A} \\
$\AOne$         & Change of basis matrix:
                  $Y_{lm}$s to polynomials              & \S\ref{sec:basis} \\
$\ATwo$         & Change of basis matrix:
                  polynomials to Green's polynomials    & \S\ref{sec:basis} \\
$\mathcal{A}_{i,u,v}$
                & Vieta's formula coefficient           & \eq{vieta} \\
$b$             & Impact parameter in units of occulted
                  body's radius                         & \S\ref{sec:lightcurves} \\
$B_{lm}^{jk}$   & Spherical harmonic normalization      & \eq{blmjk} \\

$c_{\bigdot}$   & $\cos(\bigdot)$                       & \\
$C_{pq}^k$      & Expansion coefficient for
                  $\z(\x, \y)$                          & \eq{ckpq} \\
$\bvec{D}^l$    & Rotation matrix for the
                  complex spherical harmonics
                  of degree $l$                         & \eq{dl} \\
$\bvec{D}\,\wedge$
                & Exterior derivative                   & \eq{extderiv} \\
$E(\bigdot)$    & Complete elliptic integral of the
                  second kind                           & \eq{elliptic} \\
$F$             & Total flux seen by observer           & \eq{starry} \\
$\mathcal{F}$   & Function of $b$ and $r$               & \eq{PGn_reparam2} \\
$_2F_1$         & Generalized Hypergeometric function   & \eq{IJHypergeo} \\
$_2\tilde{F}_1$ & Regularized Hypergeometric function   & \eq{Jlargek} \\
$\gbasis$       & Green's basis                         & \eq{bg} \\
$\bvec{g}$      & Vector in the basis $\gbasis$         & \\
$\bvec{G}_n$    & Anti-exterior derivative of the
                  $n^\mathrm{th}$
                  term in the Green's basis             & \eq{Gn} \\
$\mathcal{H}_{u,v}$
                & Occultation integral                  & \eq{Huv} \\
$i$             & Dummy index                           & \\
$I$             & Specific intensity, $I(\x, \y)$       & \eq{I} \\
$\mathcal{I}_{v}$
                & Occultation integral                  & \eq{KILJ} \\
$j$             & Dummy index                           & \\
$\mathcal{J}_{v}$

                & Occultation integral                  & \eq{KILJ} \\
$k$             & Elliptic parameter                    & \eq{k2} \\
                & Dummy index                           & \\
$k_c$           & $\sqrt{1 - k^2}$                      & Appendix~\ref{app:P} \\
$K(\bigdot)$    & Complete Elliptic integral of the
                  first kind                            & \eq{elliptic} \\
$\mathcal{K}_{u,v}$
                & Occultation integral                  & \eq{KILJ} \\
$l$             & Spherical harmonic degree             & \eq{lm} \\
$\mathcal{L}_{u,v}^{(t)}$
                & Occultation integral                  & \eq{KILJ} \\
$m$             & Spherical harmonic order              & \eq{lm} \\
$n$             & Surface map vector index,
                  $n = l^2 + l + m$                     & \eq{n} \\
$p$             & Dummy index                           & \\
$\bar{P}$       & Normalized associated Legendre
                  function                              & \eq{plm} \\
$\pbasis$       & Polynomial basis                      & \eq{bp} \\
$\bvec{p}$      & Vector in the basis $\pbasis$         & \\
$\bvec{P}$      & Cartesian axis-angle rotation matrix  & \eq{rotP} \\
$\mathcal{P}$   & Primitive integral along perimiter
                  of occultor                           & \eq{primitiveP} \\
$q$             & Dummy index                           & \\
$\bvec{Q}$      & Cartesian Euler angle rotation matrix & \eq{rotQ} \\
$\mathcal{Q}$   & Primitive integral along perimiter
                  of occulted body                      & \eq{primitiveQ} \\
$r$             & Occultor radius in units of occulted
                  body's radius                         & \S\ref{sec:lightcurves} \\
$\bvec{r}$      & Phase curve solution vector           & \eq{rn} \\
$\bvec{R}$      & Rotation matrix for the real
                  spherical harmonics                   & \eq{rblockdiag} \\
$\bvec{R}^l$    & Rotation matrix for the real
                  spherical harmonics of degree $l$     & \eq{rl} \\
$s_{\bigdot}$   & $\sin(\bigdot)$                       & \\
$\bvec{s}$      & Occultation light curve solution
                  vector                                & \eq{rn} \\
$u$             & Dummy index                           & \\
$u_1, u_2$      & Quadratic limb darkening coefficients & \eq{quadraticld} \\
$\bvec{u}$      & Unit vector corresponding to the
                  axis of rotation                      & \S\ref{app:axisangle} \\
$\bvec{U}$      & Complex to real spherical harmonics
                  transform matrix                      & \eq{U} \\
$v$             & Dummy index                           & \\
$\x$            & Cartesian coordinate                  & \eq{xyz} \\
$\y$            & Cartesian coordinate                  & \eq{xyz} \\
$Y_{l,m}$       & Spherical harmonic of degree $l$
                  and order $m$                         & \eq{ylm0} \\
$\ybasis$       & Spherical harmonic basis              & \eq{by} \\
$\bvec{y}$      & Vector in the basis $\ybasis$         & \\
$\z$            & Cartesian coordinate,
                  $z = \sqrt{1 - \x^2 - \y^2}$          & \eq{xyz} \\
$\alpha$        & Euler angle ($\zhat$ rotation)        & Appendix~\ref{app:euler} \\
$\beta$         & Euler angle ($\yhat$ rotation)        & Appendix~\ref{app:euler} \\
$\gamma$        & Euler angle ($\zhat$ rotation)        & Appendix~\ref{app:euler} \\
$\Gamma$        & Gamma function                        & \\
$\delta$        & Function of $b$ and $r$               & \eq{delta} \\
$\uptheta$      & Spherical harmonic polar angle        & \eq{ylmtp} \\
$\theta$        & Rotation angle                        & Appendix~\ref{app:axisangle} \\
$\Theta$        & Heaviside step function               & \eq{biglam} \\
$\kappa$       & Angular position of
                  occultor/occulted intersection point  & \eq{kappa} \\
$\lambda$       & Angular position of
                  occultor/occulted intersection point  & \eq{lambda} \\
$\Lambda$       & \citet{MandelAgol2002} function       & \eq{biglam} \\
$\mu$           & $l - m$                               & \eq{munu} \\
$\upmu$         & Limb darkening radial parameter       & \eq{quadraticld} \\
$\nu$           & $l + m$                               & \eq{munu} \\
$\Pi(\bigdot,\bigdot)$
                & Complete elliptic integral of the
                  third kind                            & \eq{elliptic} \\
$\upphi$        & Spherical harmonic azimuthal angle    & \eq{ylmtp} \\
$\phi$          & Angular position of
                  occultor/occulted intersection point  & \eq{phi} \\
$\varphi$       & Dummy integration variable            & \\
$\omega$        & Angular position of occultor          & \eq{zrot}
\end{longtable}
\end{center}

\end{document}